\documentclass[10pt, twoside, a4paper]{article}

\usepackage[dvips, pdftex]{graphicx}
\usepackage[left=2.cm,top=2.cm,right=2.cm,bottom=1.9cm]{geometry}  
\usepackage{amsmath, amsthm, amssymb}
\usepackage[backref=page]{hyperref}
\usepackage[dvipsnames]{xcolor} 
\usepackage[normalem]{ulem} 
\usepackage{aas_macros}     
\usepackage{natbib}         

\setcitestyle{authoryear,open={(},close={)}} 

\newcommand{\DeltaLOD}{\Delta\!L\!O\!D}
\newcommand{\PO}{{P\!O}}
\newcommand{\GR}{\mathrm{GR}}
\makeatletter
\newcommand*\bull{\mathpalette\bull@{1.1}}
\newcommand*\bull@[2]{\mathbin{\vcenter{\hbox{\scalebox{#2}{$\m@th#1\bullet$}}}}}
\makeatother

\begin{document}
 \title{Mars orientation and rotation angles}
\author{Marie Yseboodt$^1$, Rose-Marie Baland$^1$, S\'ebastien Le Maistre$^{1,2}$ \\ 
$^1$ Royal Observatory of Belgium, Avenue circulaire 3, Brussels, Belgium,\\
$^2$ UCLouvain, Louvain-la-Neuve, Belgium\\
m.yseboodt@oma.be\\ \\
In press, Celestial Mechanics and Dynamical Astronomy, 2023}
\maketitle

\tableofcontents

\section*{Abstract}
 
The rotation and orientation of Mars is commonly described using two different sets of angles, namely (1) the Euler angles with respect to the Mars orbit plane and (2) the right ascension, declination, and prime meridian location angles with respect to the Earth equator at J2000 (as adopted by the IAU). 
We propose a formulation for both these sets of angles, which consists of the sum of a second degree polynomial and of periodic and Poisson series. Such a formulation is shown here to enable accurate (and physically sound) transformation from one set of angles to the other. The transformation formulas are provided and discussed in this paper. In particular, we point that the quadratic and Poisson terms are key ingredients to reach a transformation precision of $0.1$ mas, even $30$ years away from the reference epoch of the rotation model (e.g.~J2000). Such a precision is required to accurately determine the smaller and smaller geophysical signals observed in the high-accuracy data acquired from the surface of Mars. 

In addition, we present good practices to build an accurate Martian rotation model over a long time span ($\pm 30$ years around J2000) or over a shorter one (e.g.~lifetime of a space mission). We recommend to consider the J2000 mean orbit of Mars as the reference plane for Euler angles. An accurate rotation model should make use of up-to-date models for the rigid (this study) and liquid \citep{Lem23} nutations, relativistic corrections in rotation \citep{Bal23}, and polar motion induced by the external torque (this study).

Our transformation model and recommendations can be used to define the future IAU solution for the rotation and orientation of Mars using right ascension, declination, and prime meridian location. In particular, thanks to its quadratic terms, our transformation model does not introduce arbitrary and non-physical terms of very long period and large amplitudes, thus providing unbiased values of the rates and epoch values of the angles.

\section{Introduction}
First observed from Earth by astronomers likes Huygens, Cassini or Herschel, the rotation of Mars has been more and more accurately determined during the last centuries. While the historical observations from telescopes only revealed the diurnal rotation and mean obliquity of Mars, the recent measurements provided by landers (e.g Viking, Pathfinder) and orbiters (e.g.~Mars Global Surveyor, Mars Odyssey) operating at Mars allowed an extremely accurate monitoring of the rotational motion of Mars, revealing its small wobbles and long trends. 
As a consequence, the Mars rotation models have been improved regularly during the last four decades, see for example \citet{Rea79, Yod97, Fol97a, Fol97b, Yod03, Roo99, Kon06, Kon11, Kon16, Kon20, Kuc14} or \citet{Bal20}. Over the past five years, the rotation of Mars has generated even more interest in the scientific community. The NASA InSight mission \citep{Ban20} has accumulated since its landing on Mars in Nov.~26th 2018, a large amount of data from RISE, a radioscience experiment designed to measure the tiniest variations in the rotation of the planet \citep{Fol18}. This enthusiasm for the rotational dynamics of Mars is also shared by Europe, which has an instrument, named LaRa \citep{Deh20}, similar to RISE and ready to fly \citep{Lem22}, which we will try to place at the surface of Mars in the coming years.
In order to interpret the rotation data from these extremely precise radioscience experiments (the frequencies at about 8.4 GHz are measured on Earth stations with a precision of less than 0.002 Hz), a very precise theoretical rotation model for Mars is needed. 
\\

Two different sets of angles are commonly used in the rotation matrix transforming the coordinates from the Martian body frame (BF) to an inertial frame (IF, e.g.~the ICRF J2000).
In this paper, we refer to them as to the \textit{IAU angles} and the \textit{Euler angles}. The first two Euler angles are the obliquity and the node longitude that define the orientation of the spin axis in space. These two angles are similarly defined for the planets and for the Earth, and commonly used in theoretical rotation modeling.
The IAU (Internal Astronomical Union) Working Group on Cartographic Coordinates and Rotational Elements recommends to use the equatorial coordinates (right ascension and declination) to describe the orientation of the spin axis of a celestial body with respect to the Earth equator, whereas radio data scientists use one or the other set.
In both angle sets, a third angle, named herein the rotation angle, defines the location of the prime meridian.
\\

Because different software used to analyse radioscience data do not all provide the same set of angle estimates, an accurate transformation between the two sets of angles is needed to properly compare different solutions of Mars rotation.
For instance, the Jet Propulsion Laboratory (JPL) MONTE program \citep{Eva18} deals with Euler angles while the Centre National d'Etudes Spatiales (CNES) GINS program \citep{Mar11, Lem13} deals with IAU angles.
The current IAU rotation model \citep{Arc18} comes from the rotation model proposed by \citet{Kuc14}. These authors inferred the IAU angles from their Euler angles estimates, following a method first used by \citet{Jac10} which involves first converting the periodic series of Euler angles estimate in the time domain, then transforming them into IAU angles times series, before fitting the IAU angles periodic series based on a frequency analysis of the time series. 
As a result, a non-physical term with a very long period of $\sim$71,000 years appears in each angle of their IAU solution. 
In addition to the estimation errors on the fitted parameters (e.g.~the secular rates, the periodic variations in rotation angles) and the modeling errors on the fixed parameters (e.g.~the rigid nutations), the accuracy of the IAU solution also depends on the transformation errors. Due to these errors, the accuracy of the IAU solution is currently low.
\\

In this paper we explain how to precisely convert the orientation and rotation angles from one set to the other using spherical geometry. The targeted precision of the transformation is 0.1 mas or smaller for each angle on an interval of about 30 years before and after J2000.
We choose this precision level because it is one order of magnitude smaller than the current precision from the observations (a few mas), in order to limit the error coming from the theory.
Our transformation method still guaranties a precision of 0.3 mas before and after 100 years.
$1$ mas corresponds to about 1.6 cm on the Martian surface at the equator.
Following e.g.~\citet{Bal20} formalism, we describe each angle as the sum of their epoch value, a linear term (associated to a precession rate or diurnal rate), a quadratic term, periodic variations (nutations or length-of-day variations), and a Poisson series (a periodic series with amplitudes changing linearly with time). To achieve such precision, we have to consider in our transformation the quadratic terms and the Poisson series of the angles, presently neglected in the studies based on actual estimates. 
\\

The paper is organised as follows. 
In Section 2, we define the angles and rotation matrices. 
In Sections 3 and 4, we provide the transformation between the orientation and rotation angles, respectively. 
In Section 5, we discuss the addition of polar motion terms in the rotation matrices. 
Section 6 gathers recommendations for good practices to accurately build a rotation model, displaying some numerical values of the angles for illustrative purposes. 
Section 7 presents our recommendation for the definition of the future IAU solution. 
A discussion and conclusion are given in Section 8. 
Note that we do not provide any new entire rotation solution as we do not perform any data analysis here, but we correct some part of existing rotation models that were not computed accurately.
Note also that we provide a code based on the equations of this paper, allowing to transform a set of angles into the other.

\section{Definition of the angles and the rotation matrices} 
\label{sec_definition}

\begin{figure}[!htb]
\centering
\includegraphics[height=9cm, width=15cm]{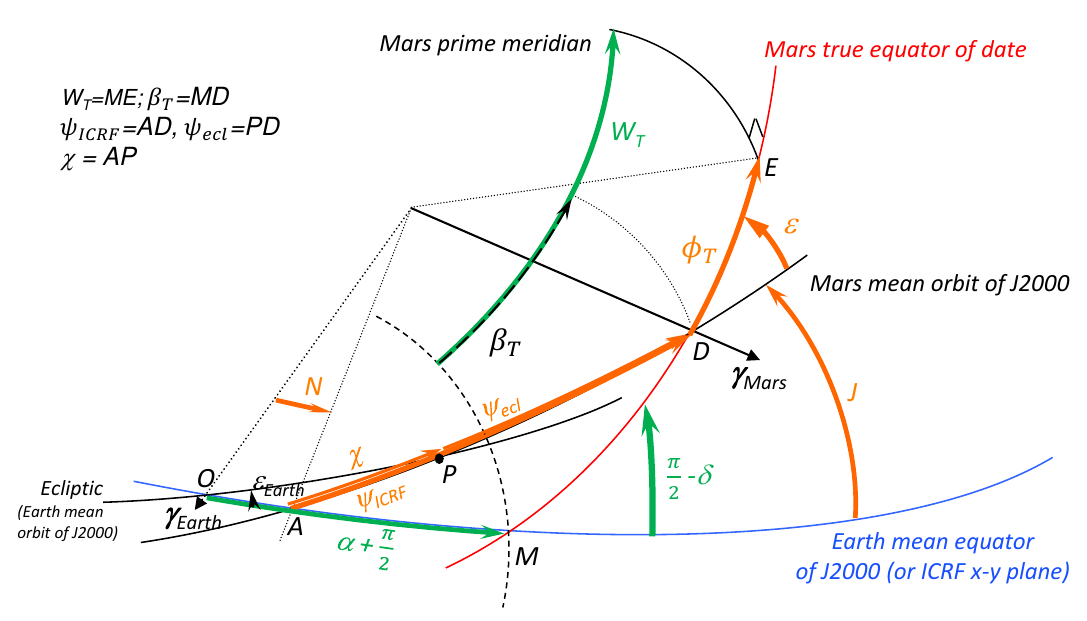}  
\caption{Mars orientation with respect to the ICRF J2000 using IAU angles ($\alpha, \delta, W_T$) or Euler angles ($\varepsilon, \psi, \phi_T$). The angles $N$ and $J$ orientate the mean orbit of Mars with respect to the ICRF J2000.}
\label{fig_ang}
\end{figure}

The transformation of a lander position in the Martian body frame $\vec r$ to the lander position in the inertial frame $\vec R$ is:
\begin{equation}
\vec R = \mathbf{M}_{BF\rightarrow IF} \; \vec r 
\label{eq_landerpos}
\end{equation}
with $\mathbf{M}_{BF\rightarrow IF}$ the rotation matrix from the body frame (BF) to the inertial frame (IF). The body frame is attached to the planet and centered at the center of mass of Mars.
Its $x$ and $y-$axes define the equator of figure of Mars, and its $z-$axis defines the figure polar axis.
The inertial frame is also centered at the center of mass but its equatorial plane is the Earth mean equator at the J2000 reference epoch (or the International Celestial Reference Frame ICRF equator).
The IF $X-$axis points to the ascending node of the ecliptic on the Earth equator (i.e.~to the vernal point $\gamma_{Earth}$ on Fig.~\ref{fig_ang}). 
\\

The transformation can be decomposed into successive transformations from the BF to a frame attached to the AM (Angular Momentum) axis and from the AM frame to the IF:
\begin{eqnarray}
\vec R &=& \mathbf{M}_{AM\rightarrow IF} \cdot \mathbf{M}_{BF\rightarrow AM} \; \vec r \label{eq_landerpos1}\\
\mathbf{M}_{BF\rightarrow AM} &=& R_X(Y_P) \cdot R_Y(X_P)
\label{eq_landerpos2}
\end{eqnarray}
$R_X$, $R_Y$ and $R_Z$ are elementary rotation matrices (see e.g.~Eqs.~(4-6) of \citealt{Yse17}).
The AM axis is very close to the spin axis, the difference is smaller than $0.01$ mas \citep{Bou99, Roo99}, so no distinction is made between them in this paper.
In the first part of this paper, we neglect the polar motion components $X_P$ and $Y_P$ (the motion of the spin axis in the BF).
The difference between the AM and figure axes can be of a few mas and is discussed in Section \ref{sec_PM} as well as the way to take that properly into account in the rotation matrix.
We detail below the two usual approaches to write the transformation matrix $\mathbf{M}=\mathbf{M}_{AM\rightarrow IF}$.

\subsection{The rotation matrix using the Euler angles}
\label{sec_rotmatrix_Euler}

The first approach uses the Euler angles, as in \citet{Fol97a} or \citet{Kon06}, that orientate a celestial body with respect to the body's mean orbit at a reference epoch (usually 1980 for Mars, as a legacy of the Viking missions, though the J2000 epoch would be better now that more than 40 years of lander observations are dispersed around J2000, see Section \ref{Section_deg2poly} for a detailed discussion).
The orientation of the $z-$axis is defined by the obliquity $\varepsilon$ and the node longitude $\psi_{ICRF}$. The prime meridian ($x$-axis of BF) is positioned by $\phi_T$, see Fig.~\ref{fig_ang}.
Using the Euler angles in the rotation matrix is very convenient, since the models (e.g.~\citealt{Rea79, Roo99, Bal20}) for the precession and nutations caused by the gravitational torques exerted on Mars generally refer the orientation of the planet to its own orbit, with respect to which the obliquity is almost constant over time.
Similarly, the models for the variation in the rotation angles refer the rotation to the node of Mars' equator on Mars orbit.
With the Euler angles, the rotation matrix is expressed as:
\begin{equation}
\mathbf{M}_{\varepsilon\psi\phi} = 
R_Z(-N) \cdot R_X(-J) \cdot R_Z(-\psi_{ICRF}) \cdot R_X(-\varepsilon) \cdot R_Z(-\phi_T).
\label{eq_ep}
\end{equation}
The last three rotations use the Euler angles.
The first two rotations are for the transformation from the body's mean orbit of epoch to the ICRF J2000.
The angle $N$ ($\gamma_{Earth} A$ in Fig.~\ref{fig_ang}) is defined from the vernal equinox 
to the node of the Mars mean orbit of epoch on the Earth mean equator at J2000.
The angle $J$ is the inclination of the Mars mean orbit of epoch relative to the Earth mean equator at J2000.
The angles $N$ and $J$ are constant over time.
\\

The obliquity $\varepsilon$ is the inclination between the Mars true equator of date and the Mars mean orbit of epoch, as defined unambiguously in the literature. Contrariwise, the node longitude angle $\psi$ is not defined the same way by all authors. 
$\psi_{ICRF}$, as used here, is the angle from the node of the Mars mean orbit of epoch on the J2000 Earth mean equator to the node of the Mars true equator of date on Mars mean orbit of epoch (angle $AD$ on figure \ref{fig_ang}).
It is the angle used by \citet{Kon06, Kon16} and its numerical value is approximately $81.975^\circ$ (see Section \ref{sec_num} for its exact value).
It is also possible to define $\psi_{ecl}$ as the angle from the node of the Mars mean orbit of epoch on the J2000 Earth ecliptic to the node of the Mars true equator of date on Mars mean orbit of epoch (the angle $P\!D$ on Fig.~\ref{fig_ang}).
This is the angle used e.g.~by \citet{Fol97a} and in \citet{Bal20}.
Its numerical value is approximately $35.497^\circ$.
If $\chi$ is the angle along the Mars mean orbit of epoch from the node with the J2000 Earth mean equator to the node with the J2000 Earth ecliptic (the angle $A\!P$ on Fig.~\ref{fig_ang}), $\psi_{ICRF}$ is the sum:
\begin{equation}
\psi_{ICRF} = \psi_{ecl} + \chi
\label{psi}
\end{equation}
Since the angle $\chi$ is constant over time, the temporal behavior (precession, nutations, quadratic and Poisson terms) of both $\psi_{ICRF}$ and $\psi_{ecl}$ is the same. In the following sections, we omit the subscript in $\psi_{ICRF}$.
\\

The numerical values of the constant angles $J, N,$ and $\chi$ are easily obtained in the triangle $O\!A\!P$ from the following equality between rotation matrices from ICRF to Mars mean orbit of epoch
\begin{equation}
R_Z(\chi) \cdot R_X(J) \cdot R_Z(N) = R_X(i_0) \cdot R_Z(\Omega_0) \cdot R_X(\varepsilon_{Earth})
\label{eqinertpl}
\end{equation}
with node longitude $\Omega_0$ (the angle $O\!P$ on Fig.~\ref{fig_ang}) and inclination $i_0$ describing the orientation of the mean orbit of Mars at epoch with respect to the J2000 Earth ecliptic.
$\varepsilon_{Earth}$ is the Earth obliquity at J2000. We refer the reader to Section \ref{Section_deg2poly} and Table \ref{tab_orbit19802000} for numerical values.
\\

In this paper, unless otherwise specified, we choose to consider only angles in the prograde direction and ascending nodes of moving planes with respect to fixed planes.
This convention differs from the one chosen by some authors who define the obliquity and/or node longitude as retrograde angles, following the habits of the Earth's rotation community (e.g.~\citealt{Roo99}).
\\

The rotation angle $\phi_T$ is measured from the ascending node of the Mars true equator of date over the Mars mean orbit of epoch to the intersection of Mars Prime Meridian on the Mars true equator of date 
(angle $D\!E$).
It is possible to define different $\phi$ angles depending on which body equator (mean, true of date, etc.) is considered (see Section \ref{sec_rotang}).

\subsection{The rotation matrix using the IAU angles}
\label{sec_rotmatrix_IAU}

Alternatively to the Euler angles, the orientation of the planetary rotation axis can also be defined by the two angles $\alpha$ (right ascension) and $\delta$ (declination) relative to the inertial Earth mean equator at J2000 (see green angles on Fig.~\ref{fig_ang}). 
The rotation around the $z-$axis is defined by the angle $W_T$ that locates the prime meridian of the body. It is measured easterly along the Mars true equator of date from the ascending node of the Mars true equator of date over the Earth mean equator at epoch and to the prime meridian.
Using IAU angles, the rotation matrix is expressed as
\begin{equation}
\mathbf{M}_{\alpha\delta W} = R_Z(-\frac{\pi}{2} - \alpha) \cdot R_X(-\frac{\pi}{2} + \delta) \cdot R_Z(-W_T)
\label{eq_ad}
\end{equation}
This expression follows IAU conventions (see \citealt{Arc18}).
\\

As for the Euler rotation angle $\phi$, it is possible to define different rotation $W$ angles depending on which equator is considered (see Section \ref{sec_rotang}). The difference between the rotation angles $W_T$ and $\phi_T$ is the angle $\beta_T$ corresponding to the angle MD on Fig.~\ref{fig_ang}. 
It is defined along the Mars true equator of date from the node with the Earth mean equator of J2000 to the node with the Mars mean orbit of epoch. 
\begin{equation}
W_T = \phi_T + \beta_T
\label{eq_W}
\end{equation}
The angle $\beta_T$ is not constant with time because of the motion of the Mars true equator of date (see Section \ref{sec_rotang} for more details).

\section{Transformation between the orientation angles \texorpdfstring{$\alpha$}{alpha}, \texorpdfstring{$\delta$}{delta} and \texorpdfstring{$\varepsilon$}{epsilon}, \texorpdfstring{$\psi$}{psi}} 
\label{sec_transf}

In this section, we describe the accurate transformation between the Euler and IAU orientation angles.
This transformation is independent of the rotation angles and is needed for operational reasons, as explained in the introduction.

\subsection{Exact transformations}
\label{sec_exacttransfo}

Since the two rotation matrices $\mathbf{M}_{\varepsilon\psi\phi}$ and $\mathbf{M}_{\alpha\delta W}$ (Eqs.~\ref{eq_ep} and \ref{eq_ad}) are equal, and using Eq.~(\ref{eq_W}), we can write: 
\begin{equation}
R_X(-J) \cdot R_Z(-\psi) \cdot R_X(-\varepsilon) = R_Z(N-\frac{\pi}{2} - \alpha) \cdot R_X(-\frac{\pi}{2} + \delta) \cdot R_Z(-\beta_T)
\label{eq_equality}
\end{equation}
By multiplying each side of last equation by the unit vector $(0,0,1)$, we obtain exact trigonometric relations from the right ascension $\alpha$ and declination $\delta$ to the obliquity $\varepsilon$ and node longitude $\psi$:
\begin{eqnarray}
\cos\varepsilon &=& \sin\delta \cos J + \cos\delta \sin J \sin(N-\alpha)
\label{eq_ce} \\
\sin\varepsilon \cos\psi &=& \cos\delta \cos J \sin(N-\alpha) - \sin\delta \sin J
\label{eq_cp} \\
\sin\varepsilon \sin\psi &=& \cos\delta \cos(N-\alpha)
\label{eq_sp} 
\end{eqnarray}
and conversely from $(\varepsilon,\psi)$ to $(\alpha,\delta)$:
\begin{eqnarray}
\sin\delta &=& \cos\varepsilon \cos J - \sin\varepsilon \sin J \cos\psi 
\label{eq_sde} \\
\cos\delta \cos(N-\alpha) &=& \sin\varepsilon \sin\psi
\label{eq_ca} \\
\cos\delta \sin(N-\alpha) &=& \cos\varepsilon \sin J + \cos J \sin\varepsilon \cos\psi
\label{eq_sa}
\end{eqnarray}
in which only the constant over time angles $N$ and $J$ are needed as additional information, since the rotation $R_Z(-\beta_T)$ simplifies to identity when multiplied by the vector $(0,0,1)$.
\\

These exact trigonometric relations can be used to exactly convert times series for the orientation angles.
In particular, they must be used to transform the angles at their epoch values ($\varepsilon_0$, $\psi_0$, $\alpha_0,$ and $\delta_0$, the value of $\varepsilon_0$ and $\psi_0$ being dependent on the choice of the reference orbit, see Section \ref{Section_deg2poly}). 
However, we aim to express each angle as a sum of different parts, including periodic series, and for that purpose, the trigonometric exact relations end up in long and not practical analytical expressions.
In particular, the representation in series for the nutation is lost.
We present below approximate but accurate transformations which preserve the usual representation of the angles.
\\

Even though $\beta_T$ does not intervene in the transformations above, we provide here expressions which will help to express the approximate transformation of Section~\ref{sec_app}. 
$\beta_T$, as defined in Eq.~(\ref{eq_W}), is on the side of the spherical triangle ADM defined by Mars true equator of date, and can be written as 
\begin{subequations}
\label{eq_D} 
\begin{eqnarray}
\cos\beta_T &=& \cos J \sin\psi \cos(N-\alpha) + \cos\psi \sin(N-\alpha)
\label{eq_cD} \\
\sin\beta_T &=& \frac{\sin J \sin\psi}{\cos\delta} 
\label{eq_sD}
\end{eqnarray}
\end{subequations}

\subsection{Approximate transformations} 
\label{sec_app}

A practical form for each orientation angles as function of the time is a sum of epoch values, of linear and quadratic terms, of a series of periodic (nutation) terms, and of a Poisson series (a periodic series with amplitudes changing linearly with time): 
\begin{subequations}
\label{eq_vartemp}
\begin{alignat}{4}
\varepsilon(t) &= \varepsilon_0 && + \dot\varepsilon_0 \, t + \Delta\varepsilon(t) &&
+ \varepsilon_Q \, t^2 + \varepsilon_\PO(t)
\label{eq_epst} \\
\psi(t) &= \psi_0 && + \dot\psi_0 \, t + \Delta\psi(t) &&
+ \psi_Q \, t^2 + \psi_\PO(t)
\label{eq_psit} \\
\alpha(t) &= \alpha_0 && + \dot\alpha_0 \, t + \Delta\alpha(t) &&
+ \alpha_Q \, t^2 + \alpha_\PO(t)
\label{eq_alphat} \\
\delta(t) &= \delta_0 && + \underbrace{\dot\delta_0 \, t + \Delta\delta(t)}_{\textrm{First order}} && 
+ \underbrace{\delta_Q \, t^2 + \delta_\PO(t)}_{\textrm{Second order}}
\label{eq_deltat}
\end{alignat}
\end{subequations} 
where $t$ is the time, starting in J2000.
In this modelization, secular and nutations terms are first order variations while quadratic terms and Poisson series are of second order.
For Mars, considering all those terms ensures the required precision ($<0.1$ mas on the interval $\pm 30$ years around J2000) on the transformations between the angles and on the definition of a rotation model (see Section \ref{sec_num}).
\\

The rate $\dot\psi_0$ is the precession rate in longitude while $\dot\varepsilon_0$ is the secular change in the Mars obliquity relative to the Mars mean orbit of epoch, whereas $\dot\alpha_0$ and $\dot\delta_0$ are the rates in equatorial coordinates.
Numerical values are provided in Section \ref{sec_num}, for illustration. \\

The nutations are the periodic variations of the orientation of the spin axis in space, as a response to the external gravitational torque. Rigid nutations series in Euler angles $\Delta\varepsilon$ and $\Delta\psi$ have been computed by different authors, see Section \ref{sec_rigidnut}. For a non-rigid planet, a transfer function taking into account the internal properties of Mars is applied to the rigid nutations (see Section \ref{sec_nonri}). As the nutation, the Euler quadratic coefficients $\varepsilon_Q$ and $\psi_Q$ and the Poisson series $\varepsilon_\PO(t)$ and $\psi_\PO(t)$ can be modeled as a response to the external torque. According to \citet[equations (85a) and (85b), and Tab.~10]{Bal20}, they are second order effects, resulting from the mean precession of the Body Frame of Mars and from the Poisson terms of the chosen ephemerides (which account for the non Keplerian motion of the planets). 
The quadratic coefficients and Poisson amplitudes are small (of the order of $0.01$~mas/y$^2$, and of $0.05$~mas/y), but can change the angles by a few mas $30$ years away from J2000 and therefore can not be neglected. 
The quadratic coefficients multiplied by a factor 2 correspond to acceleration of the angles. 
Nutations, quadratic and Poisson terms in $\alpha(t)$ and $\delta(t)$ appear after transformation (see below).
\\

To obtain the approximate expressions for $\alpha(t)$ and $\delta(t)$, assuming we have approximate expressions for $\varepsilon(t)$ and $\psi(t)$, we inject the expressions of Eqs.~(\ref{eq_epst}-\ref{eq_psit}) in the exact trigonometric expressions (\ref{eq_sde}-\ref{eq_sa}) of Section~\ref{sec_exacttransfo}, 
and express Eqs~(\ref{eq_alphat}-\ref{eq_deltat}) correct up to the second order in small terms, under the form:
\begin{subequations}
\label{eq_ad_Gamma}
\begin{eqnarray}
\alpha(t) &=& \alpha_0
+ \Gamma_{\alpha\varepsilon}\, (\dot\varepsilon_0 \, t + \Delta\varepsilon) 
+ \Gamma_{\alpha\psi} \, (\dot\psi_0 \, t + \Delta\psi) 
+ \Gamma_{\alpha\varepsilon} \, (\varepsilon_Q \, t^2 + \varepsilon_\PO) 
+ \Gamma_{\alpha\psi}\, (\psi_Q \, t^2 + \psi_\PO) \nonumber \\
&& + \Gamma_{\alpha\varepsilon\varepsilon}\,(\dot\varepsilon_0^2 \, t^2 + 2\, \Delta\varepsilon\, \dot\varepsilon_0 \, t \, + \Delta\varepsilon^2)
+ \Gamma_{\alpha\varepsilon\psi}\,(\dot\varepsilon_0 \, \dot\psi_0 \, t^2 + \Delta\varepsilon\, \dot\psi_0 \, t + \Delta\psi\, \dot\varepsilon_0 \, t + \Delta\varepsilon\, \Delta\psi) \nonumber \\
&& + \Gamma_{\alpha\psi\psi}\, (\dot\psi_0^2 \, t^2 + 2 \, \Delta\psi\, \dot\psi_0 \, t \, + \Delta\psi^2) 
\label{eq_alphat_Gamma} \\
\delta(t) &=& \delta_0 
+ \Gamma_{\delta\varepsilon}\, (\dot\varepsilon_0 \, t + \Delta\varepsilon) 
+ \Gamma_{\delta\psi}\, (\dot\psi_0 \, t + \Delta\psi)
+ \Gamma_{\delta\varepsilon}\, (\varepsilon_Q \, t^2 + \varepsilon_\PO)
+ \Gamma_{\delta\psi}\, (\psi_Q \, t^2 + \psi_\PO) 
\nonumber \\
&& + \Gamma_{\delta\varepsilon\varepsilon}\,(\dot\varepsilon_0^2 \, t^2 + 2 \, \Delta\varepsilon\, \dot\varepsilon_0 \, t \, + \Delta\varepsilon^2)
+ \Gamma_{\delta\varepsilon\psi}\,(\dot\varepsilon_0 \, \dot\psi_0\,t^2 + \Delta\varepsilon\, \dot\psi_0 \, t + \Delta\psi\, \dot\varepsilon_0 \, t + \Delta\varepsilon \, \Delta\psi) \nonumber \\
&& + \Gamma_{\delta\psi\psi}\, (\dot\psi_0^2 \, t^2 + 2\, \Delta\psi\, \dot\psi_0 \, t \, + \Delta\psi^2)
\label{eq_deltat_Gamma} 
\end{eqnarray} 
\end{subequations}
with the conversion factors:
{\allowdisplaybreaks
\begin{subequations}
\label{eq_Gammaad}
\begin{alignat}{3}
\Gamma_{\alpha\varepsilon} &= \frac{\sin\psi_0 \sin J}{\cos^2 \delta_0} 
&=& \, \frac{\sin\beta_0}{\cos\delta_0}, 
\label{eq_gamma} \\
\Gamma_{\alpha\psi} &= \frac{\cos J -\sin\delta_0 \cos\varepsilon_0}{\cos^2\delta_0}
&=& \, \frac{\sin\varepsilon_0 \cos\beta_0}{\cos\delta_0} \\
\Gamma_{\delta\varepsilon} &= 
\frac{- \cos J + \cos\varepsilon_0\sin\delta_0}{\cos\delta_0\sin\varepsilon_0} \; 
&=& \, -\cos\beta_0\\
\Gamma_{\delta\psi} &= \frac{\sin\varepsilon_0 \sin J \sin\psi_0}{\cos\delta_0} 
&=& \, \sin\varepsilon_0 \sin\beta_0 
\label{eq_gammab} \\
\Gamma_{\alpha\varepsilon\varepsilon} &= -\frac{\sin\beta_0 \cos\beta_0 \sin\delta_0} {\cos^2\delta_0} && 
\label{eq_gammac} \\
\Gamma_{\alpha\varepsilon\psi} &= \frac{\sin J (2 \cos\beta_0 \sin(N - \alpha_0) - \cos\psi_0)} {\cos^2\delta_0} && \\
\Gamma_{\alpha\psi\psi} &=
\frac{\sin\beta_0 \sin\varepsilon_0 (2 \cos\beta_0 \sin\delta_0 \sin\varepsilon_0 - \cos\delta_0 \cos\varepsilon_0)} {2 \cos^2\delta_0} && \\
\Gamma_{\delta\varepsilon\varepsilon} &= - \frac{\sin^2\beta_0 \sin\delta_0}{2\cos\delta_0} && \\
\Gamma_{\delta\varepsilon\psi} &= \frac{\sin\beta_0 \sin J \sin(N-\alpha_0)}{\cos\delta_0} && \\
\Gamma_{\delta\psi\psi} &= \frac{\cos\beta_0 \sin J \sin\varepsilon_0 \sin(N - \alpha_0)} {2 \cos\delta_0} &&
\label{eq_gammad}
\end{alignat}
\end{subequations}
}

To obtain the opposite transformation, from approximate expressions for $\alpha(t)$ and $\delta(t)$ to approximate expressions for $\varepsilon(t)$ and $\psi(t)$, we inject the expressions of Eqs.~(\ref{eq_alphat}-\ref{eq_deltat}) in the exact trigonometric expressions (\ref{eq_ce}-\ref{eq_sp}) of Section~\ref{sec_exacttransfo}, and express Eqs.~(\ref{eq_epst}-\ref{eq_psit}) correct up to the second order in small terms, as:
\begin{subequations}
\label{eq_ep_Gamma}
\begin{eqnarray}
\varepsilon(t) &=& \varepsilon_0 
+ \Gamma_{\varepsilon\alpha}\, (\dot\alpha_0 \, t + \Delta\alpha) 
+ \Gamma_{\varepsilon\delta}\, (\dot\delta_0 \, t + \Delta\delta) 
+ \Gamma_{\varepsilon\alpha}\, (\alpha_Q \, t^2 + \alpha_\PO)
+ \Gamma_{\varepsilon\delta}\, (\delta_Q \, t^2 + \delta_\PO) 
\nonumber \\
&& + \Gamma_{\varepsilon\alpha\alpha}\, (\dot\alpha_0^2 \, t^2 + 2 \, \Delta\alpha\, \dot\alpha_0 \, t \, + \Delta\alpha^2)
+ \Gamma_{\varepsilon\alpha\delta}\,(\dot\delta_0 \, \dot\alpha_0 \, t^2 + \Delta\delta\, \dot\alpha_0 \, t + \Delta\alpha\, \dot\delta_0 \, t + \Delta\delta \, \Delta\alpha) 
\nonumber \\
&& + \Gamma_{\varepsilon\delta\delta}\,(\dot\delta_0^2 \, t^2 + 2 \, \Delta\delta \, \dot\delta_0 \, t \, + \Delta\delta^2)
\label{eq_epst_Gamma} \\
\psi(t) &=& \psi_0 
+ \Gamma_{\psi\alpha} \, (\dot\alpha_0 \, t + \Delta\alpha)
+ \Gamma_{\psi\delta} \, (\dot\delta_0 \, t + \Delta\delta)
+ \Gamma_{\psi\alpha} \, (\alpha_Q \, t^2 + \alpha_\PO) 
+ \Gamma_{\psi\delta} \, (\delta_Q \, t^2 + \delta_\PO) 
\nonumber \\
&& + \Gamma_{\psi\alpha\alpha} \, (\dot\alpha_0^2 \, t^2 + 2 \, \Delta\alpha\, \dot\alpha_0 \, t \, + \Delta\alpha^2)
+ \Gamma_{\psi\alpha\delta} \, (\dot\delta_0 \, \dot\alpha_0 \, t^2 + \Delta\delta\, \dot\alpha_0 \, t + \Delta\alpha \, \dot\delta_0 \, t + \Delta\delta \, \Delta\alpha) 
\nonumber \\
&& + \Gamma_{\psi\delta\delta} \, (\dot\delta_0^2 \, t^2 + 2 \, \Delta\delta\, \dot\delta_0 \, t \, + \Delta\delta^2)
\label{eq_psit_Gamma} 
\end{eqnarray} 
\end{subequations}
with the opposite conversions factors
\begin{subequations}
\label{eqinvtot} 
\begin{eqnarray}
\Gamma_{\varepsilon\alpha} &=& \cos\delta_0 \sin\beta_0 
\label{eqinv} \\
\Gamma_{\varepsilon\delta} &=& -\cos\beta_0 \\
\Gamma_{\psi\alpha} &=& \frac{\cos\beta_0 \cos\delta_0}{\sin\varepsilon_0} \\
\Gamma_{\psi\delta} &=& \frac{\sin\beta_0}{\sin\varepsilon_0}
\label{eqinvb} \\
\Gamma_{\varepsilon\alpha\alpha} &=& \frac{\cos\beta_0 \cos\delta_0 \sin J \cos\psi_0}{2 \sin\varepsilon_0} 
\label{eqinv2} \\
\Gamma_{\varepsilon\alpha\delta} &=& \frac{\sin\beta_0 \sin J \cos\psi_0}{\sin\varepsilon_0} \\
\Gamma_{\varepsilon\delta\delta} &=& \frac{\sin^2\beta_0 \cos\varepsilon_0}{2 \sin\varepsilon_0} \\
\Gamma_{\psi\alpha\alpha} &=& \frac{\cos\delta_0 \sin\beta_0 (\sin\delta_0 \sin\varepsilon_0 -2 \cos\beta_0 \cos\delta_0 \cos\varepsilon_0)}{2 \sin^2\varepsilon_0} \\
\Gamma_{\psi\alpha\delta} &=& \frac{\sin J (\sin(N-\alpha_0) - 2 \cos\varepsilon_0 \sin\psi_0 \sin\beta_0)} {\sin^2\varepsilon_0} \\
\Gamma_{\psi\delta\delta} &=& \frac{\sin\beta_0 \cos\beta_0 \cos\varepsilon_0} {\sin^2\varepsilon_0}.
\end{eqnarray}
\end{subequations}

By identification between Eqs.~(\ref{eq_alphat}-\ref{eq_deltat}) and (\ref{eq_ad_Gamma}), the transformation from $(\varepsilon, \psi)$ to $(\alpha, \delta)$ is obtained as
\begin{subequations}
\label{eq_alphadelta}
\begin{eqnarray}
\lbrace \dot\alpha_0, \, \Delta\alpha\rbrace &=& \Gamma_{\alpha\varepsilon} \, \lbrace \dot\varepsilon_0, \, \Delta\varepsilon \rbrace 
+ \Gamma_{\alpha\psi} \, \lbrace \dot\psi_0, \, \Delta\psi \rbrace 
\label{eq_del1} \\
\lbrace \dot\delta_0, \, \Delta\delta\rbrace &=& \Gamma_{\delta\varepsilon} \, \lbrace \dot\varepsilon_0, \, \Delta\varepsilon \rbrace 
+ \Gamma_{\delta\psi} \, \lbrace \dot\psi_0, \, \Delta\psi \rbrace
\label{eq_del2} \\
\lbrace \alpha_Q, \, \alpha_\PO \rbrace &=& \Gamma_{\alpha\varepsilon} \, \lbrace \varepsilon_Q, \, \varepsilon_\PO \rbrace 
+ \Gamma_{\alpha\psi} \, \lbrace \psi_Q, \, \psi_\PO \rbrace 
\nonumber \\
&& + \Gamma_{\alpha\varepsilon\varepsilon}\, \lbrace \dot\varepsilon_0^2, \, 2 \Delta\varepsilon\, \dot\varepsilon_0 \, t \rbrace 
+ \Gamma_{\alpha\varepsilon\psi}\, \lbrace \dot\varepsilon_0 \, \dot\psi_0, \, \Delta\varepsilon\, \dot\psi_0 \, t + \Delta\psi\, \dot\varepsilon_0 \, t \rbrace 
+ \Gamma_{\alpha\psi\psi}\, \lbrace \dot\psi_0^2, \, 2 \Delta\psi\, \dot\psi_0 \, t \rbrace 
\label{eq_del3} \\
\lbrace \delta_Q, \, \delta_\PO\rbrace &=& \Gamma_{\delta\varepsilon} \, \lbrace \varepsilon_Q, \, \varepsilon_\PO \rbrace 
+ \Gamma_{\delta\psi} \, \lbrace \psi_Q, \, \psi_\PO \rbrace 
\nonumber \\
&& + \Gamma_{\delta\varepsilon\varepsilon}\, \lbrace \dot\varepsilon_0^2, \, 2 \Delta\varepsilon\, \dot\varepsilon_0 \, t \rbrace 
+ \Gamma_{\delta\varepsilon\psi}\, \lbrace \dot\varepsilon_0 \, \dot\psi_0, \, \Delta\varepsilon\, \dot\psi_0 \, t + \Delta\psi\, \dot\varepsilon_0 \, t \rbrace 
+ \Gamma_{\delta\psi\psi}\, \lbrace \dot\psi_0^2, \, 2 \Delta\psi\, \dot\psi_0 \, t \rbrace
\label{eq_del4}
\end{eqnarray}
\end{subequations}
and the opposite relations between the terms of the $\varepsilon$, $\psi$ and $\alpha$, $\delta$ angles is written as, see Eqs.~(\ref{eq_epst}-\ref{eq_psit}) and Eqs.~(\ref{eq_ep_Gamma}):
\begin{subequations}
\label{eq_epsilonpsi}
\begin{eqnarray}
\lbrace \dot\varepsilon_0, \, \Delta\varepsilon \rbrace 
&=& \; \Gamma_{\varepsilon\alpha} \, \lbrace \dot\alpha_0, \, \Delta\alpha \rbrace 
+ \Gamma_{\varepsilon\delta} \, \lbrace \dot\delta_0, \Delta\delta \rbrace \\
\lbrace \dot\psi_0, \, \Delta\psi \rbrace 
&=& \Gamma_{\psi\alpha} \, \lbrace \dot\alpha_0, \, \Delta\alpha \rbrace 
+ \Gamma_{\psi\delta} \, \lbrace \dot\delta_0, \, \Delta\delta \rbrace \\
\lbrace \varepsilon_Q, \, \varepsilon_\PO\rbrace &=& \Gamma_{\varepsilon\alpha} \, \lbrace \alpha_Q, \, \alpha_\PO \rbrace 
+ \Gamma_{\varepsilon\delta} \, \lbrace \delta_Q, \, \delta_\PO \rbrace
\nonumber \\
&& + \, \Gamma_{\varepsilon\alpha\alpha}\, \lbrace \dot\alpha_0^2, \, 2 \Delta\alpha\, \dot\alpha_0 \, t \rbrace 
+ \Gamma_{\varepsilon\alpha\delta}\, \lbrace \dot\alpha_0 \, \dot\delta_0, \, \Delta\alpha\, \dot\delta_0 \, t + \Delta\delta\, \dot\alpha_0 \, t \rbrace 
+ \Gamma_{\varepsilon\delta\delta}\, \lbrace \dot\delta_0^2, \, 2 \Delta\delta\, \dot\delta_0 \, t \rbrace 
\label{eq_eps1} \\
\lbrace \psi_Q, \, \psi_\PO\rbrace &=& \Gamma_{\psi\alpha} \, \lbrace \alpha_Q, \, \alpha_\PO \rbrace 
+ \Gamma_{\psi\delta} \, \lbrace \delta_Q, \, \delta_\PO \rbrace 
\nonumber \\
&& + \, \Gamma_{\psi\alpha\alpha}\, \lbrace \dot\alpha_0^2, \, 2 \Delta\alpha\, \dot\alpha_0 \, t \rbrace 
+ \Gamma_{\psi\alpha\delta}\, \lbrace \dot\alpha_0 \, \dot\delta_0, \, \Delta\alpha\, \dot\delta_0 \, t + \Delta\delta\, \dot\alpha_0 \, t \rbrace 
+ \Gamma_{\psi\delta\delta}\, \lbrace \dot\delta_0^2, \, 2 \Delta\delta\, \dot\delta_0 \, t \rbrace 
\label{eq_psi1}
\end{eqnarray}
\end{subequations}

In the expressions for the conversion factors (Eqs.~\ref{eq_ad_Gamma} and \ref{eqinvtot}), the $\varepsilon$, $\psi$, $\alpha$, $\delta$ and $\beta_T$ angles are evaluated at epoch.
The conversion factors are therefore constant over time but their value depend on the reference orbit chosen (see Section \ref{Section_deg2poly}).
\\

In both transformations (Eqs.~\ref{eq_alphadelta} and \ref{eq_epsilonpsi}), we have neglected the terms proportional to the product of nutation terms, as they are very small (smaller than second order terms), and do not significantly affect the transformations (they change the angles by less than 0.001 mas over 20 years).
\\

The conversion factors $\Gamma_{\alpha\varepsilon}, \Gamma_{\alpha\psi}, \Gamma_{\delta\varepsilon}$ and $\Gamma_{\delta\psi}$ of Eqs.~(\ref{eq_gamma}-\ref{eq_gammab}) or $\Gamma_{\varepsilon\alpha}, \Gamma_{\varepsilon\delta}, \Gamma_{\psi\alpha}$ and $\Gamma_{\psi\delta}$ of Eqs.~(\ref{eqinv}-\ref{eqinvb}) are sufficient to define a first order transformation, but not to define a second order transformation. An interesting consequence is that even if an input set of angles does not includes quadratic and Poisson terms, the output set will. 
The transformation of quadratic and Poisson terms described in Eqs.~(\ref{eq_alphadelta}) and (\ref{eq_epsilonpsi}) corrects the transformation given in Eq.~(24) of \citet{Bal20}, where only the first order conversion factors are considered, leading to error of the order of $10$ mas on the angles after 20 years. 
\\

Up to now, Poisson terms have never been included in radioscience data analysis.
However, neglecting them in the angle transformation leads to an error up to $4$ mas after $30$ years and makes the rotation solutions of GINS (in $\alpha/\delta$) and MONTE (in $\varepsilon/\psi$) inherently incompatible at some level. 
In order to guaranty the target precision on the transformation over a long period (for example on the 1970-2030 time interval), the software provided with this paper gives non-zero Poisson terms in the output set of angles even if the user sets the input Poisson terms to zero. 
\\

Following the same approach, we can define the angle $\beta_T$ (i.e.~the $M\!D$ arc) up to second-order.
As for the other angles, $\beta_T$ can be approximately written as the sum of epoch, linear, periodic, quadratic and Poisson terms:
\begin{equation}
\beta_T(t) = \beta_0 + \underbrace{\dot\beta_0^T \, t + \Delta\beta_T(t)}_{\textrm{First order}} + \underbrace{\beta^T_Q \, t^2 + \beta^T_\PO(t)}_{\textrm{Second order}}
\label{eq_betat}
\end{equation}
Injecting Eqs.~(\ref{eq_psit}) and (\ref{eq_alphat}) into the exact relation of Eq.~(\ref{eq_cD}) we obtain:
\begin{eqnarray}
\beta_T(t) &=& \beta_0 
+ \Gamma_{\beta\alpha}\, (\dot\alpha_0 \, t + \Delta\alpha) + \Gamma_{\beta\psi}\, (\dot\psi_0 \, t + \Delta\psi) + \Gamma_{\beta\alpha} \, (\alpha_Q \, t^2 + \alpha_\PO) + \Gamma_{\beta\psi}\, (\psi_Q \, t^2 + \psi_\PO) 
\nonumber \\
&& + \Gamma_{\beta\alpha\alpha}\,(\dot\alpha_0^2 \, t^2 + 2 \, \Delta\alpha\, \dot\alpha_0 \, t \, + \Delta\alpha^2)
+ \Gamma_{\beta\alpha\psi} \, (\dot\alpha_0 \, \dot\psi_0 \, t^2 + \Delta\alpha\, \dot\psi_0 \, t + \Delta\psi\, \dot\alpha_0 \, t + \Delta\alpha\, \Delta\psi) 
\nonumber \\
&& + \Gamma_{\beta\psi\psi}\, (\dot\psi_0^2 \, t^2 + 2 \, \Delta\psi \, \dot\psi_0 \, t \, + \Delta\psi^2) 
\label{eq_betat_Gamma} 
\end{eqnarray} 
with the constant over time conversion factors
\begin{subequations}
\label{eq_Gamma_beta}
\begin{eqnarray}
\Gamma_{\beta\alpha} &=& -\sin{\delta_0} 
\label{eq_b1} \\
\Gamma_{\beta\psi} &=& \cos{\varepsilon_0} \\
\Gamma_{\beta\alpha\alpha} &=& \frac{\cos\beta_0 \cos^2\delta_0}{2 \sin\beta_0} 
\label{eq_b2} \\
\Gamma_{\beta\alpha\psi} &=& -\frac{\cos\delta_0 \sin\varepsilon_0}{\sin\beta_0} \\
\Gamma_{\beta\psi\psi} &=& \frac{\cos\beta_0 \sin^2\varepsilon_0}{2 \sin\beta_0} 
\label{eq_b3} 
\end{eqnarray}
\end{subequations}
By identification between Eq.~(\ref{eq_betat}) and Eq.~(\ref{eq_betat_Gamma}), the transformation from $\alpha$ and $\psi$ to $\beta_T$ is obtained as (we again neglect the terms proportional to nutation product)
\begin{subequations}
\label{eq_betaT3}
\begin{eqnarray}
\lbrace \dot\beta_0^T ,\Delta\beta_T \rbrace &=& 
\Gamma_{\beta\alpha} \lbrace \dot\alpha_0, \Delta\alpha \rbrace + \Gamma_{\beta\psi} \lbrace\dot\psi_0, \Delta\psi \rbrace 
\label{eq_betaM} \\
\lbrace \beta_Q^T , \beta^T_\PO\rbrace &=& 
\Gamma_{\beta\alpha} \lbrace \alpha_Q, \alpha_\PO \rbrace + \Gamma_{\beta\psi} \lbrace\psi_Q, \psi_\PO \rbrace + \Gamma_{\beta\alpha\alpha} \lbrace \dot\alpha_0^2, \, 2 \Delta\alpha \, \dot\alpha_0 \, t \rbrace
\nonumber \\
&& + \Gamma_{\beta\alpha\psi} \lbrace \dot\alpha_0 \, \dot\psi_0, \, \Delta\alpha \, \dot\psi_0 \, t + \Delta\psi\, \dot\alpha_0 \, t \rbrace + \Gamma_{\beta\psi\psi} \lbrace \dot\psi_0^2, \, 2 \Delta\psi\, \dot\psi_0 \, t \rbrace
\label{eq_betaQ}
\end{eqnarray}
\end{subequations} 
Because the exact relation of Eq.~(\ref{eq_cD}) has been used to express $\beta$, the angle depends on both $\alpha$ and $\psi$ in Eq.~(\ref{eq_betat_Gamma}).
It is possible to obtain expressions for $\beta$ that depends either on $\alpha$ and $\delta$ or on $\varepsilon$ and $\psi$ instead, using the transformations between Euler and IAU angles.
Since $\beta_T$ links the angles $\phi_T$ and $W_T$ (see the Eq.~\ref{eq_W}), the above equation will be useful in Section \ref{sec_rotang} to write the approximate transformation between the two angles.

\subsection{The different representations of nutations} 
\label{sec_nutationrepresent}

The nutations of Mars are periodic oscillations of the orientation of the spin axis in space.
In the previous subsections, they were described as oscillations in obliquity $\Delta\varepsilon$ and in longitude $\Delta\psi$ or alternatively, as oscillations in right ascension $\Delta\alpha$ and in declination $\Delta\delta$. 
Another way to express the nutations is as a sum of prograde and retrograde motions.
Since many different representations for the nutations exist in the literature and in software codes and since the transformations between them are not straightforward, we explicitly give in this section the equations linking these representations and describe the different existing phase conventions. 
The prograde/retrograde representation has been used to compute the transfer functions in case of the existence of a liquid core \citep{Sas80}, see Section \ref{sec_nonri}.
\\

For each orientation angle, the nutations are written as periodic series, the periodicities being mostly the revolution period of Mars and its harmonics:
\begin{subequations}
\label{eq_nuttemporal}
\begin{eqnarray}
\Delta\varepsilon(t) &=& \sum_i (\varepsilon^s_i \, \sin(f_i \, t + \varphi_i^0) 
+ \varepsilon^c_i \, \cos(f_i \, t + \varphi_i^0) ) \\
\Delta\psi(t) &=& \sum_i (\psi^s_i \, \sin(f_i \, t + \varphi_i^0) 
+ \psi^c_i \, \cos(f_i \, t + \varphi_i^0) ) \\
\Delta\alpha(t) &=& \sum_i (\alpha^s_i \, \sin(f_i \, t + \varphi_i^0) 
+ \alpha^c_i \, \cos(f_i \, t + \varphi_i^0) ) \\
\Delta\delta(t) &=& \sum_i (\delta^s_i \, \sin(f_i \, t + \varphi_i^0) 
+ \delta^c_i \, \cos(f_i \, t + \varphi_i^0) )
\end{eqnarray}
\end{subequations}
For each term of the series, $f_i$ is the frequency and $\varphi_i^0$ is the phase of the nutation's argument. $\varepsilon_i^{s,c}$, $\psi_i^{s,c}$, $\alpha_i^{s,c}$, and $\delta_i^{s,c}$ are the sine and cosine amplitudes. Similar representations also apply to the Poisson terms, provided that the amplitudes are multiplied by the time.

This generic representation of nutations can be used in different ways.
In BMAN20, following RMAN99 (\citealt{Roo99}), the arguments of the trigonometric functions are linear combinations of fundamental arguments (the mean longitude of Mars and of the other planets, the node longitudes of Phobos and Deimos), so that $\varphi_i^0 \neq 0$, with the convention that $f_i$ must be positive. The BMAN20 and BMAN20RS nutation series are available at \href{https://doi.org/10.24414/h5pn-7n71}{https://doi.org/10.24414/h5pn-7n71}. BMAN20RS series are shortened series for radioscience applications, in which the nutation and Poisson terms at the same period are merged into a unique nutation term, to be used around the chosen mission epoch.

Another convention is to set $\varphi_i^0 = 0$, leading to what we call the pure frequency representation (the alternative representation described in Eq.~(28) of BMAN20, where the amplitudes are denoted with a $\,\tilde{ }\,$ to avoid confusion). The pure frequency representation is used in the GINS software. 
A third representation is the pure sine (or cosine) representation (see Eq.~29 in BMAN20), where each nutation is described by means of one amplitude and one phase for each frequency (e.g.~\citealt{Rea79}, as used in the MONTE software).
\\

At each period, the nutation corresponds to an elliptical motion of the spin axis in the inertial space. 
Therefore the nutations can also be expressed as the sum of two circular motions of opposite directions 
(one prograde, with the amplitude $\mathcal{P}_i$ and one retrograde, with the amplitude $\mathcal{R}_i$) at the same period.
It is possible to link the prograde/retrograde representation and the obliquity/longitude representation, or right ascension/declination representation by projecting the trajectory of a unit vector along the AM axis onto the equator of the J2000 mean equinox body frame\footnote{The J2000 mean equinox BF is just like the J2000 mean BF, but with the x-axis in the direction of the J2000 autumn equinox instead of the prime meridian.}, see section 3.3.2 of \citet{Bal20}:
\begin{subequations}
\label{Eqdeltaxy}
\begin{eqnarray}
\delta x = \sum_i \Big(\mathcal{P}_i \cos(f_i \, t + \pi_i) + \mathcal{R}_i \cos(-f_i \, t-\rho_i)\Big) =& 
\sin\varepsilon_0 \, \Delta\psi(t) &= \Delta\alpha(t) \cos\delta_0\cos\beta_0 + \Delta\delta(t) \sin\beta_0
\\
\delta y = \sum_i \Big(\mathcal{P}_i \sin(f_i \, t + \pi_i) + \mathcal{R}_i \sin(-f_i \, t-\rho_i)\Big) =& -\Delta\varepsilon(t) &= 
-\Delta\alpha(t) \cos\delta_0\sin\beta_0 + \Delta\delta(t) \cos\beta_0
\end{eqnarray}
\end{subequations}
These relations that express the first order relations between the periodic variations (nutations) of different representations are obtained by neglecting the precession and quadratic terms. We also neglect the Poisson terms, since the prograde/retrograde formulation is in the first place an intermediate step to obtain the non-rigid nutations in Euler or IAU angles (see Section \ref{sec_nonri}).

The following equations link the prograde $\mathcal{P}_i$ and retrograde $\mathcal{R}_i$ nutation amplitudes and the nutation amplitudes of the Euler and IAU angles:
\begin{subequations}
\label{eq_PR}
\begin{eqnarray}
\mathcal{P}_i &=& \sqrt{\left(\frac{\sin\varepsilon_0\, \psi^c_i - \varepsilon^s_i}{2}\right)^2
+ \left(\frac{\sin\varepsilon_0\, \psi^s_i + \varepsilon^c_i}{2}\right)^2} 
\label{Aproep} \\
 &=& \frac{1}{2} \sqrt{\left(\alpha^c_i \cos\delta_0 + \delta^s_i \right)^2 + 
\left(\alpha^s_i \cos\delta_0 - \delta^c_i \right)^2}
\label{Apro} \\
\mathcal{R}_i &=& \sqrt{\left(\frac{\sin\varepsilon_0\, \psi^c_i + \varepsilon^s_i}{2}\right)^2 
+ \left(\frac{\sin\varepsilon_0\, \psi^s_i - \varepsilon^c_i}{2}\right)^2} 
\label{Aretep} \\
 &=& \frac{1}{2} \sqrt{\left(\alpha^c_i \cos\delta_0 - \delta^s_i \right)^2 + 
\left(\alpha^s_i \cos\delta_0 + \delta^c_i \right)^2}
\label{Aret}
\end{eqnarray}
\end{subequations}
By definition, the amplitudes $\mathcal{P}_i$ and $\mathcal{R}_i$ are positive. When expressed in terms of the obliquity/longitude (right ascension/declination) angles, only the obliquity (declination) at epoch is needed in addition to the angles' amplitudes.
The prograde and retrograde phases $\pi_i$ and $\rho_i$ can be obtained from:
\begin{subequations}
\label{eq_phasespirho}
\begin{eqnarray}
\hspace*{-2cm}
2 \, \mathcal{P}_i \, \cos(\pi_i-\varphi_i^0) &=& \sin\varepsilon_0 \, \psi^c_i - \varepsilon^s_i 
\label{eq_phases0} \\
 &=& (\alpha^c_i \cos\delta_0 + \delta^s_i) \cos\beta_0 
+ (- \alpha^s_i \cos\delta_0 + \delta^c_i) \sin\beta_0 \\
2 \, \mathcal{P}_i \, \sin(\pi_i-\varphi_i^0) &=& -\sin\varepsilon_0 \, \psi^s_i - \varepsilon^c_i \\
 &=& (-\alpha^s_i \cos\delta_0 + \delta^c_i) \cos\beta_0 
- (\alpha^c_i \cos\delta_0 + \delta^s_i) \sin\beta_0 \\
2 \, \mathcal{R}_i \, \cos(\rho_i-\varphi_i^0) &=& \sin\varepsilon_0 \, \psi^c_i + \varepsilon^s_i \\
 &=& (\alpha^c_i \cos\delta_0 - \delta^s_i) \cos\beta_0 
+ (\alpha^s_i \cos\delta_0 + \delta^c_i) \sin\beta_0 \\
2 \, \mathcal{R}_i \, \sin(\rho_i-\varphi_i^0) &=& -\sin\varepsilon_0 \, \psi^s_i + \varepsilon^c_i \\
&=& (-\alpha^s_i \cos\delta_0 - \delta^c_i ) \cos\beta_0 
+ (\alpha^c_i \cos\delta_0 - \delta^s_i) \sin\beta_0 
\label{eq_phases1} 
\end{eqnarray}
\end{subequations}
Note that the phase $\varphi_i^0$ vanishes from the left-hand terms if the nutations are written with the pure frequency representation.
Other conventions have been used for defining the phases of the prograde and retrograde nutations, considering the direction of the node of the AM equator onto the equator of J2000 equinox body frame as reference instead of the projection of the vector along the AM axis, leading to a phase shift of 90$^\circ$ 
with respect to the phases defined in this section.
\\

If we want to express the nutations amplitudes in $\varepsilon/\psi$ or in $\alpha/\delta$ as a function of the prograde and retrograde amplitudes and phases, the opposite relations are
\begin{subequations}
\label{eq_oppositerel}
\begin{eqnarray}
\varepsilon^c_i &=& -\mathcal{P}_i \sin\left(\pi_i-\varphi_i^0\right) + \mathcal{R}_i \sin\left(\rho_i-\varphi_i^0\right) 
\label{eq_epsPR} \\
\varepsilon^s_i &=& -\mathcal{P}_i \cos\left(\pi_i-\varphi_i^0\right) + \mathcal{R}_i \cos\left(\rho_i-\varphi_i^0\right) \\
\psi^c_i &=& \frac{\mathcal{P}_i \cos\left(\pi_i-\varphi_i^0\right) + \mathcal{R}_i \cos\left(\rho_i-\varphi_i^0\right)}{\sin\varepsilon_0} \\
\psi^s_i &=& \frac{-\mathcal{P}_i \sin\left(\pi_i-\varphi_i^0\right) - \mathcal{R}_i \sin\left(\rho_i-\varphi_i^0\right)}{\sin\varepsilon_0} \\
\delta^c_i &=& \mathcal{P}_i \sin\left(\pi_i-\varphi_i^0+\beta_0\right) - \mathcal{R}_i \sin\left(\rho_i-\varphi_i^0-\beta_0\right) \\
\delta^s_i &=& \mathcal{P}_i \cos\left(\pi_i-\varphi_i^0+\beta_0\right) - \mathcal{R}_i \cos\left(\rho_i-\varphi_i^0-\beta_0\right) \\
\alpha^c_i &=& \frac{\mathcal{P}_i \cos\left(\pi_i-\varphi_i^0+\beta_0\right) + \mathcal{R}_i \cos\left(\rho_i-\varphi_i^0-\beta_0\right)}{\cos\delta_0} \\
\alpha^s_i &=& \frac{-\mathcal{P}_i \sin\left(\pi_i-\varphi_i^0+\beta_0\right) - \mathcal{R}_i \sin\left(\rho_i-\varphi_i^0-\beta_0\right)}{\cos\delta_0} 
\label{eq_alphaPR}
\end{eqnarray}
\end{subequations}

\subsection{The non-rigid nutations}
\label{sec_nonri}

Since Mars has a liquid core \citep{Yod03}, the nutation amplitudes differ from the rigid nutation amplitudes, well predicted by the models. At some frequencies, the amplitudes can be significantly modified (in particular at the semi-annual and ter-annual frequencies, see e.g.~\citealt{Deh20,Lem12}). 
The non-rigid nutation amplitudes can be obtained by applying on the rigid nutation amplitudes a transfer function that depends on the interior properties of the planet. Note that it does not apply to the geodetic nutation. A simplified form of this transfer function is given in \citet{Fol97a} for Mars with a liquid core, in the prograde/retrograde formulation. 
The transfer function includes 2 parameters: the core factor $F$ and the Free Core Nutation frequency (FCN) $\sigma_0$.
The non-rigid prograde/retrograde nutation amplitudes are:
\begin{subequations}
\label{eq_Fsig}
\begin{eqnarray}
\mathcal{P}'_i & = & \mathcal{P}_i \left(1 \, + \, F \frac{f_i}{f_i - \sigma_0} \right) 
\label{eq_Fsig1} \\
\mathcal{R}'_i & = & \mathcal{R}_i \left(1 \, + \, F \frac{f_i}{f_i + \sigma_0} \right) 
\label{eq_Fsig2}
\end{eqnarray}
\end{subequations}
Here, $\sigma_0$ is negative and $f_i$ is positive.
Although $\mathcal{P}_i$ and $\mathcal{R}_i$ were defined positive in the previous section, $\mathcal{P}'_i$ and $\mathcal{R}'_i$ can be negative here since this transfer function keeps the non-rigid nutation phases unchanged with respect to the rigid nutation phases. \\

Modified Eqs.~\eqref{eq_oppositerel}, with $'$ denoting the non-rigid nutation amplitudes, can be written for the case of non-rigid nutations.
By injecting Eqs.~(\ref{eq_Fsig}) and Eqs.~(\ref{eq_phasespirho}) into the modified Eqs.~(\ref{eq_oppositerel}), it is possible to express the non-rigid nutation amplitudes for $\varepsilon$, $\psi$, $\alpha$ and $\delta$ as a function of their rigid amplitudes and of the transfer function parameters:
{\allowdisplaybreaks
\begin{subequations}
\label{EqnutRN}
\begin{eqnarray} 
\varepsilon'^c_i &=& \varepsilon^c_i \, \mathcal{F}_i + \sin\varepsilon_0 \, \psi^s_i \, \mathcal{G}_i \\
\varepsilon'^s_i &=& \varepsilon^s_i \, \mathcal{F}_i - \sin\varepsilon_0 \, \psi^c_i \, \mathcal{G}_i \\
\psi'^c_i &=& \psi^c_i \, \mathcal{F}_i - \frac{\varepsilon^s_i}{\sin\varepsilon_0} \, \mathcal{G}_i \\
\psi'^s_i &=& \psi^s_i \, \mathcal{F}_i + \frac{\varepsilon^c_i}{\sin\varepsilon_0} \, \mathcal{G}_i \\
\alpha'^c_i &=& \alpha^c_i \, \mathcal{F}_i + \frac{\delta^s_i}{\cos\delta_0} \, \mathcal{G}_i \\
\alpha'^s_i &=& \alpha^s_i \, \mathcal{F}_i - \frac{\delta^c_i}{\cos\delta_0} \, \mathcal{G}_i \\
\delta'^c_i &=& \delta^c_i \, \mathcal{F}_i - \alpha^s_i \, \cos\delta_0 \, \mathcal{G}_i \\
\delta'^s_i &=& \delta^s_i \, \mathcal{F}_i + \alpha^c_i \, \cos\delta_0 \, \mathcal{G}_i
\end{eqnarray}
\end{subequations}
}
with 
\begin{subequations}
\begin{eqnarray} 
\mathcal{F}_i &=& 1 + F \frac{f_i^2}{f_i^2 - \sigma^2_0} \\
\mathcal{G}_i &=& F \frac{f_i \, \sigma_0}{f_i^2 - \sigma^2_0} 
\end{eqnarray}
\end{subequations}

We have neglected the Poisson terms in Eqs.~(\ref{Eqdeltaxy}), so that Eqs.~(\ref{EqnutRN}) are build for the amplitude of nutations. 
However, an accurate rotation model valid on a long time interval (like 1970-2030) may include Poisson terms besides the usual nutation terms. Remind also that the software provided with this paper returns non-zero Poisson terms in the output set of angles even if the user sets the input Poisson terms to zero (see Eqs.~(\ref{eq_alphadelta}) and (\ref{eq_epsilonpsi}), Section \ref{sec_app}). In the non-rigid case, any output Poisson terms also needs to be modified with transfer functions. The non-rigid counterparts of the rigid Poisson terms can be obtained similarly as for the nutation terms, using Eqs~(\ref{EqnutRN}), while respecting the targeted accuracy of $\sim0.1$ mas (the demonstration is not provided here but basically, any relatively small error on a second-order quantity is even smaller and can be neglected).

\section{Transformation between the rotation angles \texorpdfstring{$\phi$}{phi} and \texorpdfstring{$W$}{W}} 
\label{sec_rotang}

\begin{figure}[!ht]
\centering
\includegraphics[height=9cm, width=15cm]{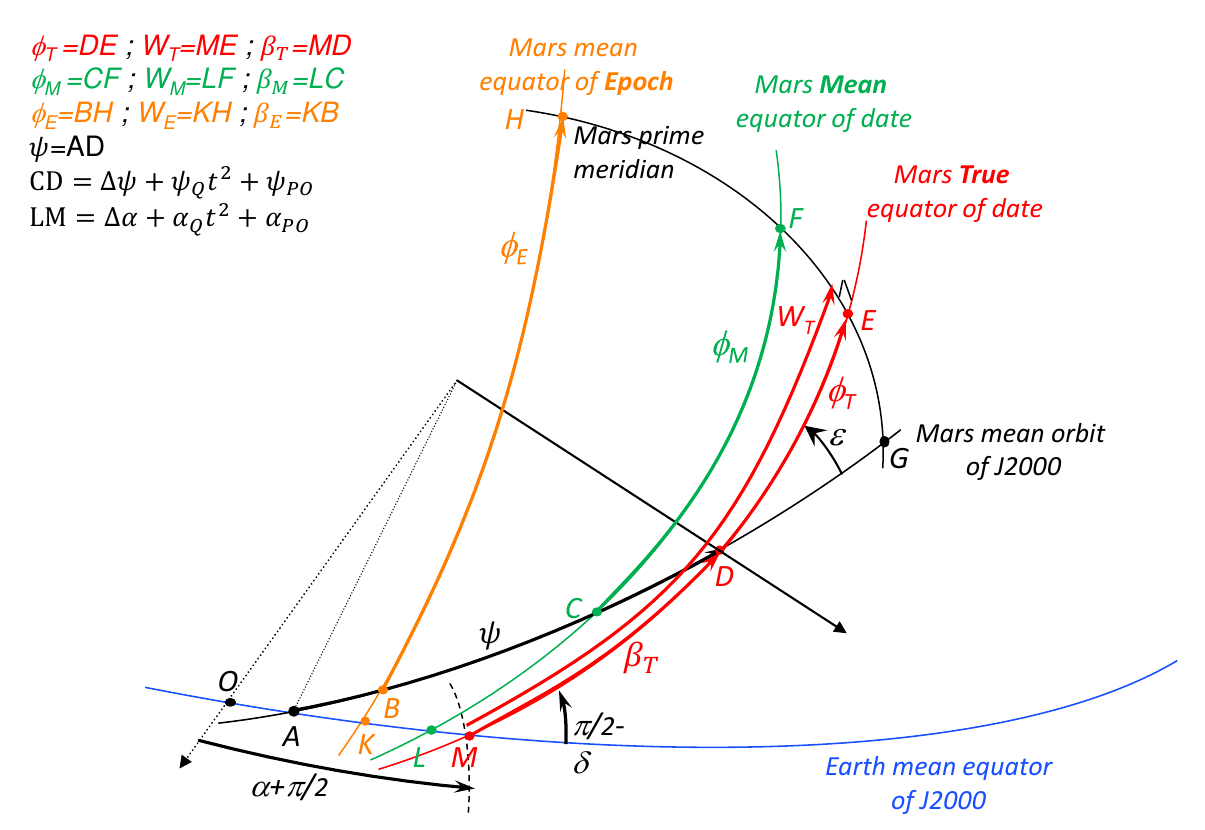} 
\caption{Rotation angles $\phi$, $W$ and $\beta$ of Mars, for different choices of the reference plane. 
The angles in red, green, and orange are related to Mars true equator of date, Mars mean equator of date, and Mars mean equator of Epoch (J2000), respectively.} 
\label{fig_ang2}
\end{figure}

The rotation angles $\phi$ and $W$ describe the location of the intersection between the prime meridian and the equator of Mars, measured from the node of the equator over Mars orbit of epoch and over the Earth mean equator of J2000, respectively. Their definition is non unique, as the equatorial plane of Mars can be defined in different ways.
Whereas in the IAU conventions, the rotation angle is measured along the True equator of date (noted $W_T$ hereafter), atmospheric scientists, modeling variations in rotation angles caused by the exchange of angular momentum between the atmosphere and the surface, usually measure the rotation angle along the Mean equator of date (noted $\phi_M$ hereafter). 
In this section, we precisely define the different rotation angles and describe the relations between them. We specially focus on the rotation angles most commonly used by the community.
They are the angles $\phi_T$ and $W_T$ (measured along the True equator of date) that intervene directly in the rotation matrices, but also $\phi_M$ (measured along the Mean equator of date) because the periodic variations in $\phi_T$ are written as function of the periodic variations in $\phi_M$, in order to be compared to the periodic variations modeled by atmospheric scientists (see section \ref{sec_LOD}).

\subsection{The different definitions of the rotation angles \texorpdfstring{$\phi$}{phi} and \texorpdfstring{$W$}{W}}
The {\bf Mars mean equator of epoch} is a fixed plane defined by the angles $\varepsilon_0$ and $\psi_0$ (or $\alpha_0$ and $\delta_0$) at epoch. 
There is no convention regarding the equator to be used to define the rotation angles.
We choose to define the {\bf Mars mean equator of date} as the plane that only follows the mean precession in longitude and in obliquity (or in right ascension and declination), meaning that the nutations, quadratic drift, and Poisson terms are not considered in that definition. 
The {\bf Mars true equator of date} is the plane that follows the full temporal variations of the angles, including the precession, the nutations, the quadratic and the Poisson terms.
The prime meridian can not be perpendicular to both the true and mean equators of date.
We choose to define the prime meridian as the plane perpendicular to the true equator of date, the intersection being the point E on Fig.~\ref{fig_ang2}.\\

$\phi_T$ is the {\bf True} Euler angle of rotation measured along the Mars {\bf True} equator of date, from the node with the Mars mean orbit of J2000 to the prime meridian of Mars. 
It corresponds to the angle $D\!E$ on Figure~\ref{fig_ang2}. 
$\phi_M$ is the rotation angle measured along the Mars {\bf Mean} equator of date ($C\!F$ on Fig.~\ref{fig_ang2}),
and $\phi_E$ is the rotation angle measured along the Mars mean equator of {\bf Epoch} ($ B\!H$ on Fig.~\ref{fig_ang2}).
Here, the starting nodes $B$, $C$ or $D$ of these rotation angles are all in the Mars mean orbit of J2000.
The mean orbit of 1980, as defined by \citet{Fol97a}, is however more often used.
Equivalently, we can define the 3 different rotation angles $W$ with the starting nodes being $K$, $L$ or $M$ in the Earth mean equator of J2000: 
$W_T$ is the angle $M\!E$ measured along the Mars {\bf True} equator of date, 
$W_M$ is the angle $L\!F$ measured along the Mars {\bf Mean} equator of date  
and $W_E$ is the angle $K\!H$ measured along the Mars mean equator of {\bf Epoch}, see Fig.~\ref{fig_ang2}. 
\\

As for the orientation angles, we assume for the rotation angles a generic form with an initial value at epoch, a linear and a quadratic term, a series of periodic perturbations, and a Poisson series. For $\phi_T$ and $W_T$, we write:
\begin{subequations}
\begin{alignat}{4}
\phi_T(t) &= \phi_0^T && + \dot\phi^T_0 \, t && + \Delta\phi_T(t)&& + \phi^T_Q \, t^2 + \phi^T_\PO(t) 
\label{eq_phiTt} \\
W_T(t) &= W_0^T && + \dot W^T_0 \, t && + \underbrace{\Delta W_T(t)}_{\textrm{First order}} && + \underbrace{W^T_Q \, t^2 + W^T_\PO(t)}_{\textrm{Second order}}
\end{alignat}
\end{subequations} 
The generic form for $\beta_T = W_T-\phi_T$ (see eq.~\ref{eq_W}) is similar and given in Eq.~(\ref{eq_betat}).
In this modelization, the periodic terms are first order variations while quadratic terms and Poisson series are of second order. The diurnal frequencies $\dot\phi^T_0$ and $\dot W^T_0$ are so large with respect to the other terms that they cannot be considered as first order variations, contrary to the linear terms in $\epsilon,\psi$, $\alpha,\delta$ and in $\beta$. 
The angles $\phi_M(t), \phi_E(t)$ and $W_M(t), W_E(t)$ have similar generic expressions as $\phi_T(t)$ and $W_T(t)$.

\subsection{Link between the rotation angles \texorpdfstring{$\phi_T$}{phiT}, \texorpdfstring{$\phi_M$}{phiM} and \texorpdfstring{$W_T$}{WT}}

This section provide the relations to be used to transform $\phi_T$ into $W_T$ and $\phi_M$, and reciprocally. \\

Following the definition of the $\beta_T$ angle (see Eqs.~\ref{eq_W} and \ref{eq_betat}) and its expression in terms of $\alpha$ and $\psi$ (Eqs.~\ref{eq_D} and \ref{eq_betaT3}), the relations between the different parts of $\phi_T$ and $W_T$ are:
\begin{subequations}
\begin{alignat}{5}
\phi_0^T &= W_0^T - \beta_0 & \\
\dot\phi_0^T &= \dot W_0^T - \dot\beta_0^T 
&=& \, \dot W_0^T + \sin\delta_0 \, \dot\alpha_0 - \cos\varepsilon_0 \, \dot\psi_0 & \\
\Delta\phi_T &= \Delta W_T - \Delta\beta_T 
&=& \, \Delta W_T + \sin\delta_0 \, \Delta\alpha - \cos\varepsilon_0 \, \Delta\psi &
\label{eq_deltaphiT} \\
\phi^T_Q &= W^T_Q - \beta^T_Q \,
&=& \, W^T_Q + \sin\delta_0 \, \alpha_Q - \cos\varepsilon_0 \, \psi_Q 
- \Gamma_{\beta\alpha\alpha} \, \dot\alpha_0^2
- \Gamma_{\beta\alpha\psi} \, \dot\alpha_0 \, \dot\psi_0
- \Gamma_{\beta\psi\psi} \, \dot\psi_0^2 
\label{eq_phiTQ} \\
\phi^T_\PO &= W^T_\PO - \beta^T_\PO & \\
&= W^T_\PO + \sin\delta_0 \, \alpha_\PO 
&& - \cos\varepsilon_0 \, \psi_\PO 
- 2 \, \Gamma_{\beta\alpha\alpha} \,\Delta\alpha \, \dot\alpha_0 \, t 
- \Gamma_{\beta\alpha\psi} \, (\Delta\alpha \, \dot\psi_0 \, t + \Delta\psi \, \dot\alpha_0 \, t)
- 2 \, \Gamma_{\beta\psi\psi} \, \Delta\psi \, \dot\psi_0 \, t \label{Eq37f}
\end{alignat}
\end{subequations}

After spherical trigonometry transformations in the $CFG$ and $DEG$ triangles, the difference $\phi_T-\phi_M$, correct up to the second order in small quantities, is given by:
\begin{subequations}
\label{eq_phiTMWTM}
\begin{equation}
\phi_T - \phi_M = -\cos\varepsilon_0 \, (\Delta\psi + \psi_Q \, t^2 + \psi_\PO)
+ \sin\varepsilon_0 \, \Delta\psi \, \dot\varepsilon_0 \, t
\label{eq_phiTM}
\end{equation}
By analogy, the difference $W_T-W_M$ is given by:
\begin{equation}
W_T - W_M = - \sin\delta_0 \, (\Delta\alpha + \alpha_Q \, t^2 + \alpha_\PO) \, 
- \cos\delta_0 \, \Delta\alpha \, \dot\delta_0 \, t 
\label{eq_WTM} 
\end{equation}
\end{subequations}
As in the previous sections, we neglect here the terms proportional to the nutations' products.
Since the obliquity rate is expected to be 0 or very close, the last term of Eq.~(\ref{eq_phiTM}), a Poisson term, is very small.
We keep it in the expression to conserve the analogy with Eq.~(\ref{eq_WTM}) where $\dot\delta_0$ cannot be neglected. 
At the first order, Eq.~(\ref{eq_phiTM}) reduces to $\phi_T-\phi_M=-\Delta\psi \cos\varepsilon_0$, and can simply be understood as minus the geometrical projection of the nutation in longitude $\Delta\psi$ on the mean equator of date.
This part is usually included in the estimated rotation models while the quadratic and Poisson terms are neglected (see for instance the Eq.~18 of \citealt{Kon06}, describing $\phi_T(t)$).
Using Eqs.~(\ref{eq_phiTt}) and (\ref{eq_phiTM}), the relations between the different parts of $\phi_T$ and $\phi_M$ are:
\begin{subequations}
\label{Eq39}
\begin{eqnarray}
\Delta\phi_T &=& \Delta\phi_M - \cos\varepsilon_0 \, \Delta\psi \label{eq_deltaphiM} \\
\label{EqphiQ}\phi^T_Q &=& \phi^M_Q - \cos\varepsilon_0 \, \psi_Q \\
\label{EqphiPO}\phi^T_\PO &=& \phi^M_\PO - \cos\varepsilon_0 \, \psi_\PO + \sin\varepsilon_0 \, \Delta\psi \, \dot\varepsilon_0 \, t 
\end{eqnarray}
\end{subequations}
The epoch value of both angles $\phi_T$ and $\phi_M$ are equal, as well as their rate.
\begin{subequations}
\begin{eqnarray}
\phi_0^T &=& \phi_0^M = W_0^T-\beta_0 \\ 
\qquad \dot\phi^T_0 &=& \dot\phi^M_0 = \dot W^T_0 - \dot\beta^T_0
\end{eqnarray}
\end{subequations}
Expressions (\ref{eq_deltaphiT}-\ref{Eq37f}) and (\ref{eq_deltaphiM}-\ref{EqphiPO}) can also be combined under the form:
\begin{subequations}
\begin{eqnarray}
\Delta W_T &=& \Delta\phi_M - \sin\delta_0 \, \Delta\alpha \label{eq_deltaWT}\\
W_Q^T&=& \phi^M_Q  - \sin\delta_0 \, \alpha_Q + \Gamma_{\beta\alpha\alpha} \, \dot\alpha_0^2
+ \Gamma_{\beta\alpha\psi} \, \dot\alpha_0 \, \dot\psi_0
+ \Gamma_{\beta\psi\psi} \, \dot\psi_0^2 \label{eq_WTQ}\\
W^T_\PO &=& \phi^M_\PO - \sin\delta_0 \, \alpha_\PO + \sin\varepsilon_0 \, \Delta\psi \, \dot\varepsilon_0 \, t + 2 \, \Gamma_{\beta\alpha\alpha} \,\Delta\alpha \, \dot\alpha_0 \, t 
+ \Gamma_{\beta\alpha\psi} \, (\Delta\alpha \, \dot\psi_0 \, t + \Delta\psi \, \dot\alpha_0 \, t)
+ 2 \, \Gamma_{\beta\psi\psi} \, \Delta\psi \, \dot\psi_0 \, t \nonumber \\ 
\label{eq_WTPo}
\end{eqnarray}
\end{subequations}

Most of the needs of the community should be covered by the relations described above. 
For completeness, we could have derived other relations that could be found of interest by our readers.
For instance, one can find an expression for $W_M$ as a function of the different parts of $W_T$, by using Eq.~(\ref{eq_WTM}).
$\beta_M$ can be obtained as a function of the different parts of the orientation Euler and IAU angles, from $\beta_M=W_M-\phi_M = \beta_T-(W_T-W_M) + (\phi_T-\phi_M)$ and Eq.~(\ref{eq_betat_Gamma}) and Eq.~(\ref{eq_phiTMWTM}). We do not provide those relations here.
\\

For later use, expressions for $\phi_T-\phi_E$ and $W_T-W_E$ are given here correct up to the first order:
\begin{subequations}
\label{Eq42}
\begin{eqnarray}
\phi_T - \phi_E & \simeq& - \cos\varepsilon_0 \, (\dot\psi_0 \,t + \Delta\psi), 
\label{eq_phiTE} \\
W_T - W_E & \simeq& - \sin\delta_0 \, (\dot\alpha_0 \,t + \Delta\alpha).
\label{eq_WTE} 
\end{eqnarray}
\end{subequations}
These projection terms include both a precession and a nutation part.
Note also that the angle $\beta_E = W_E - \phi_E = \beta_T - (W_T-W_E) + (\phi_T-\phi_E)$ is constant over time and equal to $\beta_0$, which is easily understood from the fact that the mean equator of epoch is a fixed plane.

\subsection{The 3 diurnal rotation rates}
\label{sec_rotrat}
So far, two quantities have been used to characterise the diurnal spin rate of Mars, $\dot\phi_0$ and $\dot W_0$. The former positions the prime meridian relatively to the node of Mars mean orbit of epoch and Mars equator of date, while the latter is relative to the node of the Earth mean equator of J2000 and Mars equator of date. Since the difference between the true and the mean equator of date only involves periodic, quadratic, and Poisson variations (no rates), we have $\dot\phi_0 = \dot\phi_0^M = \dot\phi_0^T$ and $\dot W_0 = \dot W_0^M = \dot W_0^T$.
However, $\dot\phi_0\neq\dot W_0$ because of the precession motion of Mars equator of date that is not performed at the same rate ($\dot\psi_0$ or $\dot\alpha_0$) neither with the same inclination ($\varepsilon_0$ or $\pi/2-\delta_0$), depending on the chosen reference plane (Mars mean orbit of epoch or Earth J2000 equator), see Eq.~(\ref{eq_betaM}) for $\dot W_0-\dot\phi_0=\dot\beta^T_0=\dot\beta^M_0$. 
On the other hand, when expressed relatively to the node of Mars mean equator of epoch over either reference plane, the spin rate of Mars 
becomes independent from the set of angles used, since then both reference frames are inertial, with a constant offset $\beta_0$ between the nodes over the reference planes. 
In (any) inertial frame, the spin rate of Mars is constant and we can thus define a third quantity, $\Omega = \dot\phi_0^E = \dot W_0^E$, which corresponds to the stellar rate of rotation of Mars.
$\Omega$ is linked to $\dot W_0$ and $\dot\phi_0$ through, see Eq.~(\ref{Eq42}):
\begin{equation}
\Omega = \dot W_0 + \sin\delta_0 \, \dot\alpha_0 = \dot\phi_0 + \cos\varepsilon_0 \, \dot\psi_0
\label{eq_phipWp}
\end{equation}
This expression is the Eq.~(19) of \citet{Kon06}. 
For Mars, we have numerically
\begin{equation}
\Omega < \dot W_0 < \dot\phi_0
\end{equation}
By analogy with the Earth, we define the period $2\pi/\Omega$ as the stellar day (measured relative to the stars). The period $2\pi/\dot\phi_0$ is the sidereal day (measured relative to the precessing Martian equinox). 
For Mars, the sidereal day is shorter than the stellar day by $1.3$ ms, because of the precession of the Martian equinox (see Section \ref{Sec_RotationRate}). The period $2\pi/\dot W_0$ has no fancy name (it is simply referred to as a ``rotation period'' in the IAU conventions, while $\dot W_0$ is called "IAU spin rate" in \citealt{Kon06}). This ``IAU day'' is shorter than the stellar day by $0.6$ ms and larger than the sidereal day by $0.7$ ms. The stellar, the ``IAU'', and the sidereal days are all shorter than the solar day by about 2.2 minutes.

\subsection{The relativistic corrections to the rotation angle} 
\label{sec_rel}

The time scale (Barycentric Dynamical Time) used to analyse radioscience data differs from Mars proper time.
A relativistic correction $\left[\phi\right]_\GR\!(t)$ is included in the model for the rotation angle $\phi_T(t)$ of Eq.~(\ref{eq_phiTt}), see also Eq.~(\ref{eq_phiMF}) and Eq.~(18) of \citet{Kon06}.
This correction is due to the velocity of Mars in its motion around the Solar system barycenter and to the varying distance of Mars with respect to the Sun.
\citet{Bal23} have updated the estimation of the relativistic correction first provided by \citet{Yod97}:
\begin{equation}
[\phi]_\GR(t) = \dot\phi_\GR \, t + \Delta\phi^{Rel} = \dot\phi_\GR \, t + \sum_i \phi^r_i \sin(f_i \, t + \varphi_i^0),
\end{equation}
with $\phi^r_i$ the amplitudes, $f_i$ the frequencies, and $\varphi_i^0$ the phases of the relativistic correction.
This correction includes seasonal terms, at the same periods as the rotation variations $\Delta\phi_M^{Atm}$ (see next subsection), other periodic terms (e.g.~a term at the synodic period between Jupiter and Mars), and a secular term with the rate $\dot\phi_\GR$ which adds to the local sidereal rotation rate ($\dot\phi_0 = \dot\phi_0^{local} + \dot\phi_\GR)$. 
These relativistic corrections affect the rotation angles $W_T$ and $\phi_M$ the same way as for $\phi_T$.
Some numerical values are given in Section \ref{sec_ampLOD}.

\subsection{Periodic variations of the rotation angles}
\label{sec_LOD}

Mars experiences significant length-of-day (LOD) variations, caused by the global-scale seasonal exchanges of mass (mostly carbon dioxide) and angular momentum between the atmosphere and the solid body (the ice caps). 
The amplitudes of these variations can be predicted with Global Circulation Models (GCM) (e.g.~\citealt{Def00, Van02, Kar06}).
The assumption behind the GCM-based computations is that Mars is a gravitationally isolated rotating body, with the Sun as a source of heat. The body does not experience any variation in orientation, rotating regularly, without any nutation or precession.
Therefore the GCM-computed series of periodic variations in rotation correspond to the periodic variations of a rotation angle referred to an inertial plane, for example the Mars mean equator at J2000.
This means that GCMs provide rotation variations corresponding to $\Delta\phi_E$.
As the periodic variations $\Delta\phi_M$ and $\Delta\phi_E$ are the same (see Eqs.~\ref{eq_phiTM} and \ref{eq_phiTE}), the output of the GCM-based model also corresponds to a periodic series associated to the angle $\phi_M$, located on the Mars mean equator of date.
\\

The total periodic part of $\phi_M$ consists of the sum of the atmospheric and periodic relativistic terms 
(see Section \ref{sec_rel}):
\begin{equation}
\Delta\phi_M = \Delta\phi_M^{Atm} + \Delta\phi^{Rel} 
\label{eq_phiMF}
\end{equation}
$\Delta\phi_M^{Atm}$ is usually written (e.g.~\citealt{Kon06}) as a series of 4 terms with periods being the harmonics of the orbital period of Mars:
\begin{equation}
\Delta\phi_M^{Atm} = \sum_{i=1}^4 (\phi^c_i \cos i\,l'(t) + \phi^s_i \sin i\,l'(t))
\end{equation}
where $\phi^c_i$ and $\phi^s_i$ are the cosine and sine amplitudes and $l'(t)$ is the mean anomaly of Mars. 
Using Eqs.~(\ref{eq_deltaphiM}) and (\ref{eq_deltaWT}), we see that the periodic parts $\Delta\phi_T$ and $\Delta W_T$, relative to the true equator of Mars, include nutation terms in addition to the atmospheric and relativistic terms present in $\Delta\phi_M$:
\begin{subequations}
\label{eq_deltaphi}
\begin{eqnarray}
\Delta\phi_T &=& \Delta\phi_M - \cos\varepsilon_0 \, \Delta\psi
 = \Delta\phi_M^{Atm} + \Delta\phi^{Rel} - \cos\varepsilon_0 \, \Delta\psi \\
\Delta W_T &=& \Delta W_M - \sin\delta_0 \, \Delta\alpha
 = \Delta\phi_M^{Atm} + \Delta\phi^{Rel} - \sin\delta_0 \, \Delta\alpha
\end{eqnarray}
\end{subequations}
with $\Delta W_M = \Delta\phi_M$. In Section \ref{sec_ampLOD}, we provide recommendations to express $\Delta\phi_T$ and $\Delta W_T$.
\\

The periodic variations of atmospheric origin in the rotation of Mars are sometimes also expressed as Length-Of-Day variations ($\DeltaLOD$).
These variations are proportional to the time derivative of the periodic variations $\Delta\phi_M^{Atm}$ and are written as 
\begin{equation}
\DeltaLOD = -\frac{2 \pi}{\Omega^2} \, \frac{d\Delta\phi_M^{Atm}}{dt}=\sum_{i=1}^4 \Big(\DeltaLOD^c_i \cos i\,l'(t) + \DeltaLOD^s_i \sin i\,l'(t)\Big)
\label{eqLOD}
\end{equation}
with 
\begin{subequations}
\begin{eqnarray}
\DeltaLOD^c_i &=& - \frac{2\pi \, i\, n}{\Omega^2} \, \phi^s_i
\label{eqLODC} \\
\DeltaLOD^s_i &=& + \frac{2\pi \, i\, n}{\Omega^2} \, \phi^c_i
\label{eqLODS}
\end{eqnarray}
\end{subequations}
where $n$ is the mean motion ($n=d \, l'/d \, t$).

\section{The polar motion in the transformation}
\label{sec_PM}

In Section \ref{sec_definition}, we have decomposed the transformation from the Martian BF to the IF into two successive transformations (see Eq.~\ref{eq_landerpos1}). The subsequent sections were devoted to the Euler or IAU angles intervening in the second transformation $(\mathbf{M}_{AM\rightarrow IF})$ from the AM frame to the IF and how to switch from one to the other formulation. We now discuss the parameters involved in the first transformation $(\mathbf{M}_{BF\rightarrow AM})$ from the BF to the AM frame, namely the polar motion components $X_P$ and $Y_P$.
The parametrization of the polar motion is the same whether we work with Euler angles or IAU angles in the second transformation. \\

The components of the rotation vector of a body are often written in the coordinates of the Body Frame as
\begin{subequations}
\begin{equation}
\label{Omegaa}
{\bf \Omega} =
\Omega
\left(\begin{array}{c}
  m_1\\
  m_2\\
 1 + m_3
\end{array}\right),
\end{equation} 
\end{subequations}
with $\Omega$ the stellar rate of rotation (see Section \ref{sec_rotrat}).
The polar motion is the projection $(m_1,m_2)$ of the unit vector along the rotation axis on the BF equator and is often denoted $(X_P,-Y_P)$ with the minus sign in front of $Y_P$ coming from the convention used for the Earth. The polar motion of Mars is the combination of 2 motions: 
\begin{subequations}
\label{eq_pmSum}
\begin{eqnarray}
X_P &=& X_P^{Atm} + X_P^{Ext},\\
Y_P &=& Y_P^{Atm} + Y_P^{Ext},
\end{eqnarray}
\end{subequations}
the first part caused by the atmosphere and the ice caps dynamics (a few tens of mas, \citealt{Def00}, \citealt{Van02}, \citealt{Kon20}), 
and the second part caused by the external (mainly solar) torque (about 5 cm at the surface of Mars or 3 mas, \citealt{Rea79}). 
The decomposition of the transformation from the BF to the IF into two successive transformations is convenient because of the relatively large polar motion of atmospheric origin. Otherwise, it would be equivalent to perform the transformation with angles describing directly the orientation of the BF with respect to the IF. In any case, the proper modeling of the motion between the spin and figure axes induced by the external torque is required.\\

If $\hat p$ is the unit vector along the perpendicular to the inertial J2000 mean BF equator, then its components $(p_x,p_y,p_z)$ in the coordinates of the BF are obtained, at first order in small variations, as (more details in Section 3.7 of \citealt{Bal20})
\begin{eqnarray}
 \label{Eqp} \hat p &=& \mathbf{R_z}(\phi)\cdot\left(
\begin{array}{c}
 -\delta x_{fig} \\
 -\delta y_{fig} \\
 1 \\
\end{array}
\right),
\end{eqnarray}
with ($\delta x_{fig},\delta y_{fig}$) defined as in Eq.~(\ref{Eqdeltaxy}), but for the figure axis instead of the AM/spin axis. Since ($\delta x_{fig},\delta y_{fig}$) can be described in space as series of prograde and retrograde circular motions with amplitudes $\mathcal{P}_i^{fig}$ and $\mathcal{R}_i^{fig}$, the projection $(p_x,p_y)$ can be expressed in the BF as a series of retrograde quasi diurnal circular motions with opposite amplitudes ($-\mathcal{P}_i^{fig}$ and $-\mathcal{R}_i^{fig}$).
By definition, any circular motion of the figure axis in space at a given frequency is the combination of one circular motion of the AM axis with the opposite of one circular polar motion (sometimes called "Oppolzer term" when seen from space) at the same frequency, so that we note $\mathcal{P}_i^{fig}=\mathcal{P}_i-m_P$ and $\mathcal{R}_i^{fig}=\mathcal{R}_i-m_R$ with $m_P$ and $m_R$ amplitudes of circular (retrograde quasi diurnal in the BF) motions. Making use of the Euler kinematic relation linking precession of the BF in space and polar motion in the BF, we also have \citep[Section 3.7]{Bal20}:
\begin{subequations}
\begin{eqnarray}
 m^P_i &=& -\mathcal{P}^{fig}_i\,\frac{ f_i }{\Omega},\\
 m^R_i &=&  \mathcal{R}^{fig}_i\,\frac{ f_i }{\Omega}.
\end{eqnarray}
\end{subequations}
It follows that the polar motion components can be obtained as 
\begin{eqnarray}
\label{EqPMext}\left\lbrace \begin{array}{c}
  m_1\\
  m_2
 \end{array}\right\rbrace &=& \sum_i \left(m^P_i \left\lbrace 
 \begin{array}{c}\cos\\
 \sin\end{array}\right\rbrace (-\phi+f_i t +\pi_{i})+m^R_i \left\lbrace \begin{array}{c}\cos\\
 \sin\end{array}\right\rbrace(-\phi-f_i t-\rho_{i})\right)
\end{eqnarray}
with 
\begin{subequations}
\label{EqPMext2}
 \begin{eqnarray}
 m^P_i &=& -\mathcal{P}_i\frac{f_i}{\Omega-f_i} \\
 m^R_i &=&  \mathcal{R}_i\frac{f_i}{\Omega+f_i}
\end{eqnarray}
\end{subequations}
and with the prograde and retrograde amplitudes of the AM nutation series ($\mathcal{P}_i$ and $\mathcal{R}_i$) obtained as in Eq.~(\ref{eq_PR}). Here the rotation angle $\phi$ is approximated in the cosine and sine arguments by $\simeq \phi_0 + \dot\phi_0 \, t$.
The numerical values of the series $m_1$, $m_2$ will be given in Section \ref{sec_PMnum}.

\section{Recommendations for the rotation model of Mars}
\label{sec_num}

Mars scientists need a rotation model to orientate the planet body frame in space, for radioscience or for cartographic applications for example. Such a model consists in both an analytical expression of the angles and the numerical values (estimated or modeled) of their different terms. Our goal in this section is not to discuss every single value present in the literature, nor to define a new IAU rotation model (see also Section \ref{sec_IAU}), but rather to show how to avoid making mistakes when providing new rotation model of Mars and discuss some oddities of the literature. We also numerically quantify the errors produced by simplifying assumptions or omitting terms. 
As the precession/nutation modelisation that we choose (a corrected version of the BMAN20 model, \citealt{Bal20})
is built to be consistent with the estimated solution of \cite{Kon16}, we will mainly refer to the latter. This does not mean that we recommend the solution of \cite{Kon16} compared to another. 
More recent estimates are also available, and more are yet to come.
\\ 

Following the definitions presented in the first part of the paper (Sections \ref{sec_definition}-\ref{sec_rotang}), the Euler and IAU orientation and rotation angles can be written as a sum of a degree-two polynomial, a periodic and Poisson series: \\
\begin{subequations}
\label{Eqmodel}
\begin{alignat}{5}
\varepsilon(t) &= \varepsilon_0 && + \dot\varepsilon_0 \, t && + \varepsilon_Q \, t^2 && + \Delta\varepsilon(t) && + \varepsilon_\PO(t) 
\label{modeeps} \\
\psi(t) &= \psi_0 && + \dot\psi_0 \, t && + \psi_Q \, t^2 && + \Delta\psi(t) && + \psi_\PO(t) 
\label{modepsi} \\
\phi_T(t) &= \phi_0 && + \dot\phi_0 \, t && + \phi^T_Q \, t^2 && + \underbrace{\Delta\phi_M^{Atm}(t) + \Delta\phi^{Rel}(t) - \cos\varepsilon_0 \, \Delta\psi (t)}_{\Delta\phi_T(t)} && + \phi^T_\PO(t) 
\label{modephi} \\
\alpha(t) &= \alpha_0 && + \dot\alpha_0 \, t && + \alpha_Q \, t^2&& + \Delta\alpha(t) && + \alpha_\PO(t) \label{modealpha} \\
\delta(t) &= \delta_0 && + \dot\delta_0 \, t && + \delta_Q \, t^2&& + \Delta\delta(t) && + \delta_\PO(t) \\
W_T(t) &= W_0 && + \dot W_0 \, t&& +W^T_Q \, t^2 && + \underbrace{\Delta\phi_M^{Atm}(t) + \Delta\phi^{Rel}(t) - \sin\delta_0 \, \Delta\alpha(t) }_{\Delta W_T(t)} && + W^T_\PO(t)
\label{modeW}
\end{alignat}
\end{subequations} 
where $t$ is the time, starting from J2000. 
Eqs.~\ref{modeeps}-\ref{modephi} replaces Eqs.~(14) and (18) of \cite{Kon06}, where there were no quadratic and Poisson terms. 
The quadratic and Poisson terms in the orientation angles are predicted by the theory (see Sections \ref{Section_deg2poly} and \ref{sec_rigidnut}). The quadratic terms in rotation angles $\phi_Q^T$ and $W_Q^T$ (see Eqs.~\ref{EqphiQ} and \ref{eq_WTQ}) are function of the quadratic terms in orientation angles, of ``rate x rate'' terms, and of the quadratic term $\phi_Q^M$. The Poisson terms in rotation angles $\phi^T_\PO$ and $W^T_\PO$ (see Eqs.~\ref{EqphiPO} and \ref{eq_WTPo}) are function of the Poisson terms in orientation angles, of ``nutation x rate'' terms, and of the Poisson term $\phi_\PO^M$. Explicit expressions for a rotation model centered at J2000 and precise to the 1 mas level, where all parameters can be replaced by estimated or modeled numerical values, are provided in Appendix \ref{sec_appen}. To complete the rotation model, expressions for the polar motion components $X_P$ and $Y_P$ (see Section \ref{sec_PM}) are also needed. They will be expressed as periodic series.
\\

Estimated and modeled parts of the rotation angles of Mars are more often expressed in Euler angles than in IAU angles. The estimated solution of \cite{Kon16}, used as a main reference in this section, is no exception to the rule. To obtain the corresponding expressions in IAU angles, we will below make use of the transformation presented in the first part of the paper to obtain the corresponding IAU angles.
For this, we use the Mars oRientation and Rotation Software (MARRS), available at \href{http://gitlab-as.oma.be/mars-tools/marrs}{gitlab-as.oma.be/mars-tools/marrs}.
This python code makes the transformation between Euler and IAU angles, including the epoch, secular, quadratic terms and the periodic and Poisson series. 
Comparison can be made between the transformation matrices obtained using the different (Euler or IAU) formulations and between the transformation matrices obtained using different published rotation models.\\

We now describe the successive terms of the rotation model, after having defined a local version of the rotation model, and specified the needed accuracy on each term to meet the targeted level of precision.

\subsection{Local rotation model}
\label{Sec_local_model}

For practical reasons (e.g.~software implementation, model complexity, etc.), the reader may need to simplify the rotation model of Eq.~(\ref{Eqmodel}), while keeping a high level of model accuracy for a limited period of time (e.g.~over the lifetime of a given mission). In such a case, we recommend to use a model, which incorporates the Poisson terms in the periodic amplitudes, neglects the small "nutation $\times$ rate" contributions to the Poisson terms in Eqs.~(\ref{EqphiPO} and \ref{eq_WTPo}), but keeps the quadratic terms (we insist that the polynomials of any local model are identical to the polynomials of the global model). Omitting the latter would affect the J2000 epoch values and rates, rapidly resulting in large errors on the planet rotation and orientation as discussed in more details in section~\ref{Section_deg2poly}. Such a simplified rotation model, which we call a local rotation model, would write
\begin{subequations}
\label{Eqmodellocal}
\begin{eqnarray}
\varepsilon(t) &=& \varepsilon_0 + \dot\varepsilon_0 \, t + \varepsilon_Q \, t^2 + \Delta\varepsilon_{\bull}(t) \label{modeepsRS} \\
\psi(t) &=& \psi_0 + \dot\psi_0 \, t + \psi_Q \, t^2 + \Delta\psi_{\bull}(t) \label{modepsiRS} \\
\phi_T(t) &=& \phi_0 + \dot\phi_0 \, t + \phi^T_Q \, t^2 + \Delta\phi_{M,\bull}^{Atm}(t) + \Delta\phi^{Rel}(t) - \cos\varepsilon_0 \, \Delta\psi_{\bull} (t) \label{modephiRS} \\
\alpha(t) &=& \alpha_0 + \dot\alpha_0 \, t + \alpha_Q \, t^2 + \Delta\alpha_{\bull}(t) \label{modealphalocal}\\
\delta(t) &=& \delta_0 + \dot\delta_0 \, t + \delta_Q \, t^2 + \Delta\delta_{\bull}(t) \\
W_T(t) &=& W_0 + \dot W_0 \, t + W_Q^T \, t^2 + \Delta\phi_{M \, {\bull}}^{Atm}(t) + \Delta\phi^{Rel}(t) - \sin\delta_0 \, \Delta\alpha_{\bull}(t) \label{modeWlocal}
\end{eqnarray}
\end{subequations}
where subscript "$_{\bull}$" indicates a local periodic term, obtained by evaluating the amplitude of each Poisson terms at the chosen time (e.g.~in middle of a space mission) and adding it to the amplitude of the periodic term of same period. 
The local periodic terms in orientation angles ($\Delta\varepsilon_{\bull}, \Delta\psi_{\bull}, \Delta\alpha_{\bull}$, and $\Delta\delta_{\bull}$) are readily obtained using the nutation theory that we recommend in Section \ref{sec_rigidnut} that includes both periodic and Poisson terms.
The local periodic spin variations $\Delta\phi_{M \, {\bull}}^{Atm}$ are here formally and by analogy a merger between the periodic $\Delta\phi_M^{Atm}$ and Poisson $\phi^M_\PO$ terms, though such rotation Poisson terms, or any other kind of time-variation in LOD amplitudes, are not predicted by any theory that we know of (i.e.~we expect $\phi^M_\PO = 0$).
\\

The $0.1$ mas accuracy of the local (in time) model as defined above 
is guaranteed only in the close vicinity ($\pm$ 3 years) of the date at which the Poisson terms are evaluated. This makes such a local model good enough for the analysis of the 2.5 years of Viking landers or of the 4 years for InSight separately. A similar approach is used in \citet{Lem23} for a joint analysis of Viking and RISE data, in which two local rotation models, which differ only by the amplitudes of the periodic variations in rotation, are used. 
\\

\subsection{Needed accuracy in the angles}

To ensure the targeted level of precision in the transformed angle, the numerical values of the rotation model in input must be provided with a precision basically equivalent to the target precision. Thus, a precision level of 0.1 mas on the input angles is required to ensure 0.1 mas precision on the output angles. Propagated in the unit relevant for each separated parameter, the precision needed on each type of input parameters of the model (see Eqs.~\ref{Eqmodel}) to ensure 0.1 mas accuracy between 1970 and 2030 in the transformed angles is reported in Tab.~\ref{tab_precision}. If another precision is targeted, one can simply rescale the values of this table by applying a common factor to all quantities since they are all proportional to the transformation precision. \\

\begin{table}[!ht]
\caption{Precision needed in the different quantities in the rotation model in order to have a precision of $0.1$~mas after $30$ years.
The precision on the frequency of a periodic term is computed assuming an amplitude smaller than $1000$ mas, which is valid for the nutation and spin variation amplitudes.}
\label{tab_precision}
\begin{center}
\begin{tabular}{ll} 
\hline\noalign{\smallskip}
Parameter & Target precision = 0.1 mas over +/- 30yrs \\
\noalign{\smallskip}
\hline\noalign{\smallskip}
angle epoch value & $2 \, 10^{-8}$ deg \\ 
rate & $0.003$ mas/y or $2 \, 10^{-12}$ deg/d \\
quadratic coefficient & $0.0001$ mas/y$^2$ \\
annual angular frequency & $9 \, 10 ^{-9}$ rad/d \\
annual period & $0.0007$ d \\
\noalign{\smallskip}\hline
\end{tabular}
\end{center}
\end{table}

\subsection{The degree two polynomials} 
\label{Section_deg2poly}

\begin{table}[!ht]
\caption{Numerical values of the orientation and rotation angles polynomials.
We adapt the solution of \cite{Kon16} to define the epoch values and rates of the angles of Mars. 
The quadratic terms in obliquity and longitude are obtained from the BMAN20.1 model. 
The different terms of the IAU angles are obtained with our transformation and do not depend on the chosen reference orbit. 
\\
}
\label{tab_orbit19802000}
\begin{tabular}{crrr}
\hline\noalign{\smallskip}
 & referred to the 1980 orbit & referred to the J2000 orbit & diff.~orb.~J2000/1980\\
\noalign{\smallskip}\hline\noalign{\smallskip}
\multicolumn{4}{l}{Constant angles (a):}\\
$i_0$ & $1^\circ.85137000$ & $\;\;1^\circ.84972607$ & $\phantom{0}-5,918.1$ mas \\
$\Omega_0$ & $49^\circ.61669995$ & $49^\circ.55807197$ & $-211,060.7$ mas \\
$\varepsilon_{Earth}$ & $23^\circ.43928110$ & $23^\circ.43928093$ & $\phantom{00000}-0.6$ mas \\
$\chi$ & $46^\circ.53072031$ & $46^\circ.47755461$ & $-191,396.5$ mas \\ 
$J$ & $24^\circ.67682669$ & $24^\circ.67706841$ & $\phantom{000}+870.2$ mas \\
$N$ & $\;\; 3^\circ.37919183$ & $\;\; 3^\circ.37321423$ & $\phantom{}-21,519.4$ mas \\
&&& \\
\multicolumn{4}{l}{Euler angles: }\\
$\varepsilon_0$ (b) & $25^\circ.1893823 -1.4$ mas & $25^\circ.19181935$ & $+8774.8$ mas \\
 & $= 25^\circ.18938191$ & & \\
$\psi_0$ & $81^\circ.9683988\;\;$ & $81^\circ.97508039$ & $+24053.7$ mas \\
$\phi_0$ & $133^\circ.386277\;\;\;\;$ & $133^\circ.38489575$ & $-4972.5$ mas \\
$\dot\varepsilon_0$ & $-2.0$ mas/y & $-2.078$ mas/y & $-0.078$ mas/y \\
$\dot\psi_0$ & $-7608.3$ mas/y & $-7607.612$ mas/y & $+0.688$ mas/y \\
$\dot\phi_0$ & $350.891985307$ deg/d & $350.891985306422$ deg/d & $-0.760$ mas/y \\
$\varepsilon_Q$ & $0.0020$ mas/$y^2$ & $0.0020$ mas/$y^2$ & $\sim 0$ \\
$\psi_Q$ & $-0.0144$ mas/$y^2$ & $-0.0144$ mas/$y^2$ & $\sim 0$ \\
$\phi_Q^T$ & free parameter & free parameter & $\sim 0$\\
 \\
\multicolumn{4}{l}{IAU angles (independent from the reference orbit): }\\
$\alpha_0$ & $317^\circ.68111503$ & & \\
$\delta_0$ & $\;\; 52^\circ.88635277$ & & \\
$W_0$ & $176^\circ.63189634$ & & \\
$\dot\alpha_0$ & $-3911.410$ mas/y & & \\
$\dot\delta_0$ & $-2217.109$ mas/y & & \\
$\dot W_0$ & $350.891982443147$ deg/d & & \\
$\alpha_Q$ & $-0.0108$ mas/y$^2$ & & \\
$\delta_Q$ & $0.0159$ mas/y$^2$ & & \\
$W_Q^T$    & $\phi_Q^T - 0.0171$ mas/y$^2$ & & \\
\noalign{\smallskip}\hline
\end{tabular} \\
Euler angles are referred either to the Martian orbit of 1980 or to the Martian orbit of J2000 (differences in the last column). IAU angles are independent from the chosen reference orbit. The constant angles that orientate the reference orbits with respect to the ICRF, and allow to transform between Euler and IAU angles, are also provided. We here respect the precision levels listed in Table~\ref{tab_precision}. \\
(a) The relations between the constant angles are obtained from Eq.~(\ref{eqinertpl}). For the 1980 orbit, we take the values of $\epsilon_{Earth}$, $J$, and $N$ of Table~5 of \cite{Kon06}, and recompute the values of $i_0$, $\Omega_0$, and $\chi$. The latter angles were given with a poor accuracy in \cite{Kon06}. Note that the value of $i_0$ in \cite{Kon06} is affected by a typo. For the J2000 orbit, we start from the values for $i_0$, $\Omega_0$ and $\varepsilon_{Earth}$ from \cite{Sim13} as listed here, and compute $\chi$, $J$ and $N$.\\
(b) The nutations series used in \cite{Kon16} include a constant term that can be seen as a shift in the obliquity value at epoch referred to the J1980 orbit (see \citealt{Kon16,Rea79}). The epoch value referred to the J2000 orbit reported here includes the effect of this shift. 
\end{table}

\begin{table}[!ht]
\caption{Gamma coefficients, computed from Eqs.~\ref{eq_Gammaad}, \ref{eqinvtot} and \ref{eq_Gamma_beta} and values of Table~\ref{tab_orbit19802000}.
These coefficients, to be used in the conversion equations between Euler and IAU angles, are inferred from the orientation values of \cite{Kon16}.}
\label{tab_Gammas}
\begin{tabular}{cll} 
\hline\noalign{\smallskip}
 & referred to the 1980 orbit & referred to the J2000 orbit \\
\noalign{\smallskip}\hline\noalign{\smallskip}
\multicolumn{3}{l}{From Euler to IAU orientation angles: } \\
 $\Gamma_{\alpha\varepsilon}$   &  \;1.1354485 &  \;1.1354776  \\
 $\Gamma_{\alpha\psi}$          &  \;0.5137993 &  \;0.5138341  \\
 $\Gamma_{\delta\varepsilon}$   & -0.7284234 & -0.7284068  \\
 $\Gamma_{\delta\psi}$          &  \;0.2915981 &  \;0.2916320  \\
 $\Gamma_{\alpha\varepsilon\varepsilon}$& -1.0931 & -1.0931 \\
 $\Gamma_{\alpha\varepsilon\psi}$       &  \;1.0354 &  \;1.0353 \\
 $\Gamma_{\alpha\psi\psi}$              & -0.0206 & -0.0206 \\
 $\Gamma_{\delta\varepsilon\varepsilon}$& -0.3102 & -0.3102 \\
 $\Gamma_{\delta\varepsilon\psi}$       &  \;0.3393 &  \;0.3392 \\
 $\Gamma_{\delta\psi\psi}$              &  \;0.0768 &  \;0.0768 \\
 && \\
\multicolumn{3}{l}{From IAU to Euler orientation angles: } \\
 $\Gamma_{\varepsilon\alpha}$   & \;0.4134044 & \;0.4134150 \\
 $\Gamma_{\varepsilon\delta}$   & -0.7284234  & -0.7284068  \\
 $\Gamma_{\psi\alpha}$          & \;1.0327001 & \;1.0325833 \\
 $\Gamma_{\psi\delta}$          & \;1.6097477 & \;1.6096434 \\
 $\Gamma_{\varepsilon\alpha\alpha}$ & \;0.0301 & \;0.0301  \\
 $\Gamma_{\varepsilon\alpha\delta}$ & \;0.0939 & \;0.0938  \\
 $\Gamma_{\varepsilon\delta\delta}$ & \;0.4990 & \;0.4990  \\
 $\Gamma_{\psi\alpha\alpha}$        & -0.5204  & -0.5203   \\
 $\Gamma_{\psi\alpha\delta}$        & -1.1803  & -1.1804   \\
 $\Gamma_{\psi\delta\delta}$        & \;2.4931 & \;2.4926  \\
 && \\
\multicolumn{3}{l}{Coefficients for the $\beta$ angle ($W=\phi+\beta$): }\\
$\Gamma_{\beta\alpha}$       & -0.7974402  & -0.7974402 \\
$\Gamma_{\beta\psi}$         & \;0.9049059 & \;0.9048878 \\
$\Gamma_{\beta\alpha\alpha}$ & \;0.1935    & \;0.1935    \\
$\Gamma_{\beta\alpha\psi}$   & -0.3748     & -0.3749     \\
$\Gamma_{\beta\psi\psi}$     & \;0.0963    & \;0.0963    \\
\noalign{\smallskip}\hline
\end{tabular}
\\
Some $\Gamma$'s are obtained from the $\beta$ angle, computed from Eq.~(\ref{eq_sD}).
We have $\beta_0 = 43^\circ.2456193$ and $\beta_0 = 43^\circ.2470006$ for the reference mean orbits of 1980 and J2000, respectively.
\end{table}

In this section we give some numerical values of the polynomial for each orientation and rotation angle describing the motion of the AM axis, both for the Euler angles and in IAU formulation. 
We demonstrate the importance of the chosen reference plane (Mars mean orbit of J980 versus J2000). We also discuss the reasons for the differences between the values chosen here and those of \cite{Bal20} for the orientation angles. Finally, we stress out the numerical importance of the quadratic terms.

\subsubsection*{The choice of the reference plane}
Euler angles are referred to the mean orbit of Mars, which is orientated with respect to the ICRF thanks to the angles $N$ and $J$ (see Section \ref{sec_rotmatrix_Euler}).
The mean orbit of reference used by \cite{Kon16} is Mars mean orbit of 1980 as first used by \cite{Fol97a}, historically chosen at the Viking epoch.
Since the present orbit of Mars has shifted with respect to the one at the Viking time, it would be appropriate now to use the mean orbit of J2000 as a reference.
\\

The values of $N$ and $J$ for the two orbits are listed in Table~\ref{tab_orbit19802000}, along with the epoch values, rates, and quadratic terms of the Euler angles referred to both orbits. The Euler epoch values and rates are adapted from \cite{Kon16}. 
The different terms of the IAU angles polynomials are obtained with our transformation and do not depend on the chosen reference orbit. 
The $\Gamma$ coefficients of the transformation are given in Table~(\ref{tab_Gammas}). \\

Changing of reference orbit is not anecdotal as it introduces important changes in the epoch values of the Euler angles. The changes in $\varepsilon_0$, $\psi_0$, and $\phi_0$ (ranging from $\sim5000$ to $\sim25,000$ mas) are much larger than the respective uncertainties on these angles (from $\sim10$ to $\sim70$ mas). 
The changes in rates of up to $\sim1$ mas per year are below their present uncertainties. For instance the difference in precession rate 
is only a third of the present uncertainty. However, using obliquity and longitude rates associated with the wrong reference plane leads to angle differences after $30$ years of about $2$ and $20$ mas, respectively. 
Expressing the rotation rate $\dot\phi_0$ with respect to the wrong reference plane would result in a difference of about $10^{-4}$ ms in the length-of-day, or $23$ mas in the rotation angle ($0.4$ m) after $30$ years. 
Note that while an estimated rotation rate is used to build the rotation model of Mars, a relativistic contribution of $7.3088$~mas/d should be removed from the estimated rotation rate to obtain the proper Mars' rotation rate \citep{Bal23}. 
Considering our targeted accuracy, the quadratic terms are left unchanged by the choice between the Mars mean orbit of J1980 or J2000 as a reference plane.

\subsubsection*{Update of the polynomials in orientation angles}

The polynomials for the orientation angles in Table~\ref{tab_orbit19802000} differ from those of BMAN20 for different reasons. First the angles' epoch values of BMAN20 are correctly referred to the J2000 orbit, but not their rates (wrongly referred to the 1980 orbit). 
Using the node longitude rate referred to the J2000 mean orbit, the estimated moment of inertia inferred by the BMAN20 theory shifts by $0.00003$ (1/3 of the error bar) to $C/m_a r_e^2 = 0.36370 \pm 0.00010$. $m_a=6.41712\,10^{23}$~kg and $r_e=3396$ km are the mass and radius of Mars, respectively (\citealt[MRO120D solution]{Kon16}).
Second, we here include in the obliquity epoch value a correction of $-1.4$ mas, corresponding to the constant term of the nutations series used in \cite{Kon16} that can be seen as a shift in the obliquity value at epoch (see \citealt{Kon06,Rea79}). 
Third, $\dot\varepsilon_0$ is set to $0$ in BMAN20, since the gravitational torques acting on the flattened figure of Mars mainly cause precession in longitude. However, actual estimates (e.g.~\citealt{Kon16,Kon20}, \citealt{Kah21} or \citealt{Lem23}) tend to converge to a non-zero value.  
Though there is no physical reason for $\dot\varepsilon_0$ being as large as estimated, a non-zero rate can be considered in the rotation model if desired, as done here. 
Finally the quadratic terms of BMAN20 for the IAU angles were wrongly computed with $\Gamma_{\alpha\varepsilon\varepsilon} = \Gamma_{\alpha\varepsilon\psi} = \Gamma_{\alpha\psi\psi} = 0$.

\subsubsection*{Importance of the quadratic terms}

The values for the quadratic terms in obliquity and longitude reported here in Table \ref{tab_orbit19802000} are obtained from the BMAN20.1 model, the corrected version of the BMAN20 model \citep{Bal20}. 
$\varepsilon_Q$, $\psi_Q$, $\alpha_Q$ and $\delta_Q$ are presently known with a precision below about $0.1\%$. Therefore the numerical values of the quadratic coefficients as provided here are almost sufficiently precise to be considered as well known parameters in rotation determination studies. Only the last digits of the numerical values of Tables~\ref{tab_orbit19802000} and \ref{tab_Gammas} may change if another paper than \cite{Kon16} is chosen as reference for the precession and obliquity rates. As for the spin angle, there is no a priori reasons to have a quadratic coefficient $\phi_Q^M$ bigger than  $-0.0002$ mas/y$^2$, which is the combined contribution of the Sun and Phobos tides. In other words, we expect a priori that $\phi^T_Q\simeq-\cos\varepsilon_0 \, \psi_Q=0.0130$ mas/y$^2$ (see Eq.~\ref{EqphiQ}). Nevertheless, we leave $\phi^T_Q$ as a free parameter in order to accommodate for a possible not predicted phenomenon (see \citealt{Lem23}).
\\

Quadratic terms are an addition to the rotation model commonly used by the community. As such, we can suspect that Mars scientists would be reluctant to include them in their rotation model during data analysis, despite their importance, especially if the data to be analyzed covers a short time span.
Up to 2 or 3 years away from J2000, omitting the quadratic terms may be a good approximation for a lot of applications (errors $\lesssim 0.1$ mas), but that would produce errors as large as $15$ mas $30$ years away from J2000. 
Regarding radioscience data analysis, omitting the quadratic terms would even spoil the determination of the angles' epoch values and rates since the angles polynomial, $a + b\, t + c \, t^2$, would then reduce to their local tangent, i.e.~$(a - c \, t_m^2)+(b+2\, c \, t_m)\, t $ if evaluated at the mission time $t_m$. In the case of RISE, which acquired data around 2020, i.e.~at $t_m=20$ years from the reference epoch of J2000, fitting a first degree polynomial instead of a parabola would introduce a bias on the J2000 epoch value of $-400 c$ ($6$ mas in longitude for instance) and of $40 c$ on the rate ($0.6$ mas/y).

\subsection{Rotation rate - length of day for Mars}
\label{Sec_RotationRate}

Depending on the software used, and therefore on the set of angles chosen, radioscience analyses provide a value of the rotation rate either in terms of $\phi$ ("sidereal rate"), or in terms of angle $W$ ("IAU rate", see Section \ref{sec_rotrat}). Some published values for $\dot\phi_0$ and $\dot W_0$ are gathered in Table~\ref{tab_ratevalue}.
In Section \ref{sec_rotrat} we defined the "stellar rate" $\Omega$, which is independent of the set of angles chosen. The relation between the three rates is given by Eq.~(\ref{eq_phipWp}). For accurate applications, the three rates cannot be used in place of each other. \\

\begin{table}[!ht]
\caption{Numerical values of different rotation rates (non-exhaustive selection).}
\label{tab_ratevalue}
\begin{tabular}{ll} 
\hline\noalign{\smallskip}
 Value ($^\circ$/day) & Reference \\
\noalign{\smallskip}\hline\noalign{\smallskip}
\multicolumn{2}{l}{IAU rate $\dot W_{0}$ } \\
$350.891\,982\,443\,297$ & \citet{Kuc14, Arc18} \\
$350.891\,982\,430\,062$ & \citet{Kuc14, Arc18} with the long periods integrated \\
$350.891\,982\,443\,147$ & Computed from the rotation rate of \citet{Kon16}, see Table \ref{tab_orbit19802000} \\
$350.891\,982\,5$        & \citet{Kah21} \\
 \\
\multicolumn{2}{l}{Sidereal rate $\dot\phi_0$, referred to the 1980 orbit } \\
$350.891\,985\,307$ & \citet{Kon16} \\
$350.891\,985\,377$ & \citet{Kah21} \\
$350.891\,985\,339$ & \citet{Lem23} \\
\\
\multicolumn{2}{l}{Stellar rate $\Omega$ } \\
$350.891\,980\,071$ & Computed from the rotation rate of \citet{Kon16} and Eq.~(\ref{eq_phipWp}) \\
\\
\multicolumn{2}{l}{Uncertainties } \\
$000.000\,000\,002$ & \citet{Kon20} \\
$000.000\,000\,2$   & \citet{Kah21}\\
\noalign{\smallskip}\hline
\end{tabular}\\
The numerical value reported in \cite{Kah21} for the rotation rate (i.e.~$350.8919825$ deg/day) is actually $\dot W_0$ and not $\dot\phi_0$. The correct value for $\dot\phi_0$ is $350.891985377$ deg/day is given in \cite{Lem23}.
\end{table}

In general for Mars, $\dot W_0 < \dot\phi_0 $ and the two rates differ by about $2.9\times 10^{-6}$ deg/day, resulting in a difference of about $113,500$ mas ($0.03$ deg, $1900$ m at the surface) between the two angles after 30 years. Our transformation between the two rates is accurate to the $2.5\times 10^{-12}$ deg/day level (error of less than $0.1$ mas after $30$ years). The transformation of \cite{Kuc14}, as used by \cite{Arc18} to define the IAU Mars rotation model, is less accurate (error of about $500$ mas after $30$ years) because part of the IAU rotation rate is in fact absorbed by the long-period term in their series for $W$ (see section \ref{sec_IAU}). 
To take advantage of the precision (0.1~mas) of the transformation proposed in this paper, a large number of decimals ($12$ digits after the decimals point, in deg/day, see Table~\ref{tab_precision}) must be provided with any  rotation rate estimate. \\

\begin{table}[!ht]
\caption{Numerical values for the different definitions of the Martian day, computed from the rotation rate of \citet{Kon16}.}
\label{tab_day}
\centering
\begin{tabular}{ll} 
\hline\noalign{\smallskip}
& Period (sec) \\
\noalign{\smallskip}\hline
Sidereal day & $88\,642.662\,991\,5$ \\
IAU day      & $88\,642.663\,715\,0$ \\
Stellar day  & $88\,642.664\,314\,3$ \\ 
\\
For comparison: Solar day & $88\,775.244$ \\
\noalign{\smallskip}\hline
\end{tabular}
\\
\end{table}

The ``IAU day'' is larger than the sidereal day by $0.7$ ms. 
$\Omega$ is the lowest rate of the three and the associated stellar day is $0.6$ ms and $1.3$ ms larger than the IAU and sidereal days, respectively (see Table \ref{tab_day}). 
In comparison, the uncertainties of the rotation rates of \cite{Kah21} and \cite{Kon20} correspond to only $0.05$ ms and $0.0005$ ms, respectively, meaning that a distinction between the sidereal and IAU rotation rate is possible and necessary in radioscience analysis.

\subsection{New nutation model and associated Poisson terms}
\label{sec_rigidnut}

The numerical values of the nutation terms are obtained as those of a rigid nutation model multiplied by transfer functions (see Section~\ref{sec_nonri}). Until recently, the transfer functions' parameters had not yet been estimated and default values of $F = 0.07$ and $\sigma_0 = -2\pi/240$ rad/day proposed by \cite{Fol97a} were commonly used (e.g.~\citealt{Kuc14}). We use and recommend here the recently estimated values $F = 0.061 \pm 0.006$ and $\sigma_0 = -2\pi/(243.0 \pm 3.3)$ rad/day from \cite{Lem23}. \\

As for the rigid nutation, several models are available in the literature \citep{Rea79, Borderies1980a, Borderies1980b, Groten, Hilton, Bou99, Roo99}. 
\cite{Bal20} did a review and proposed a new rigid nutation model (BMAN20) that makes a synthesis between the different components needed to achieve an accuracy of about $0.1$ mas level in the time domain. 
A shortened version, called BMAN20RS (RS for radioscience), was proposed to be used in a locally defined rotation model (see Section~\ref{Sec_local_model}). BMAN20RS groups the main periodic terms with the Poisson terms and with small amplitude terms of period close to that of the main terms. The so-obtained limited subset of terms reproduces at best the behavior of the BMAN20 solution at the time of RISE observations.
The work of the present paper, and in particular the definition of a second-order transformation between Euler and IAU angles, allowed us to update and correct the precession/nutation model of \cite{Bal20}. The updated BMAN20.1 model (and its associated local versions for the RISE and Viking epochs for a total of three models), in both Euler and IAU angles, can be found at \href{https://doi.org/10.24414/h5pn-7n71}{https://doi.org/10.24414/h5pn-7n71}. With a truncation criterion of $0.025$ mas in prograde and/or retrograde amplitude, BMAN20.1 includes 31 periodic nutation terms. We recommend this rigid nutation model to define the Mars rotation model, and to rescale it to any preferred value of the dynamical flattening $H_D$ if deemed necessary. $H_D$=$(C-\bar A)/C$ is the ratio of the difference between the polar and the mean equatorial moments of inertia over the polar moment of inertia and contributes to the nutation amplitudes (except the geodetic one). 
Note that all BMAN20 rigid nutation series are referred to the Mars J2000 orbit reference plane. To express the nutations with respect to the J1980 reference plane instead, a projection has to be applied, which leads to amplitude corrections below $\sim 0.1$ mas.\\

\begin{figure}[!htb]
\centering
\includegraphics[height=7.5cm]{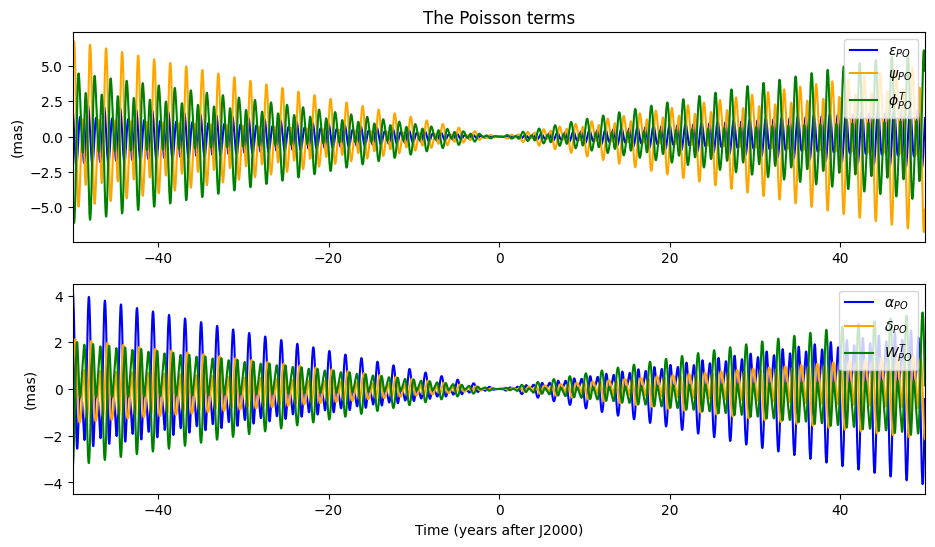} 
\caption{Temporal evolution of the Poisson terms for the 6 orientation and rotation angles over 100 years.}
\label{fig_poi}
\end{figure}

The BMAN20.1 model also includes $5$ Poisson terms in the orientation angles. Their amplitude increases with time, so neglecting them can quickly lead to significant errors, as shown on Fig.~\ref{fig_poi}, where the sum of Poisson terms already reaches 4 mas after $30$ years. With its $31$ nutation and $5$ Poisson terms, the BMAN20.1 model accuracy over $30$ years from J2000 is of $0.1$ mas in prograde/retrograde formulation (corresponding to $0.1$ mas in obliquity and $0.2$ mas in longitude), compared to the complete semi-analytical series obtained before truncation.
\\

As a conclusion, we recommend the use of BMAN20.1 as explained in details in \cite{Bal20} and here above. The use of the rigid nutation model of \cite{Rea79}, as still done by default in MONTE, introduces significant errors (at the level of tens of mas) mostly due to the absence of Phobos and Deimos nutations in \cite{Rea79}, and, to a lesser extent, because the latter positions the axis of figure of Mars and not the spin axis.

\subsection{Series for the periodic variations in rotation angles $\phi$ and $W$} 
\label{sec_ampLOD}

\subsubsection*{Different ways to express the periodic rotation variations}
The periodic variations of the rotation of Mars inform us on its atmosphere and ice caps dynamics.  The amplitudes of these periodic variations can be very different depending on the chosen formulation (e.g.~phases in or out of the sine and cosine arguments) on whether one considers $\phi$ or $W$ as the rotation angle, on whether one considers the variations along the true or mean equator of date, and also on whether one includes or not the relativistic corrections (see Section \ref{sec_LOD} for the definitions and relations between $\Delta\phi_T, \Delta\phi_M, \Delta W_T$, $\Delta W_M$, $\Delta\phi_M^{Atm}$, and $\Delta\phi^{Rel}$).\\

This is illustrated in Table~\ref{tabphi} where the components of the periodic variations of the rotation are numerically compared for the semi-annual wave. One can see for instance that, in the formulation where the phase $2\,l'_0$ is in the arguments, the total cosine semi-annual amplitudes in $\Delta\phi_T$ and $\Delta W_T$ are particularly different, with amplitudes of $-734$ mas and $-14$~mas, respectively. That difference comes entirely from the nutation terms $-\Delta\psi \cos\varepsilon_0$ and $-\Delta\alpha \sin\delta_0$ since the other components are the same (see Eqs.~\ref{eq_deltaphi}): $\Delta\phi_M = \Delta W_M = -103$ mas. Still in the same formulation, the sine components of $\Delta\phi_M$ and $\Delta W_M$ can be split into atmospheric ($\Delta\phi_M^{Atm}$, $-93$~mas) and relativistic ($\Delta\phi^{Rel}$, $-8$~mas) contributions.
A final source of differences between numerical amplitudes of the periodic variations in rotation could be related to the nutation model used. 
One should systematically use a non-rigid nutation model, since the contribution of the liquid core is large (about $35$ mas here in $\phi_T$). Note that Table~\ref{tabphi} is given for illustrative purpose only. 
\\

Now that we made clear the fact that a same model of spin variations can be provided numerically under various forms, we present below recommendations for building periodic series in rotation for Mars.
\\

\begin{table}[!ht]
\caption{Semi-annual amplitudes (in mas) of the periodic series included in the rotation angles $\phi_T$ and $W_T$, see Eqs. (\ref{eq_deltaphi}, \ref{modephi}, and \ref{modeW}), computed from the values of \citet{Kon16}.}
\label{tabphi}
\begin{tabular}{lllll} 
\hline\noalign{\smallskip}
Angle/correction & $\cos{2 \, (n \, t + l'_0)}$ & $\sin{2 \, (n \, t + l'_0)}$ & $\cos{2 \, n\, t}$ & $\sin{2 \, n \, t}$ \\
\noalign{\smallskip}\hline
$\Delta\phi_M^{Atm}$ & -103. & -93. & -138.5 & -8.1 \\
$\Delta\phi^{Rel}$   &    0. & -8.2 &  -5.1 & -6.4 \\
$\Delta\phi_M = \Delta W_M = \Delta\phi_M^{Atm} + \Delta\phi^{Rel}$ & -103. & -101.2 & -143.7 & -14.5 \\
$\Delta\psi \cos\varepsilon_0$ & 634.9 & -849.9 & --36.8 & -1060.2 \\
$\Delta\alpha \sin\delta_0$ & -89.4 & -678.0 & -494.0 & -472.8 \\
$\Delta\phi_T = \Delta\phi_M - \Delta\psi \cos\varepsilon_0$ & -737.9 & 748.7 & -106.9 & 1045.7 \\
$\Delta W_T = \Delta\phi_M - \Delta\alpha \sin\delta_0$ & -13.6 & 576.8 & 350.4 & 458.4 \\
\noalign{\smallskip}\hline
\end{tabular}
\\
Although we do not recommend the relativistic and nutation corrections used by \cite{Kon06, Kon16}, we use them here for the sake of consistency. 
Thus, the amplitudes of $\Delta\phi^{Rel}$ are taken from \cite{Yod97} and the nutation amplitudes of $\Delta\psi$ and $\Delta\alpha$ assume a non rigid model with core parameters $F = 0.07$ and $\sigma_0 = -1.5^\circ$/d based on \cite{Rea79}, with $q=142^\circ$.
We consider both the notations conventions where the phase $2\,l'_0$ is in (left columns, as in \citealt{Kon16}) or out (right columns, pure frequencies representation, see Section \ref{sec_nutationrepresent}) of the cosine and sine arguments.
We use the expression of Table \ref{tab_arg} or Eq.~\ref{eq_l} for $l'$, $l'_0$ being the epoch value.
\end{table}

\subsubsection*{Extra periodic terms in the rotation angle from nutation and relativistic effects}

In general the published series (e.g.~\citealt{Kuc14, Kon20, Lem23}) are those of $\Delta\phi_M^{Atm}$, following the habits introduced in \cite{Kon06}.
Although only the amplitudes of the annual, semi-annual, ter-annual and quater-annual waves are estimated\footnote{The estimated amplitudes of the other periods are compatible with zero \citep{Lem23}} in $\Delta\phi_M^{Atm}$, the model of rotation to implement in the software used to process the radioscience data must include  more than $4$ terms to accurately define $\Delta\phi_M=\Delta W_M$, $\Delta\phi_T$, and $\Delta W_T$,  as detailed below. 
\\

In \cite{Kon06} and the subsequent studies, the nutation series includes only 6 harmonic terms, and so does $\Delta\phi_T$. If only the first four harmonic terms of the nutation series are considered in $\Delta\phi_T$ or $\Delta W_T$ to take the same periods as in $\Delta\phi_M^{Atm}$, this causes an error of $3$~mas if compared to a model with the 6 harmonics terms.
Assuming that a local rotation model is considered and to reach the targeted accuracy of $0.1$~mas, we recommend to define the different rotation angles by using one of the RS/local versions of the nutation model BMAN20.1 as described in Section \ref{sec_rigidnut}, which include 31 terms at various periods, and in particular 2 large terms induced by Phobos and Deimos, with periods corresponding to their respective precession period (826 and 20,000 days). If only the first six harmonic terms of the nutation series are considered in the rotation angle $\Delta\phi_T$ or $\Delta W_T$, this causes an error of $7$~mas if compared to the complete model.
\\

The series for $\Delta\phi^{Rel}$ used in \cite{Kon06} and in the subsequent studies is that of \cite{Yod97}, rounded ($-176 \, \sin l' - 8 \, \sin 2l' - 1 \sin 3l'$). We recommend instead the series of \cite{Bal23}, with amplitudes given here in mas: 
\begin{eqnarray}
\Delta\phi^{Rel}(t) &=& -166.954 \sin l' - 7.783 \sin 2l' - 0.544 \sin 3l' \nonumber \\
&& \quad + \, 0.567 \sin\left(\frac{2\pi}{816.441} t + 320^\circ.997\right) + 0.102 \sin\left(\frac{2\pi}{733.833} t + 303^\circ.752\right), 
\label{eqrel} \\
\mathrm{with}\ \ l'&=& 19^\circ.37276563 + 0.009145886245716337\, t. 
\label{eq_l}
\end{eqnarray}
$t$ is the time in Julian days starting in J2000. The difference in the annual term with respect to \cite{Yod97} is of $9$ mas. There are also in addition two periodic terms at the Mars-Jupiter (2.24 years) and Mars-Saturn (2.01 years) synodic periods. Since the period of the latter are close the orbital period of Mars (1.88 year), not taking them into account in the rotation model would likely affect the estimate of the annual term at the mas level. Note that at least one digit after the decimal point in the amplitudes must be taken into account to guarantee the precision of 0.1 mas.\\

\subsubsection*{Recommended values for the arguments of the periodic rotation variations arguments}

\begin{table}[!ht]
\caption{Selected arguments of the periodic series, taken from \cite{Bal20, Bal23}.}
\label{tab_arg}
\begin{tabular}{llc} 
\hline\noalign{\smallskip}
Argument & Recommended value (in rad) & Period (in days) \\ 
\noalign{\smallskip}\hline
\multicolumn{2}{l}{Mean anomalies (for $\Delta\phi_M^{Atm}$ and $\Delta\phi^{Rel}$)} & \\
$l_{S\!a}$ &\,$5.53304687684+213.2002152909 \, T$ & 10764.22 \\
$l_{J\!u}$ &\,$0.52395267692+529.6533496052 \, T$ & 4332.897 \\
$l'$       &\,$0.3381185455+3340.5349512479 \, T$ & 686.9958 \\
& \\
\multicolumn{2}{l}{Mean longitudes (for nutations)} \\
$S\!a$& $0.87401678345 + 213.2990797783\,T$ & 10759.23 \\
$J\!u$& $0.59954667809 + 529.6909721118\,T$ & 4332.589 \\
$M\!a$& $6.20349959869 + 3340.6124347175\,T$ & 686.9799 \\
\noalign{\smallskip}\hline
\end{tabular}
\\
In $\Delta\phi_M^{Atm}$ and $\Delta\phi^{Rel}$, the arguments are multiple of the mean anomaly of Mars ($l'$) for the harmonics terms, or derived from the mean anomalies of Saturn ($l_{S\!a}$), Jupiter ($l_{J\!u}$) and Mars for the synodic terms. 
Note that in Eq.~(\ref{eqrel}), the phases of the synodic terms are not the phases difference of the mean anomalies. 
In the nutation series, the arguments are linear combinations of fundamental arguments which include the mean longitudes of Saturn ($S\!a$), Jupiter ($J\!u$), and Mars ($M\!a$). $T$ is the time measured in thousands of Julian years from J2000.
\end{table}

\begin{figure}[!htb]
\centering
\includegraphics[width=15cm]{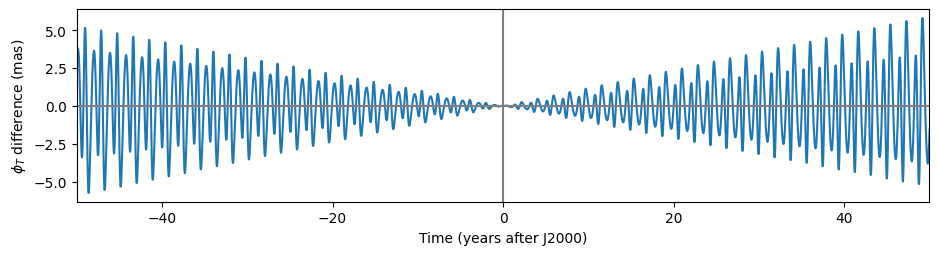}
\caption{Error in the rotation angle $\phi_T$ resulting from an error of about 0.016 day in the revolution period of Mars. This illustrates the impact of misusing the rate of the mean longitude $M\!a$ (686.9799 days, see table \ref{tab_arg}) instead of the rate of the mean anomaly $l'$ (686.9958 days) when evaluating $\Delta\phi^{Atm}_M$ and  $\Delta\phi^{Rel}$ to compute $\phi_T$.}
\label{fig_n}
\end{figure}

On the one hand, Mars seasonal atmospheric phenomenon (e.g.~surface pressure variations) are driven by the Sun-Mars angle with respect to the Northern spring equinox, the so-called true aerocentric solar longitude denoted $L_s$, which is usually expressed as a function of the mean anomaly $l'$ ($L_S\simeq 251^\circ+l'+11^\circ \sin l'$, \citealt{Kon06}). As a result, the atmospheric seasonal variations in rotation are modeled in terms of $l'$ (e.g.~\citealt{Yod97,Fol97a}). The relativistic corrections use also the mean anomaly of Mars as main argument \citep{Bal23}. On the other hand, the periodic nutations are now modeled in terms of the mean longitude of Mars, here denoted $M\!a$, along other arguments, since the position of the Sun in Mars BF is obtained from ephemerides using mean longitudes as argument instead of mean anomalies \citep{Roo99,Bal20}. Therefore the epoch values of the arguments (or phase) of the harmonic and synodic terms of $\Delta\phi_M^{Atm}$ and $\Delta\phi^{Rel}$ differ from those of the recommended nutations series (see Table \ref{tab_arg}). For instance the annual terms in $\Delta\phi_M^{Atm}$ has a phase of $19^\circ.37$ whereas the annual nutation terms have a phase of $355^\circ.43$. 
The rates of the arguments (or frequency) also slightly differ from each other. One has for instance $0^\circ.004/$year difference between Mars Mean anomaly $l'$ (for the atmospheric and the relativistic series) and Mars Mean longitude $M\!a$ (for the nutation series), corresponding to a difference of 0.016 day in the period. 
Because of these slight frequency differences, good practice would be to keep the series for the periodic variations in rotation separated from the nutation corrections series (i.e.~not adding the amplitude of the annual LOD to that of the annual nutation correction). Mixing the two (e.g.~using mean longitude rate instead of mean anomaly rate for LOD), leads to an error increasing with time and reaching $\sim$4 mas after 30 years, see Fig.~\ref{fig_n}.
In the end the series for $\Delta\phi_T$ and $\Delta W_T$ should thus include 37 terms (31 from nutation, 4 from atmosphere and 2 extra relativistic terms) to meet the targeted accuracy of $0.1$ mas.
\\

The python code provided to the user transforms the different series and takes into account the nutations and the relativistic corrections in the transformations for all the frequencies.

\subsection{Series for the polar motion}
\label{sec_PMnum}

As explained in Section \ref{sec_PM}, the polar motion (PM) is decomposed into two parts caused by the atmosphere and the ice caps dynamics ($X_P^{Atm}, Y_P^{Atm}$), and by the external torque ($X_P^{Ext}, Y_P^{Ext}$), see Eq.~(\ref{eq_pmSum}). The first part has been modeled \citep{Yod97,Def00,Van02} and recently partly measured from landers and orbiters \citep{Kon20}. The second part can be obtained as a byproduct of the nutation theory (Section \ref{sec_PM}).\\

$X_P^{Atm}$ and $Y_P^{Atm}$ are usually written (e.g.~\citealt{Kon06}) as a series of 4 terms with periods being the harmonics of the orbital period of Mars and of one term at the Chandler wobble (CW) period:
\begin{subequations}
\label{Eq_PMAtm}
\begin{eqnarray}
X_P^{Atm} &=& (X^c_{CW} \cos f_{CW} + X^s_{CW} \sin f_{CW})+ \sum_{i=1}^4 (X^c_i \cos i\,l'(t) + X^s_i \sin i\,l'(t))\\
Y_P^{Atm} &=& (Y^c_{CW} \cos f_{CW} + Y^s_{CW} \sin f_{CW})+ \sum_{i=1}^4 (Y^c_i \cos i\,l'(t) + Y^s_i \sin i\,l'(t)).
\end{eqnarray}
\end{subequations}
The argument of the CW terms, $f_{CW}$, should be taken equal to $3.3188\, l'_0 + (2\pi/206.9)\, t$, with $t$ the time in Julian days starting in J2000 \citep{Kon20}. Similarly to the rotation periodic variations (see table~\ref{tab_arg}), the arguments of the polar motion are functions of the Martian mean anomaly, $l'(t)$, and not the mean longitude. The amplitudes of the CW and of the harmonic terms are given in Table~\ref{tab_PM}. Note that the amplitudes of the harmonic terms reported in \cite{Kon20} are combinations of polar motion and seasonal changes in the even degree and order one gravity coefficients which cannot be separated by spacecraft data. 
\\

\begin{table}[!ht]
\caption{Polar motion amplitudes (mas). For the atmospheric PM, the numerical values come from \cite{Kon20}. 
The values for the external PM are computed here starting from the nutation model BMAN20.1 as recommended in Section \ref{sec_rigidnut} for the rigid case. 
Note that in fact the amplitudes from \cite{Kon20} are a combination of atmospheric polar motion and time varying gravity contributions.}
\label{tab_PM}
\begin{tabular}{lllll} 
\hline\noalign{\smallskip}
Period (days) & X cos & X sin & Y cos & Y sin \\ 
\noalign{\smallskip}\hline
\multicolumn{5}{l}{Atmospheric PM, amplitudes in the representation of Eq.~(\ref{Eq_PMAtm})}  \\
$206.9$ (CW) & -1.7  & 6.5  & 5.1  &  1.2 \\
$686.995786$ & -17.6 & 23.3 & -8.6 &  0.6 \\
$343.497893$ & -10.9 &  3.4 & -1.9 & -0.4 \\
$228.998595$ & -0.6  &  0.9 & -6.8 & -2.0 \\
$171.748946$ & -7.3  &  1.7 & -3.8 &  1.3 \\
& & & &\\
\multicolumn{5}{l}{Atmospheric PM, amplitudes in the pure frequency representation} \\
$206.9$ (CW) & 5.1  & 4.4  & 3.3  & -4.1 \\
$686.995786$ & -8.9 & 27.8 & -7.9 & 3.4 \\
$343.497893$ & -6.4 &  9.5 & -1.7 & 0.9 \\
$228.998595$ & 0.4  &  1.0 & -5.3 & 4.7 \\
$171.748946$ & 0.1  &  7.5 &  0.4 & 4.0 \\
& & & &\\
\multicolumn{5}{l}{External PM, amplitudes in the pure frequency representation} \\
0.341986	&-0.005	&-0.031	& 0.031	&-0.005\\
1.021381	& 0.011	&-0.018	& 0.018	& 0.011\\
1.022901	& 0.097	&-0.035	& 0.011	& 0.001\\
1.024427	&-0.096	& 0.181	&-0.181	&-0.096\\
1.027491	&-0.152	& 0.020	&-0.020	&-0.152\\
1.029030	&-1.164	& 1.013	&-0.991	&-1.066\\
1.030574	&-0.449	& 0.190	&-0.190	&-0.449\\
1.032122	&-0.110	& 0.007	&-0.007	&-0.110\\
\noalign{\smallskip}\hline
\end{tabular}
\\
The series for the external Polar Motion in the rigid and non rigid cases are available at \href{https://doi.org/10.24414/h5pn-7n71}{https://doi.org/10.24414/h5pn-7n71}.
\end{table}

\begin{figure}[!htb]
\centering
\includegraphics[height=8cm]{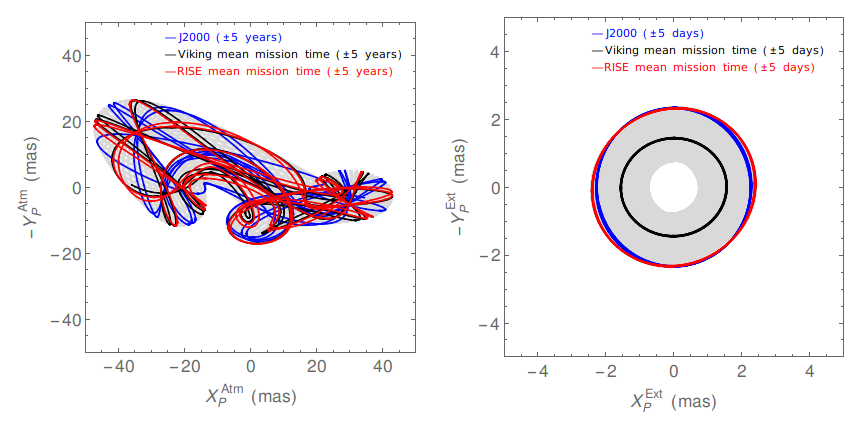} 
\caption{Atmospheric ($X_P^{Atm}$ and $Y_P^{Atm}$) and external ($X_P^{Ext}$ and $Y_P^{Ext}$) polar motion components projected on the equatorial plane for different epochs. 
The gray area shows the polar motion on the 1950-2050 period.
Even if the external polar motion is mostly diurnal,
its amplitude varies between 0.7 and 2.4 mas over a Martian year, which corresponds to the beating period of the 2 largest terms of table \ref{tab_PM}.
Note that the amplitudes from \cite{Kon20} are a combination of polar motion and time varying gravity contributions.}
\label{fig_PM}
\end{figure}

To express $X_P^{Ext}$ and $Y_P^{Ext}$, we consider the BMAN20.1 nutation model and start from Eqs.~(\ref{EqPMext}-\ref{EqPMext2}) in which we replace the retrograde and prograde nutation amplitudes $\mathcal{P}_i$ and $\mathcal{R}_i$ by their non rigid counterparts $\mathcal{P}'_i$ and $\mathcal{R}'_i$, see Eq.~(\ref{eq_Fsig}). Then we express $X_P^{Ext}$ and $Y_P^{Ext}$ in a pure frequency formulation and keep the 8 largest terms in order to ensure the $0.1$ mas accuracy level in the time domain:
\begin{subequations}
\begin{eqnarray}
X_P^{Ext} &=& \sum_{i=1}^8 (X^c_i \cos f_i(t) + X^s_i \sin f_i(t))\\
Y_P^{Ext} &=& \sum_{i=1}^8 (Y^c_i \cos f_i(t) + Y^s_i \sin f_i(t))
\end{eqnarray}
\end{subequations}
The amplitudes are given in Table~\ref{tab_PM}. The total external polar motion reaches about $2$ mas (see Fig.~\ref{fig_PM}, right panel), the effect of the liquid core being just below $0.1$ mas (not shown). Fig \ref{fig_PM}, left panel, shows the trajectory of the atmospheric polar motion projected on the equatorial plane at different epochs, which can reach about $50$ mas. Considering the external polar motion in the rotation matrix of Eq.~(\ref{eq_landerpos}) is necessary if the chosen nutation model is for the AM/spin axis, as is the case of BMAN20.1 (note that the nutation series of \citealt{Rea79} is for the figure axis). \\

\subsection{Transformation accuracy of the method}
\label{Section68}

\begin{figure}[!htb]
\centering
\includegraphics[width=16cm, height=9cm]{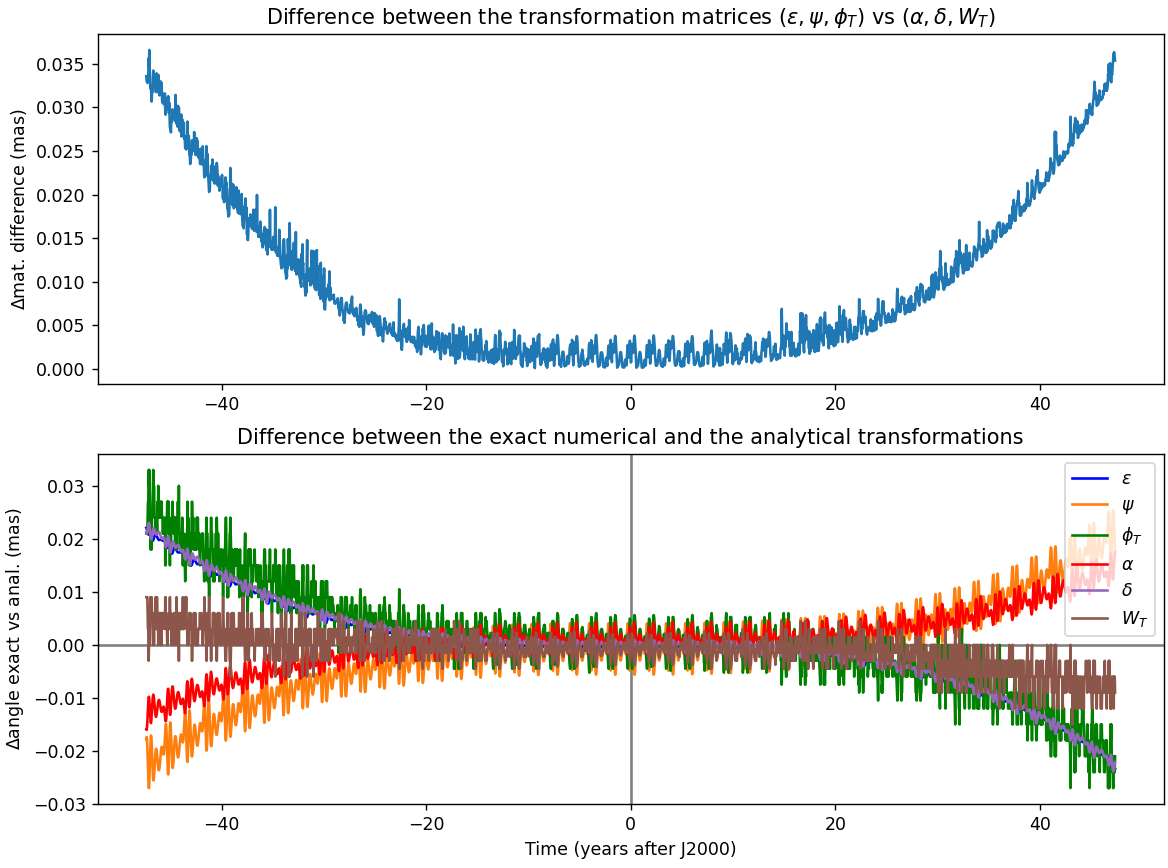} 
\caption{Difference in mas between (top) the transformation matrices computed with the Euler angles and with the IAU angles and (bottom) between the exact numerical transformation and the analytical model for the 6 angles separately.}
\label{fig_diff}
\end{figure}

In this section we quantify the level of precision of our analytical transformed model as a function of time. As shown in Figure \ref{fig_diff}, the analytical transformation between the two sets of angles described in Sections~\ref{sec_transf} and \ref{sec_rotang} with the prescriptions given in this section differs by less than the 0.1 mas target precision with respect to the exact numerical transformation, between 1970 and 2030. The differences mostly show a third-degree (cubic) behavior and tiny periodic oscillations.
By design, the difference between the transformation matrices obtained with the different sets of angles (top panel of Fig.~\ref{fig_diff}) is at the same level as the differences on the angles separately (bottom panel of Fig.~\ref{fig_diff}).
\\

On the long run (more than $50$ years or so), the approximate transformation described in this paper, even with the quadratic terms included, will not accurately account for the temporal evolution of the Martian angles, if the targeted accuracy is of $0.1$ mas, because this transformation is designed to be precise around J2000. 
Far away from that epoch of reference, the neglected higher order terms (like the third order terms of the transformation) become larger than $0.1$ mas, even though our transformation method still guaranties a precision of 0.3 mas before and after 100 years.
The astronomical reference epoch will change at some point (presumably from J2000 to J2050), and a transformation designed to be precise around J2050 could be used then, to analyze the most recent data.
\\

This shows the range of validity of our analytical transformation, which will not introduce significant bias in the manipulation of the Martian orientation and rotation parameters (MOPs), compared with the current or future uncertainty on their estimates, largely dominated by the noise on the radiometric data.
Our transformation model is at least one order of magnitude more accurate than the current precision on the MOPs estimates. For example, the rotation rate $\dot\phi_0$ of \citet{Kon20} is known to $2 \, 10^{-9}$ deg/d, corresponding to 3 orders of magnitude larger than the targeted precision. \\

We numerically checked that the process is bijective: after applying the transformation process and its inverse, the difference stays below the chosen level of precision.

\section{Recommendations for the future IAU solution}
\label{sec_IAU}

The IAU Working Group on Cartographic Coordinates and Rotational Elements (WGCCRE) regularly updates the rotational elements of the planets, describing the direction of the North pole of rotation and the position of the prime meridian, see \citet{Arc18} for the last report. 
These models are very useful, for example to make cartographic products. 
In this section we discuss the IAU-like representations published within the last 15 years 
and propose suggestions on how to construct a future IAU recommended solution for the rotational elements of Mars, in a consistent and precise manner.
We build on the discussion started in \citet[section 5]{Bal20} regarding the limitations of the current IAU recommended solution and provide additional arguments.
\\

Up to now, three papers provide a solution for the rotational elements of Mars in the IAU formulation including periodic variations in addition to the first-degree polynomials used so far.
The first one is \citet{Jac10}, which provides IAU angles derived from the Euler angles solution estimated by \cite{Kon06}.
The author provides for each angle an expression with an epoch value, a linear trend, four or five short-period terms and a long period modulation.
The second paper is \citet{Kuc14}. It provides both Euler angles, estimated from radioscience data, and IAU angles, inferred from the former using \citet{Jac10} method, which correspond to the most recent solution recommended by the IAU for Mars \citep{Arc18}.
The third paper \citep{Jac18} gives an "IAU" model based on the "Euler" solution of \citet{Kon16}, still using the same conversion method as in \citet{Jac10}.

\subsection{The physical irrelevance of the long period terms}
\label{Section71}
In the three studies cited above, the authors use exact geometric relations (see Sec.~\ref{sec_exacttransfo}) to numerically transform time series of the Euler angles into time series of the IAU angles, from which they retrieve the trigonometric series using a frequency analysis. 
They all find, for each angle, one term characterized by a very long period (about $71,000$ years in \citealt{Jac10}, and \citealt{Kuc14}, and $170,000$ years in \citealt{Jac18}) and a large amplitude (from $0.6$ degrees for $W$ in \citealt{Jac10}, and \citealt{Kuc14}; up to $8.8$ degrees for $\delta$ in \citealt{Jac18}).\\

These large very-long period terms are artifacts inherent in their method since they are not predicted by any models of atmosphere dynamics or celestial mechanics and the method does not take into account the physical meaning of the planet's rotational dynamics.
The nutation series of BMAN20 \citep{Bal20}, for both the Euler angles $(\varepsilon,\psi)$ and the IAU angles $(\alpha,\delta)$, include mainly terms with periods below $\lesssim 50$ years. The only four terms with longer periods that exist in BMAN20 have periods ranging from 800 to $364,164$ years. These long period terms have small amplitudes ($<0.5$ mas or $10^{-7}$ degree) and cannot be observed on a short time scale (i.e.~few decades).
Even if detected, they would be perceived as secular terms, that is to say as contributions to the precession.
Similarly for the angle $W$, no long periods appear in the rotation models derived from Global Circulation Models of Mars atmosphere (e.g.~\citealt{Def00, Van02, Kar06, Kar11}), which only give periodic variations at the revolution period of Mars and its sub-harmonics.
It is therefore not expected to see periodic terms in IAU angles with periods and amplitudes as large as those of the long period terms of \citet{Jac10}, \citet{Kuc14}, and \citet{Jac18}.
\\

We actually claim that the long-period terms fitted by these authors are used to absorb the quadratic signal resulting from the geometric transformation 
(see section \ref{sec_app}, where we have identified the quadratic signal in an analytical transformation) because even when there are no quadratic terms in the initial Euler angles, quadratic terms appear in the corresponding IAU angles (see Eqs.~\ref{eq_del3}, \ref{eq_del4}, \ref{eq_phiTQ} with $\varepsilon_Q = \psi_Q = \phi^T_Q = 0$). \\

On an interval of a few tens of years around the reference epoch, a very long period signal with well chosen values for the amplitude, phase and frequency can well mimic a quadratic behavior.
As shown below for $\alpha$, the sum of a degree 1 polynomial and of a long-period term can be well approximated near the time origin ($t=0$) by a degree 2 polynomial in the time $t$:
\begin{subequations}
\begin{alignat}{4}
\alpha_0' + \dot\alpha_0' \, t + A \sin(f\, t + \varphi_0) & \approx
(\alpha_0' + A \sin\varphi_0) && + (\dot\alpha_0' + A \, f \, \cos\varphi_0) \, t && - \frac{A \, f^2 \sin\varphi_0}{2} \, t^2 \\
& \approx \alpha_0 && + \dot\alpha_0 \, t && + \alpha_Q \, t^2 
\label{eq_LP}
\end{alignat}
\end{subequations}
As we can see in Eq.~(\ref{eq_LP}), adding a long period modulation with a large amplitude $A$ instead of a quadratic term largely and artificially alters the epoch value (the difference being $A \sin\varphi_0$) and rate ($A \, f \, \cos\varphi_0$). In the three aforementioned studies, the difference can be up to about $9$ degrees for the epoch value of the angles and up to $100$ mas/year for their rate, see the table \ref{tab_IAU} for the numerical values corresponding to \cite{Kuc14}. 
Besides, using a long period term instead of a quadratic term requires to provide 3 additional numerical quantities (amplitude, phase and frequency) for each angle instead of one. The set of numerical values for ($\alpha_0', \dot\alpha_0',A,f,\varphi_0$) is therefore not unique. Assumptions can be made (for instance the long period is arbitrarily fixed and identical for the 3 angles), but they lack of meaning as explained above. 

\begin{table}[!ht]
\caption{IAU2009 and IAU2015 Rotation models for the $\alpha, \delta, W_T$ angles of \cite{Kuc14} as published (with long-period terms) and as computed here (with quadratic terms replacing the long-period terms).}
\label{tab_IAU}
\begin{tabular}{llll} 
\hline\noalign{\smallskip}
Angle & Epoch Value & Rate & Long Period/Quadratic term \\ 
\noalign{\smallskip} \hline
\multicolumn{4}{l}{IAU2009}\\
$\alpha$ & $317.68143^\circ$ & $-10.45 \textrm{ mas } t = -0.1061^\circ \, T$\\ 
$\delta$ & $52.88650^\circ$  & $-6.002 \textrm{ mas } t = -0.0609^\circ \, T$\\
$W_T$    & $176.630^\circ$   & $+ 350.891\,982\,26^\circ \, t$\\
&&&\\
\multicolumn{4}{l}{IAU2015 \citep{Kuc14,Arc18} solution with long-period term}\\
$\alpha$ & $317.269\,202^\circ$ & $- 10.770\,478 \textrm{ mas } t$ 
& $+ 0.419057^\circ \sin(79.398797^\circ + 0.5042615^\circ \, T)$ \\ 
$\delta$ & $\; 54.432\,516^\circ$ & $\; - 5.743\,348 \textrm{ mas } t$ 
& $+ 1.591274^\circ \cos(166.325722^\circ + 0.5042615^\circ \, T)$ \\
$W_T$    & $176.049\,863^\circ$ & $+ 350.891\,982\,443\,297^\circ \, t$ 
& $+ 0.584542^\circ \sin(95.391654^\circ + 0.5042615^\circ \, T)$ \\
&&&\\
\multicolumn{3}{l}{\cite{Kuc14} solution with quadratic terms (this study)}\\
$\alpha$ & $317.681\,106\,09^\circ$ & $ - 10.702\,654 \textrm{ mas } t$ & $- \; 58 \textrm{ mas } T^2$ \\
$\delta$ & $\; 52.886\,346\,21^\circ$ & $\; - 6.070\,351 \textrm{ mas } t$ & $+ 215 \textrm{ mas } T^2$ \\
$W_T$    & $176.631\,817\,55^\circ$ & $+ 350.891\,982\,430^\circ \, t$ & $- \; 81 \textrm{ mas } T^2$ \\
\noalign{\smallskip}\hline
\end{tabular}

$T$ is the time in century, $t$ the time in days. 
\end{table}

\subsection{Accuracy}
\label{Section72}

In addition to the lack of physical meaning, the precision of the current IAU solution is uncertain. IAU solutions are given without information as to their accuracy, that depends both on the accuracy of the initial Euler estimates and modeling, and on the accuracy of the transformation to IAU angles. A detailed error budget analysis for the transformed IAU angles should be carried on for the future IAU reference solution. \\

The current solution adopted by the WGCCRE for Mars rotation and orientation is based on the one hand on estimates for the epoch values and rates of the Euler angles as obtained from a fit to the radioscience data. The uncertainties of several tens of mas on the epoch values and of several mas/y on the rates translate into similar level of uncertainties in the IAU angles epoch values and rates, respectively. 
On the other hand, the periodic nutations series in $\varepsilon$ and $\psi$ come from the model of \cite{Rea79}, transformed into periodic series of the IAU angles, introducing a cumulative bias of the order of 10 mas or more. First the original series is for the figure axis and not the rotation axis (the difference being of the order of 3 mas, see Section \ref{sec_PM}). Second the series does not include the nutations induced by Phobos and Deimos ($10$ mas and $4$ mas amplitude). Third, the transfer function parameters ($F$ and $\sigma_0$, see Section \ref{sec_nonri}) are fixed to values slightly different from that recently measured by \cite{Lem23}.
\\

\begin{figure}[!htb]
\includegraphics[height=6cm, width=18cm]{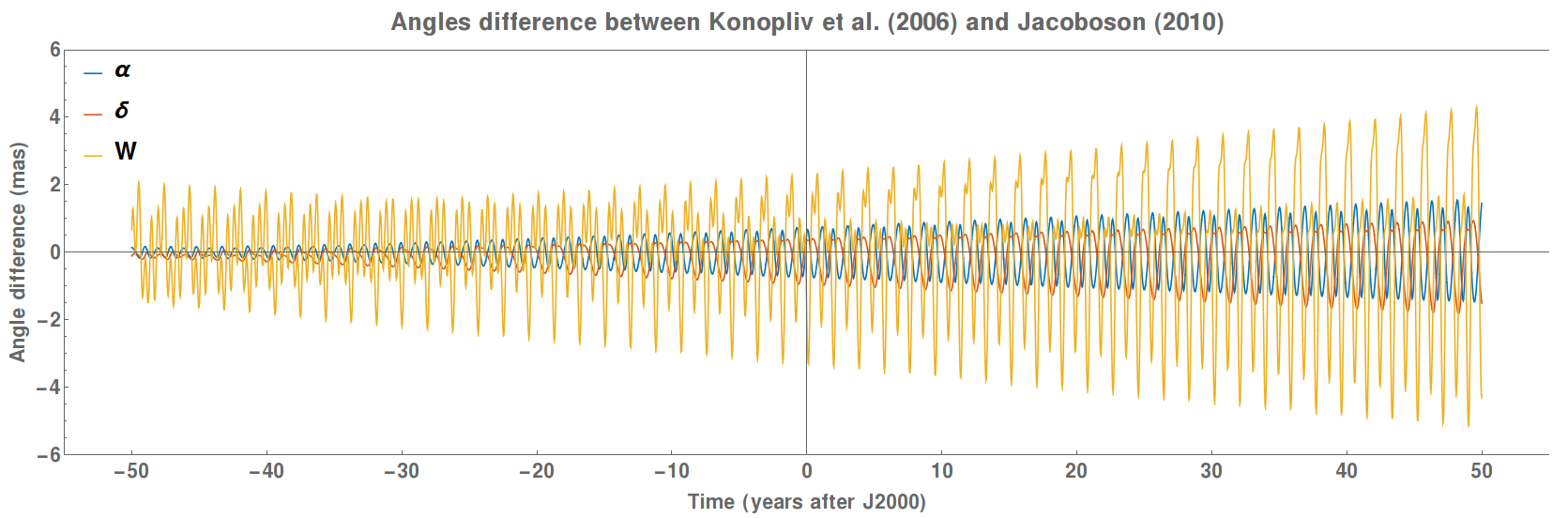} 
\caption{Temporal comparison of the IAU angles of \citet{Jac10} and the Euler angles of \citet{Kon06} converted into $\alpha, \delta, W_T$ using exact trigonometric relations.}
\label{fig_Jacob}
\end{figure}

The published Euler angles of \cite{Kuc14} are given with too low precision to allow us assessing independently the accuracy of their transformation into IAU angles\footnote{For instance the rate in obliquity is given without any decimal, but it is evident from a backward transformation that \cite{Kuc14} used a value with at least one decimal (and close to $-1.3$ mas/y) to perform the transformation to the IAU angles.}. 
\cite{Kuc14} claim a difference of about $50$ mas over an interval spanning $\pm1000$ years around the J2000 epoch. This level of difference results from the cubic signal of the transformation which is not absorbed in the fit (we obtain similar difference with our approximated transformation, where the cubic terms are neglected), and does not inform on the accuracy in a shorter period of interest around J2000. As the IAU solution of \cite{Kuc14} was computed by readjusting the amplitudes and phases of the solution of \cite{Jac10}, we assess the accuracy of \cite{Jac10} solution over $\pm 50$ years around J2000 (covering the time span of space missions) as a proxy of the accuracy of Kuchynka's transformation. We find a small difference of the order of $0.1$ to $0.2$ mas when all the  signals at short periods are discarded\footnote{We note that such a difference was computed using $\alpha_0 = 317.25168883$~deg, corrected from the published value in \citet{Jac10}, which has a typo on the third decimal. 
Note also that the constant term of $-1.4$ mas in the nutation series for $\varepsilon$ as taken in \citet{Kon06} must be considered as a correction to the epoch obliquity for such comparison.
Note finally that the IAU $W$ angle of \citet{Jac18} is likely affected by a bias of 1.2 mas because this constant term was omitted in the transformation.}.
However, when the short-period signals are considered, the difference can reach $6$ mas (see Fig.~\ref{fig_Jacob}).
We use for this comparison the values for $l_0$, $n$, and $q$ as given in the MONTE software documentation, which are essentially identical to those used by both \cite{Kon06} and \cite{Jac10} (Konopliv, personal communication). Using the same values as \cite{Kon06} and \cite{Jac10} is critical for the comparison, since a difference of just $0.02$ days in the revolution period of Mars translates into a difference of about $10$~mas in the IAU angles after $50$ years (see Fig.~\ref{fig_n}, see also Section 3.3.4 of \citealt{Bal20}). 
The numerical values chosen for the epoch mean anomaly $l_0$ and the phase angle $q$ of the nutation series of \citet{Rea79} also affect the comparison, but at a lower level (about $0.1$ mas for a phase change of $0.01$ deg).
\\

\subsection{Recommendations}
\label{Section73}

The future solution to be adopted by the WGCCRE for Mars rotation and orientation will be expressed in Earth equatorial coordinates ($\alpha, \delta, W_T$) and most likely be transformed from estimates of the different parts of Euler angles $(\varepsilon, \psi, \phi)$
obtained from a fit to the radioscience data. 
In order to guarantee both the accuracy and physical relevance, we recommend that the future WGCCRE solution uses the method described in this paper to transform the original Euler angles into IAU angles (Sections \ref{sec_transf} and \ref{sec_rotang}). \\

We thus recommend the WGCCRE to adopt a formalism that includes quadratic terms in all angles instead of the non-physical long period terms (see Section \ref{Section71}). We also recommend the use of nutation model for the rotation axis (not the figure axis) that also includes
the periodic nutations induced by Phobos and Deimos (BMAN20.1, see Section \ref{sec_rigidnut}). \\

Depending on the measurement and targeted accuracy, the future IAU orientation and rotation model adopted can be either in the form of a global model, as in Eqs.~(\ref{modealpha}-\ref{modeW}), or in the form of a local model (i.e.~without Poisson terms), as in Eqs.~(\ref{modealphalocal}-\ref{modeWlocal}), defined around J2000. 
If the targeted accuracy is $0.1$ mas, the 31 periodic nutation terms and 5 Poisson terms of the nutation theory must be included in the definition of the IAU angles. However, since the uncertainties in the angles parameters estimate is of the order of the mas\footnote{and this should not improve much even after a mission like LaRa \cite{Deh20}.} \citep{Lem23}, one could decide to relax the targeted model precision to an equivalent level of $1$ mas 30 years after the reference epoch. Guidance on this is provided in Appendix \ref{sec_appen}, where explicit expressions for a rotation model centered at J2000 and precise to the 1 mas level, derived from Eqs.~(\ref{Eqmodel}), are provided. In that case, the IAU solution would have to include the $9$ largest nutation terms and the $2$ largest Poisson terms\footnote{These terms correspond to a truncation criterion of $0.110$ mas in prograde and/or retrograde amplitude, instead of the $0.025$ mas criterion of the BMAN20.1 model. In the BMAN20.1 file available at \href{https://doi.org/10.24414/h5pn-7n71}{https://doi.org/10.24414/h5pn-7n71}, these terms are at lines 5, 6, 7, 9, 14, 19, 20, 23, and 31 for the nutations, including geodetic, Phobos and Deimos terms, and at lines 14 and 20 for the Poisson terms. }. A local model with only the largest $9$ (or $7$) nutation terms would still guaranty a precision of about $3$ (or $4$) mas after $30$ years.
\\

As a final comment, we would like to point out that the future realisation of the IAU standard model, whether it is the form of a global or a J2000 local model, could be obtained from different local rotation models centered at various epochs as obtained from a joint radioscience data analysis involving different landers and/or orbiters.
First, remind that the polynomials of all the local models are identical by construction.
Second, provided that the recommendations of Section \ref{sec_rigidnut} are followed, the different local periodic series in orientation (i.e.~nutations) would not be independent from each other, because they would be obtained from the same rigid nutation theory and from common transfer function parameters. The estimated transfer function parameters would then be used jointly with the rigid nutation theory to build nutation series (for IAU global or local models) and Poisson terms (for IAU global model only) in orientation angles.
Finally, the case for periodic variations in rotation angles is more subject to interpretation. Since the different local series for the spin periodic variations $\Delta\phi_{M \, {\bull}}^{Atm}$ would be set independently from each other (see Section \ref{sec_ampLOD}), one could choose for the IAU solution the periodic series which is thought to be the most representative (or accurate) of Mars rotation, or perform a weighted average of the different estimated series.

\section{Discussion and conclusion}

This paper presents a detailed model of Mars rotation and orientation angles. We first provide a new analytical model to transform one set of Martian orientation and rotation angles (e.g "Euler angles") into another (e.g.~"IAU angles", as defined in Section \ref{sec_definition}) and reciprocally, while introducing minimal conversion errors. Second, we explain how to build a consistent and accurate Martian rotation model. 
It is important to mention that we do not provide numerical values of a new entire rotation model. Most of the numerical values provided in the paper are for illustrative purposes only. 
\\

Because the Martian rotation and orientation parameters are determined from radioscience analysis performed with different orbitography software which use either the Euler or IAU angles, and because the present (and likely future) IAU preferred rotation model for Mars is transformed from an Euler angles rotation model, an accurate and physically sound transformation between both sets of angles is needed. That is why we have developed an analytical transformation between the orientation (Section \ref{sec_transf}) and rotation angles (Section \ref{sec_rotang}), assuming that each angle is expressed as the sum of a second degree polynomial and of periodic and Poisson series.
Exact trigonometric relations must be used to transform the angles epoch values. For the other parts, the transformation has been derived as an approximation of the exact relations between the two set of angles and is accurate to $0.1$ mas $30$ years before and after J2000 (Section \ref{Section68}). The quadratic and Poisson terms are key ingredients to achieve such a precision. Such a precision is needed to ensure that the errors related to the transformation model remain significantly lower than that coming from the data used to estimate the rotation (i.e. the numerical values of the model parameters). Indeed, the present Martian radioscience data have such a high precision ($<$2 mHz at 60s integration time on frequency measurements at about 8 GHz) that smaller and smaller physical effects can nowadays be detected, like the few mas effect of the liquid core on the nutation amplitudes \citep{Lem23}. These small geophysical effects should not be mixed up with transformation errors.   
\\

In Section \ref{sec_num}, we provided several recommendations regarding the definition of a rotation model for Mars by walking through the definition of each term of the orientation and rotation angles, pointing out past bad habits, and proposing better practices for future applications. We made a distinction between a global rotation model (see Eqs.~\ref{Eqmodel}), meant to be accurate to the $0.1$ mas level over a long time span ($\pm 30$ years around J2000) and local rotation models (see Eqs.~\ref{Eqmodellocal}, Section \ref{Sec_local_model}), meant to be accurate over a shorter time range (such as the lifetime of a space mission). Our main recommendations are the following:
\begin{itemize}
    \item In Euler angles rotation models, we recommend to consider the J2000 mean orbit of Mars as the reference plane, instead of the J1980 orbit as defined in \cite{Fol97a} at the time of Viking missions. The choice of the reference orbit has strong implications for the definition of the epoch and rate values of Euler angles (see Section \ref{Section_deg2poly}). We understand the desire for continuity with previous studies, but we now have radioscience data equally distributed with respect to J2000), which makes it the manifest reference epoch. 
    \item In both global and local rotation models, we recommend to include the quadratic terms. Omitting them in a global model leads to errors of about $10$ mas $20$ years away from J2000 in the orientation of Mars body fixed frame with respect to the stars (Section \ref{sec_app}). Omitting them in a local model defined about e.g.~the RISE mission time would spoil the determination of angles' epoch values and rates beyond tolerance (Section \ref{Section_deg2poly}). The only instance where they can be neglected is in a local model defined around J2000 ($\pm$ 3 years). The quadratic coefficients in orientation angles can be predicted within the limits of our targeted accuracy (Section \ref{Section_deg2poly}) and, as a result, can be fixed to their modeled values. 
    \item Since the Poisson terms can be as large as $4$ mas $30$ years away from J2000 (Section \ref{sec_app}), we recommend to keep them in a global rotation model. In local models, they should be merged with the periodic terms, except in the case of a local model defined around J2000, in which they would cancel by definition (Section \ref{Sec_local_model}). 
    \item We noted that for accurate applications, the rotation rates in Euler and IAU angles cannot be used in place of another (Section \ref{Sec_RotationRate}).
    \item We recommend the use of a model for the rigid nutation of the spin axis of Mars which guarantees the targeted accuracy. To the best of our knowledge, only the BMAN20.1 model, as described in Section \ref{sec_rigidnut}, and which is an update of the BMAN20 model of \cite{Bal20}, meets the criteria, thanks to an accurate definition of both the amplitudes and arguments resulting from the use of ephemerides to describe the motion of Mars and of the other planets. The model used by habits in the community, and therefore implicitly by the IAU WGCCRE, is the one of \cite{Rea79}, developed at the time of Viking missions. Although this study was an important first stone in building the Mars nutation model, we have shown that it is no longer precise enough. It only includes an handful of nutation terms at harmonics of the annual period of Mars, does not include the nutations induced by Phobos and Deimos, and is defined for the figure axis instead of the rotation axis, resulting in errors of the order of $10$ mas or more.  Regarding the function transfer parameters, we recommend the recently estimated values from \citet{Lem23}.
    \item Regarding the series for the periodic variations in rotation angles, we recommend to use the relativistic corrections of \cite{Bal23}, which update and correct the still in used corrections of \cite{Yod97}. We recommend that the nutation corrections are computed from the rigid nutation model BMAN20.1, for consistency with what precedes. This implies that the periodic series in rotation should includes $37$ terms, instead of the $6$ terms usually considered (Section \ref{sec_ampLOD}). 
    \item To locate a point at the surface (e.g.~a lander position), the rotation model must include polar motion components (see Eq.~\ref{eq_landerpos2}). The polar motion expressions are identical in both Euler and IAU formulations and consists in one part related to the atmosphere/caps dynamics and another part related to the external gravitational forcing (see Section \ref{sec_PM}). In Section \ref{sec_PMnum}, we propose numerical values for the terms due to the external forcings. Although they are one order of magnitude smaller than those related to the atmosphere, they can be greater than 2 mas in amplitude. We thus recommend to include them in the polar motion model.
    \item The input Euler angles should be provided with digits beyond the measurement uncertainties, so that the transformation error does not contribute significantly to the total error budget on the output IAU angles and that the transformation could be carried out back and forth without error. 
\end{itemize}

A particular case of rotation model is the IAU preferred model as defined by the WGCCRE, and provided in the so called "IAU angles". In section \ref{sec_IAU}, we discussed the last attempts to define such preferred model and propose recommendations to build any future reference model for the community. We showed how the previous attempts lack of physical relevance (Section \ref{Section71}) and (likely) of accuracy (Section \ref{Section72}) when Euler rotation models are transformed into IAU rotation models. This is why we recommend (Section \ref{Section73}) to use our transformation model, that includes quadratic terms, to infer the IAU angles from the estimated Euler angles. The terms with large amplitudes and arbitrary long periods included in the previously published IAU-like solutions can be easily and advantageously replaced by quadratic terms. This guarantees that the epoch value and rate of the angles are not artificially altered, and that the physical meaning of each term of the transformed solution is preserved. Depending on the targeted precision, the most compact IAU standard orientation and rotation model that we recommend should include at least 7 (better 9) nutation terms and possibly 2 Poisson terms.\\

Finally, a python routine is provided with this paper (MARRS Mars oRientation and Rotation Software, available at \href{http://gitlab-as.oma.be/mars-tools/marrs}{gitlab-as.oma.be/mars-tools/marrs}), along with an input configuration file and pre-coded numerical values from the literature.  \\

Note that the transformation developed in this paper could in theory be applied to another body for which Euler angles have to be converted into IAU angles, or vice-versa. However a careful revision of the different assumptions made here would then be needed because they may be specific to Mars.

\appendix

\section{Explicit expressions for a rotation model centered at J2000 and precise to the 1 mas level}
\label{sec_appen}

Here we put together the equations needed to define a Martian rotation model centered around J2000, in both Euler and IAU angles. Unlike the rest of the paper, a degraded precision of 1 mas is targeted here over the 1970-2030 time interval. We are looking for explicit expressions where all parameters can be replaced by estimated or modeled numerical values. We start from Eqs.~(\ref{Eqmodel}) that we reproduce below for convenience:
\begin{subequations}
\begin{alignat}{5}
\varepsilon(t) &= \varepsilon_0 && + \dot\varepsilon_0 \, t && + \varepsilon_Q \, t^2 && + \Delta\varepsilon(t) && + \varepsilon_\PO(t) \\
\psi(t) &= \psi_0 && + \dot\psi_0 \, t && + \psi_Q \, t^2 && + \Delta\psi(t) && + \psi_\PO(t)  \\
\phi_T(t) &= \phi_0 && + \dot\phi_0 \, t && + \phi^T_Q \, t^2 && + \Delta\phi_M^{Atm}(t) + \Delta\phi^{Rel}(t) - \cos\varepsilon_0 \, \Delta\psi (t) && + \phi^T_\PO(t) \\
\alpha(t) &= \alpha_0 && + \dot\alpha_0 \, t && + \alpha_Q \, t^2&& + \Delta\alpha(t) && + \alpha_\PO(t)  \\
\delta(t) &= \delta_0 && + \dot\delta_0 \, t && + \delta_Q \, t^2&& + \Delta\delta(t) && + \delta_\PO(t) \\
W_T(t) &= W_0 && + \dot W_0 \, t&& +W^T_Q \, t^2 && + \Delta\phi_M^{Atm}(t) + \Delta\phi^{Rel}(t) - \sin\delta_0 \, \Delta\alpha(t)  && + W^T_\PO(t)
\end{alignat}
\end{subequations} 
where $t$ is the time starting from J2000. The epoch values, secular rate, and quadratic coefficients are to be estimated or modeled. \\

The spin variations due to the atmosphere/caps dynamics and relativistic effects are given by:
\begin{subequations}
\begin{eqnarray}
\Delta\phi_M^{Atm}(t)   &=& \sum_{i=1}^4 (\phi^c_i \cos i\,l' + \phi^s_i \sin i\,l')\\
\Delta\phi^{Rel}(t)     &=& (-166.954 \sin l' - 7.783 \sin 2l' - 0.544 \sin 3l' \label{deltaphirel2}) \, \textrm{mas} \nonumber \\ 
&&  + \, 0.567 \, \textrm{mas} \, \sin\left(\frac{2\pi}{816.441} t + 320^\circ.997\right)  + 0.102 \, \textrm{mas} \, \sin\left(\frac{2\pi}{733.833} t + 303^\circ.752\right) \\
\mathrm{with}\ \ l'&=& 19^\circ.37276563 + 0.009145886245716337\, t
\end{eqnarray}    
\end{subequations}
Above, the time $t$ is given in days. The amplitudes $\phi^c_i$ and $\phi^s_i$ are to be estimated or modeled. The ter-annual term and the two synodic terms of $\Delta\phi^{Rel}(t)$, see Eq.~(\ref{deltaphirel2}), are below 1 mas but kept here, since taken together in the time domain they add up to the mas level, and since they are fixed by the modelling and easy to consider in the rotation model anyway.\\

The Poisson terms in $\phi$ and $W$ must be replaced by:
\begin{subequations}
\label{Eq66}
\begin{eqnarray}
\phi^T_\PO(t)   &=& \phi^M_\PO - \cos\varepsilon_0 \, \psi_\PO (t) + \sin\varepsilon_0 \, \Delta\psi (t) \, \dot\varepsilon_0 \, t \\
\nonumber W^T_\PO(t)   &=& \phi^M_\PO - \sin\delta_0 \, \alpha_\PO (t) + \sin\varepsilon_0 \, \Delta\psi (t) \, \dot\varepsilon_0 \, t +  \, \frac{\cos\beta_0 \cos^2\delta_0}{\sin\beta_0} \,\Delta\alpha (t) \, \dot\alpha_0 \, t \\
&& - \, \frac{\cos\delta_0 \sin\varepsilon_0}{\sin\beta_0} \, (\Delta\alpha (t) \, \dot\psi_0 \, t + \Delta\psi (t) \, \dot\alpha_0 \, t)
+ \, \frac{\cos\beta_0 \sin^2\varepsilon_0}{\sin\beta_0} \, \Delta\psi (t) \, \dot\psi_0 \, t 
\end{eqnarray}    
\end{subequations}
where the Poisson terms in the rotation angle expressed with respect to the mean equator of date $\phi^M_\PO$ can be set to zero as long as nothing indicates the opposite. Then, the Poisson terms in $\phi$ depend only on different parts of the Euler angles. 
Note that the term proportional to the obliquity rate $\dot\varepsilon_0$ in Eq.~(\ref{Eq66}) can be neglected in practice since the obliquity rate is expected to be zero or very close (see also the discussion between Eqs.~\ref{eq_phiTMWTM} and \ref{Eq39}). 
Given these simplifications, a compact expression for $\phi_T$ can be obtained:
\begin{equation}
\phi_T(t) = \phi_0 + \dot\phi_0 \, t + \phi^T_Q \, t^2 + \Delta\phi_M^{Atm}(t) + \Delta\phi^{Rel}(t) - \cos\varepsilon_0 \left( \Delta\psi (t) +\psi_\PO(t)\right). 
\end{equation}

The Poisson terms in $W$ depend on the different parts of both Euler and IAU angles. To obtain explicit expressions in IAU angles only, those Euler parts must be transformed into IAU angles conterparts. We leave to the interested reader this transformation, which is made explicit in the first part of the paper. 
\\

The remaining parts of the rotation model which require explicit expansions are the nutation and Poisson terms in orientation angles. In obliquity angle, they are expressed as:
\begin{subequations}
\begin{eqnarray}
\Delta\varepsilon(t) &=& \sum_i (\varepsilon'^s_i \, \sin(f_i \, t + \varphi_i^0) + \varepsilon'^c_i \, \cos(f_i \, t + \varphi_i^0) ) \\
\varepsilon_\PO(t)   &=& T \sum_i (\varepsilon'^s_{\PO,i} \, \sin(f_i \, t + \varphi_i^0) 
+ \varepsilon'^c_{\PO,i} \, \cos(f_i \, t + \varphi_i^0) ) 
\end{eqnarray}
\end{subequations}
where $T$ is the time measured in thousand of years since J2000. The series expansions for the other orientation angles $\psi$, $\alpha$ and $\delta$ are similar. The non-rigid nutation amplitudes are obtained with the equations of Section~\ref{sec_nonri} which can also be applied to the Poisson amplitudes (although in view of a targeted 1~mas precision here, the rigid Poisson terms could be used as is for simplicity). The parameters of the transfer function are to be estimated or modeled. The rigid amplitudes and arguments $(f_i \, t + \varphi_i^0)$ of the nutation and Poisson terms are given by the BMAN20.1 model. Only the 9 largest nutation terms and the 2 largest Poisson terms are needed to reach a precision of $1$ mas in the time domain. In the BMAN20.1 file available at \href{https://doi.org/10.24414/h5pn-7n71}{https://doi.org/10.24414/h5pn-7n71}, these terms are at lines $i=5, 6, 7, 9, 14, 19, 20, 23$, and $31$ for the nutations, including geodetic, Phobos and Deimos terms, and at lines 14 and 20 for the Poisson terms. We list them here in Tables \ref{nutations} and \ref{Poisson}. The arguments are multiple of the mean longitude of Mars and of the nodes of Phobos and Deimos:
\begin{subequations}
\begin{eqnarray}
 M\!a&=&6.20349959869 + 3340.6124347175 \, T,\\
 N_{P\!h}&=&2.13055663363-2779.4193805084 \, T,\\
 N_{D\!e}&=&0.20283841509-114.7466716724 \, T.
\end{eqnarray}    
\end{subequations}
where $T$ is again the time measured in thousand of years since J2000. Note that the geodetic term ($i=19$) is not affected by the transfer function.

\begin{table}[!ht]
\caption{Largest periodic terms of the BMAN20.1 nutations series, in Euler (longitude/obliquity) and IAU (right ascension/declination) representation. The first column gives the position of each term in the BMAN20.1 nutation series. The arguments are multiple of the mean longitude of Mars and of the nodes of Phobos and Deimos. The periods (in days) are here rounded for convenience. The amplitudes are given in mas.}
\label{nutations}
\begin{center}
\begin{tabular}{lrrrrrrrrrr} 
\hline\noalign{\smallskip}
 $i$ & $(f_i t+\varphi_i^0)$ & $2\pi/f_i$(d) & $\psi^c_i$ & $\psi^s_i$ & $\varepsilon^c_i$ & $\varepsilon^s_i$ & $\alpha^c_i$ & $\alpha^s_i$ & $\delta^c_i$ & $\delta^s_i$  \\
\noalign{\smallskip}
\hline\noalign{\smallskip}
5 & 6 $M\!a$ &  114.5  &   -0.898   &  0.255  &  0.118  & 0.421  &  -0.327  &  0.609  & -0.348  & -0.232\\
6 & 5 $M\!a$ &  137.4 &    -6.291  &  -0.889  & -0.429 &  2.941  & -3.719  &  2.883  & -1.523  & -2.402\\
7 & 4 $M\!a$ &  171.7  &   -34.995  & -21.764 & -10.257  &16.268  & -29.628   & 7.289  & -2.734  &-18.197\\
9 & 3 $M\!a$  &  229.0 &  -137.874 & -200.892  &-93.903 & 63.037 & -177.469 & -31.648  & 28.191 &-104.503\\
14 & 2 $M\!a$ & 343.5   &  -222.354 &-1113.594 &-509.803 & 89.074 & -693.124 &-471.061 & 306.499& -389.642\\
19 & $M\!a$ & 687.0    &     0.229 &    0.516  &  0.000 &  0.000  &   0.118  &  0.265  &  0.067   & 0.151\\
20 & $M\!a$ & 687.0   &   -283.816 & -480.020  & 47.893 & 11.968  & -91.453 &-233.061 &-117.656 &-148.707\\
23 &-$N_{P\!h} $ & 825.7   &    0.000  &  10.126  & -4.310 &  0.000  & -4.894 &   5.203 &   3.139  &  2.953\\
31 &-$N_{D\!e} $ & 20000.0   &     0.000  &   3.532  & -1.503  & 0.000   & -1.707 &   1.815   & 1.095  &  1.030\\
\noalign{\smallskip}\hline
\end{tabular}
\end{center}
\end{table}

\begin{table}[!ht]
\caption{Largest Poisson terms of the BMAN20.1 nutations series, in Euler (longitude/obliquity) and IAU (right ascension/declination) representation. The first column gives the position of each term in the BMAN20.1 Poisson series. The arguments are multiple of the mean longitude of Mars. The periods (in days) are here rounded for convenience. The amplitudes are given in mas/ky and are to be multiplied by $T$, the time measured in thousand of years since J2000. }
\label{Poisson}
\begin{center}
\begin{tabular}{lrrrrrrrrrr} 
\hline\noalign{\smallskip}
 $i$ & $(f_i t+\varphi_i^0)$ & $2\pi/f_i$(d) & $\psi^c_{\PO,i}$ & $\psi^s_{\PO,i}$ & $\varepsilon^c_{\PO,i}$ & $\varepsilon^s_{\PO,i}$ & $\alpha^c_{\PO,i}$ & $\alpha^s_{\PO,i}$ & $\delta^c_{\PO,i}$ & $\delta^s_{\PO,i}$ \\
\noalign{\smallskip}
\hline\noalign{\smallskip}
14 & 2 $M\!a$ &    343.5    &   -75.785   &  4.642  &  4.397 & 37.443   & -14.819 &  39.804 & -17.667 & -20.729\\
20 & $M\!a$   &    687.0    &    56.596   &-22.641  &  2.620 &-6.712   &  29.795 & -20.443 &  15.605 &   0.855\\
\noalign{\smallskip}\hline
\end{tabular}
\end{center}
\end{table}

\section*{Acknowledgments}
We thank Alex Konopliv for his constructive comments and suggestions that have helped to improve our paper. This work was financially supported by the Belgian PRODEX program managed 
by the European Space Agency in collaboration with the Belgian Federal 
Science Policy Office. 
This is InSight contribution ICN 323. 

\bibliography{MARRSpaper_rev2} 

\begin{thebibliography}{38}
\providecommand{\natexlab}[1]{#1}
\providecommand{\url}[1]{\texttt{#1}}
\expandafter\ifx\csname urlstyle\endcsname\relax
  \providecommand{\doi}[1]{doi: #1}\else
  \providecommand{\doi}{doi: \begingroup \urlstyle{rm}\Url}\fi

\bibitem[{Archinal} et~al.(2018){Archinal}, {Acton}, {A'Hearn}, {Conrad},
  {Consolmagno}, {Duxbury}, {Hestroffer}, {Hilton}, {Kirk}, {Klioner},
  {McCarthy}, {Meech}, {Oberst}, {Ping}, {Seidelmann}, {Tholen}, {Thomas}, and
  {Williams}]{Arc18}
B.~A. {Archinal}, C.~H. {Acton}, M.~F. {A'Hearn}, A.~{Conrad}, G.~J.
  {Consolmagno}, T.~{Duxbury}, D.~{Hestroffer}, J.~L. {Hilton}, R.~L. {Kirk},
  S.~A. {Klioner}, D.~{McCarthy}, K.~{Meech}, J.~{Oberst}, J.~{Ping}, P.~K.
  {Seidelmann}, D.~J. {Tholen}, P.~C. {Thomas}, and I.~P. {Williams}.
\newblock {Report of the IAU Working Group on Cartographic Coordinates and
  Rotational Elements: 2015}.
\newblock \emph{Celestial Mechanics and Dynamical Astronomy}, 130\penalty0
  (3):\penalty0 22, Feb. 2018.
\newblock \doi{10.1007/s10569-017-9805-5}.

\bibitem[{Baland} et~al.(2020){Baland}, {Yseboodt}, {Le Maistre}, {Rivoldini},
  {Van Hoolst}, and {Dehant}]{Bal20}
R.-M. {Baland}, M.~{Yseboodt}, S.~{Le Maistre}, A.~{Rivoldini}, T.~{Van
  Hoolst}, and V.~{Dehant}.
\newblock {The precession and nutations of a rigid Mars}.
\newblock \emph{Celestial Mechanics and Dynamical Astronomy}, 132\penalty0
  (9):\penalty0 47, Sept. 2020.
\newblock \doi{10.1007/s10569-020-09986-0}.

\bibitem[{Baland} et~al.(2023){Baland}, {Hees}, {Yseboodt}, {Bourgoin}, and {Le
  Maistre}]{Bal23}
R.-M. {Baland}, A.~{Hees}, M.~{Yseboodt}, A.~{Bourgoin}, and S.~{Le Maistre}.
\newblock {Relativistic contributions to the rotation of Mars}.
\newblock \emph{Astronomy and Astrophysics}, 670:\penalty0 A29, Feb. 2023.
\newblock \doi{10.1051/0004-6361/202244420}.

\bibitem[{Banerdt} et~al.(2020){Banerdt}, {Smrekar}, {Banfield}, {Giardini},
  {Golombek}, {Johnson}, {Lognonn{\'e}}, {Spiga}, {Spohn}, {Perrin},
  {St{\"a}hler}, {Antonangeli}, {Asmar}, {Beghein}, {Bowles}, {Bozdag}, {Chi},
  {Christensen}, {Clinton}, {Collins}, {Daubar}, {Dehant}, {Drilleau},
  {Fillingim}, {Folkner}, {Garcia}, {Garvin}, {Grant}, {Grott}, {Grygorczuk},
  {Hudson}, {Irving}, {Kargl}, {Kawamura}, {Kedar}, {King}, {Knapmeyer-Endrun},
  {Knapmeyer}, {Lemmon}, {Lorenz}, {Maki}, {Margerin}, {McLennan}, {Michaut},
  {Mimoun}, {Mittelholz}, {Mocquet}, {Morgan}, {Mueller}, {Murdoch},
  {Nagihara}, {Newman}, {Nimmo}, {Panning}, {Pike}, {Plesa}, {Rodriguez},
  {Rodriguez-Manfredi}, {Russell}, {Schmerr}, {Siegler}, {Stanley},
  {Stutzmann}, {Teanby}, {Tromp}, {van Driel}, {Warner}, {Weber}, and
  {Wieczorek}]{Ban20}
W.~B. {Banerdt}, S.~E. {Smrekar}, D.~{Banfield}, D.~{Giardini}, M.~{Golombek},
  C.~L. {Johnson}, P.~{Lognonn{\'e}}, A.~{Spiga}, T.~{Spohn}, C.~{Perrin},
  S.~C. {St{\"a}hler}, D.~{Antonangeli}, S.~{Asmar}, C.~{Beghein}, N.~{Bowles},
  E.~{Bozdag}, P.~{Chi}, U.~{Christensen}, J.~{Clinton}, G.~S. {Collins},
  I.~{Daubar}, V.~{Dehant}, M.~{Drilleau}, M.~{Fillingim}, W.~{Folkner}, R.~F.
  {Garcia}, J.~{Garvin}, J.~{Grant}, M.~{Grott}, J.~{Grygorczuk}, T.~{Hudson},
  J.~C.~E. {Irving}, G.~{Kargl}, T.~{Kawamura}, S.~{Kedar}, S.~{King},
  B.~{Knapmeyer-Endrun}, M.~{Knapmeyer}, M.~{Lemmon}, R.~{Lorenz}, J.~N.
  {Maki}, L.~{Margerin}, S.~M. {McLennan}, C.~{Michaut}, D.~{Mimoun},
  A.~{Mittelholz}, A.~{Mocquet}, P.~{Morgan}, N.~T. {Mueller}, N.~{Murdoch},
  S.~{Nagihara}, C.~{Newman}, F.~{Nimmo}, M.~{Panning}, W.~T. {Pike}, A.-C.
  {Plesa}, S.~{Rodriguez}, J.~A. {Rodriguez-Manfredi}, C.~T. {Russell},
  N.~{Schmerr}, M.~{Siegler}, S.~{Stanley}, E.~{Stutzmann}, N.~{Teanby},
  J.~{Tromp}, M.~{van Driel}, N.~{Warner}, R.~{Weber}, and M.~{Wieczorek}.
\newblock {Initial results from the InSight mission on Mars}.
\newblock \emph{Nature Geoscience}, 13\penalty0 (3):\penalty0 183--189, Feb.
  2020.
\newblock \doi{10.1038/s41561-020-0544-y}.

\bibitem[{Borderies}(1980)]{Borderies1980a}
N.~{Borderies}.
\newblock {Theory of Mars rotation in Euler angles}.
\newblock \emph{\aap}, 82:\penalty0 129--141, Feb. 1980.

\bibitem[{Borderies} et~al.(1980){Borderies}, {Balmino}, {Castel}, and
  {Moynot}]{Borderies1980b}
N.~{Borderies}, G.~{Balmino}, L.~{Castel}, and B.~{Moynot}.
\newblock {Study of Mars dynamics from lander tracking data analysis}.
\newblock \emph{Moon and Planets}, 22:\penalty0 191--200, Apr. 1980.
\newblock \doi{10.1007/BF00898430}.

\bibitem[{Bouquillon} and {Souchay}(1999)]{Bou99}
S.~{Bouquillon} and J.~{Souchay}.
\newblock {Precise modeling of the precession-nutation of Mars}.
\newblock \emph{\aap}, 345:\penalty0 282--297, May 1999.

\bibitem[{Defraigne} et~al.(2000){Defraigne}, {de Viron}, {Dehant}, {Van
  Hoolst}, and {Hourdin}]{Def00}
P.~{Defraigne}, O.~{de Viron}, V.~{Dehant}, T.~{Van Hoolst}, and F.~{Hourdin}.
\newblock {Mars rotation variations induced by atmosphere and ice caps}.
\newblock \emph{\jgr}, 105\penalty0 (E10):\penalty0 24563--24570, Oct. 2000.
\newblock \doi{10.1029/1999JE001227}.

\bibitem[{Dehant} et~al.(2020){Dehant}, {Le Maistre}, {Baland}, {Bergeot},
  {Karatekin}, {P{\'e}ters}, {Rivoldini}, {Ruiz Lozano}, {Temel}, {Van Hoolst},
  {Yseboodt}, {Mitrovic}, {Kosov}, {Valenta}, {Thomassen}, {Karki}, {Al
  Khalifeh}, {Craeye}, {Gurvits}, {Marty}, {Asmar}, {Folkner}, and {LaRa
  Team}]{Deh20}
V.~{Dehant}, S.~{Le Maistre}, R.-M. {Baland}, N.~{Bergeot}, {\"O}.~{Karatekin},
  M.-J. {P{\'e}ters}, A.~{Rivoldini}, L.~{Ruiz Lozano}, O.~{Temel}, T.~{Van
  Hoolst}, M.~{Yseboodt}, M.~{Mitrovic}, A.~S. {Kosov}, V.~{Valenta},
  L.~{Thomassen}, S.~{Karki}, K.~{Al Khalifeh}, C.~{Craeye}, L.~I. {Gurvits},
  J.-C. {Marty}, S.~W. {Asmar}, W.~M. {Folkner}, and {LaRa Team}.
\newblock {The radioscience LaRa instrument onboard ExoMars 2020 to investigate
  the rotation and interior of mars}.
\newblock \emph{\planss}, 180:\penalty0 104776, Jan. 2020.
\newblock \doi{10.1016/j.pss.2019.104776}.

\bibitem[{Evans} et~al.(2018){Evans}, {Taber}, {Drain}, {Smith}, {Wu},
  {Guevara}, {Sunseri}, and {Evans}]{Eva18}
S.~{Evans}, W.~{Taber}, T.~{Drain}, J.~{Smith}, H.-C. {Wu}, M.~{Guevara},
  R.~{Sunseri}, and J.~{Evans}.
\newblock {MONTE: the next generation of mission design and navigation
  software}.
\newblock \emph{CEAS Space Journal}, 10\penalty0 (1):\penalty0 79--86, Mar.
  2018.
\newblock \doi{10.1007/s12567-017-0171-7}.

\bibitem[{Folkner} et~al.(1997{\natexlab{a}}){Folkner}, {Kahn}, {Preston},
  {Yoder}, {Standish}, {Williams}, {Edwards}, {Hellings}, {Eubanks}, and
  {Bills}]{Fol97a}
W.~M. {Folkner}, R.~D. {Kahn}, R.~A. {Preston}, C.~F. {Yoder}, E.~M.
  {Standish}, J.~G. {Williams}, C.~D. {Edwards}, R.~W. {Hellings}, T.~M.
  {Eubanks}, and B.~G. {Bills}.
\newblock {Mars dynamics from Earth-based tracking of the Mars Pathfinder
  lander}.
\newblock \emph{\jgr}, 102\penalty0 (E2):\penalty0 4057--4064, Jan.
  1997{\natexlab{a}}.
\newblock \doi{10.1029/96JE02125}.

\bibitem[{Folkner} et~al.(1997{\natexlab{b}}){Folkner}, {Yoder}, {Yuan},
  {Standish}, and {Preston}]{Fol97b}
W.~M. {Folkner}, C.~F. {Yoder}, D.~N. {Yuan}, E.~M. {Standish}, and R.~A.
  {Preston}.
\newblock {Interior Structure and Seasonal Mass Redistribution of Mars from
  Radio Tracking of Mars Pathfinder}.
\newblock \emph{Science}, 278:\penalty0 1749, Dec. 1997{\natexlab{b}}.
\newblock \doi{10.1126/science.278.5344.1749}.

\bibitem[{Folkner} et~al.(2018){Folkner}, {Dehant}, {Le Maistre}, {Yseboodt},
  {Rivoldini}, {Van Hoolst}, {Asmar}, and {Golombek}]{Fol18}
W.~M. {Folkner}, V.~{Dehant}, S.~{Le Maistre}, M.~{Yseboodt}, A.~{Rivoldini},
  T.~{Van Hoolst}, S.~W. {Asmar}, and M.~P. {Golombek}.
\newblock {The Rotation and Interior Structure Experiment on the InSight
  Mission to Mars}.
\newblock \emph{\ssr}, 214\penalty0 (5):\penalty0 100, Aug. 2018.
\newblock \doi{10.1007/s11214-018-0530-5}.

\bibitem[{Groten} et~al.(1996){Groten}, {Molodenski}, and {Zharkov}]{Groten}
E.~{Groten}, S.~M. {Molodenski}, and V.~N. {Zharkov}.
\newblock {On the Theory of Mars' Forced Nutation}.
\newblock \emph{\aj}, 111:\penalty0 1388, Mar. 1996.
\newblock \doi{10.1086/117885}.

\bibitem[{Hilton}(1991)]{Hilton}
J.~L. {Hilton}.
\newblock {The motion of Mars' pole. I - Rigid body precession and nutation}.
\newblock \emph{\aj}, 102:\penalty0 1510--1527, Oct. 1991.
\newblock \doi{10.1086/115977}.

\bibitem[{Jacobson}(2010)]{Jac10}
R.~A. {Jacobson}.
\newblock {The Orbits and Masses of the Martian Satellites and the Libration of
  Phobos}.
\newblock \emph{\aj}, 139\penalty0 (2):\penalty0 668--679, Feb. 2010.
\newblock \doi{10.1088/0004-6256/139/2/668}.

\bibitem[{Jacobson} et~al.(2018){Jacobson}, {Konopliv}, {Park}, and
  {Folkner}]{Jac18}
R.~A. {Jacobson}, A.~S. {Konopliv}, R.~S. {Park}, and W.~M. {Folkner}.
\newblock {The rotational elements of Mars and its satellites}.
\newblock \emph{\planss}, 152:\penalty0 107--115, Mar. 2018.
\newblock \doi{10.1016/j.pss.2017.12.020}.

\bibitem[{Kahan} et~al.(2021){Kahan}, {Folkner}, {Buccino}, {Dehant}, {Le
  Maistre}, {Rivoldini}, {Van Hoolst}, {Yseboodt}, and {Marty}]{Kah21}
D.~S. {Kahan}, W.~M. {Folkner}, D.~R. {Buccino}, V.~{Dehant}, S.~{Le Maistre},
  A.~{Rivoldini}, T.~{Van Hoolst}, M.~{Yseboodt}, and J.~C. {Marty}.
\newblock {Mars precession rate determined from radiometric tracking of the
  InSight Lander}.
\newblock \emph{\planss}, 199:\penalty0 105208, May 2021.
\newblock \doi{10.1016/j.pss.2021.105208}.

\bibitem[{Karatekin} et~al.(2006){Karatekin}, {Van Hoolst}, and
  {Dehant}]{Kar06}
{\"O}.~{Karatekin}, T.~{Van Hoolst}, and V.~{Dehant}.
\newblock {Martian global-scale CO$_{2}$ exchange from time-variable gravity
  measurements}.
\newblock \emph{Journal of Geophysical Research (Planets)}, 111\penalty0
  (E6):\penalty0 E06003, June 2006.
\newblock \doi{10.1029/2005JE002591}.

\bibitem[{Karatekin} et~al.(2011){Karatekin}, {de Viron}, {Lambert}, {Dehant},
  {Rosenblatt}, {Van Hoolst}, and {Le Maistre}]{Kar11}
{\"O}.~{Karatekin}, O.~{de Viron}, S.~{Lambert}, V.~{Dehant}, P.~{Rosenblatt},
  T.~{Van Hoolst}, and S.~{Le Maistre}.
\newblock {Atmospheric angular momentum variations of Earth, Mars and Venus at
  seasonal time scales}.
\newblock \emph{\planss}, 59\penalty0 (10):\penalty0 923--933, Aug. 2011.
\newblock \doi{10.1016/j.pss.2010.09.010}.

\bibitem[{Konopliv} et~al.(2006){Konopliv}, {Yoder}, {Standish}, {Yuan}, and
  {Sjogren}]{Kon06}
A.~S. {Konopliv}, C.~F. {Yoder}, E.~M. {Standish}, D.-N. {Yuan}, and W.~L.
  {Sjogren}.
\newblock {A global solution for the Mars static and seasonal gravity, Mars
  orientation, Phobos and Deimos masses, and Mars ephemeris}.
\newblock \emph{\icarus}, 182\penalty0 (1):\penalty0 23--50, May 2006.
\newblock \doi{10.1016/j.icarus.2005.12.025}.

\bibitem[{Konopliv} et~al.(2011){Konopliv}, {Asmar}, {Folkner}, {Karatekin},
  {Nunes}, {Smrekar}, {Yoder}, and {Zuber}]{Kon11}
A.~S. {Konopliv}, S.~W. {Asmar}, W.~M. {Folkner}, {\"O}.~{Karatekin}, D.~C.
  {Nunes}, S.~E. {Smrekar}, C.~F. {Yoder}, and M.~T. {Zuber}.
\newblock {Mars high resolution gravity fields from MRO, Mars seasonal gravity,
  and other dynamical parameters}.
\newblock \emph{\icarus}, 211\penalty0 (1):\penalty0 401--428, Jan. 2011.
\newblock \doi{10.1016/j.icarus.2010.10.004}.

\bibitem[{Konopliv} et~al.(2016){Konopliv}, {Park}, and {Folkner}]{Kon16}
A.~S. {Konopliv}, R.~S. {Park}, and W.~M. {Folkner}.
\newblock {An improved JPL Mars gravity field and orientation from Mars orbiter
  and lander tracking data}.
\newblock \emph{\icarus}, 274:\penalty0 253--260, Aug. 2016.
\newblock \doi{10.1016/j.icarus.2016.02.052}.

\bibitem[{Konopliv} et~al.(2020){Konopliv}, {Park}, {Rivoldini}, {Baland}, {Le
  Maistre}, {Van Hoolst}, {Yseboodt}, and {Dehant}]{Kon20}
A.~S. {Konopliv}, R.~S. {Park}, A.~{Rivoldini}, R.-M. {Baland}, S.~{Le
  Maistre}, T.~{Van Hoolst}, M.~{Yseboodt}, and V.~{Dehant}.
\newblock {Detection of the Chandler Wobble of Mars From Orbiting Spacecraft}.
\newblock \emph{\grl}, 47\penalty0 (21):\penalty0 e90568, Nov. 2020.
\newblock \doi{10.1029/2020GL090568}.

\bibitem[{Kuchynka} et~al.(2014){Kuchynka}, {Folkner}, {Konopliv}, {Parker},
  {Park}, {Le Maistre}, and {Dehant}]{Kuc14}
P.~{Kuchynka}, W.~M. {Folkner}, A.~S. {Konopliv}, T.~J. {Parker}, R.~S. {Park},
  S.~{Le Maistre}, and V.~{Dehant}.
\newblock {New constraints on Mars rotation determined from radiometric
  tracking of the Opportunity Mars Exploration Rover}.
\newblock \emph{\icarus}, 229:\penalty0 340--347, Feb. 2014.
\newblock \doi{10.1016/j.icarus.2013.11.015}.

\bibitem[{Le Maistre}(2013)]{Lem13}
S.~{Le Maistre}.
\newblock \emph{The rotation of Mars and Phobos from Earth-based radio-tracking
  observations of a lander}.
\newblock Phd dissertation, available on dial.uclouvain.be, Universit\'e
  Catholique de Louvain, Belgium, 2013.

\bibitem[{Le Maistre} et~al.(2012){Le Maistre}, {Rosenblatt}, {Rivoldini},
  {Dehant}, {Marty}, and {Karatekin}]{Lem12}
S.~{Le Maistre}, P.~{Rosenblatt}, A.~{Rivoldini}, V.~{Dehant}, J.-C. {Marty},
  and O.~{Karatekin}.
\newblock {Lander radio science experiment with a direct link between Mars and
  the Earth}.
\newblock \emph{\planss}, 68\penalty0 (1):\penalty0 105--122, Aug. 2012.
\newblock \doi{10.1016/j.pss.2011.12.020}.

\bibitem[{Le Maistre} et~al.(2022){Le Maistre}, {Dehant}, {Baland}, {Beuthe},
  {Caldiero}, {Filice}, {Goli}, {P{\'e}ters}, {Steenput}, {Rivoldini},
  {{\"U}mit}, {Van Hoolst}, and {Yseboodt}]{Lem22}
S.~{Le Maistre}, V.~{Dehant}, R.-M. {Baland}, M.~{Beuthe}, A.~{Caldiero},
  V.~{Filice}, M.~{Goli}, M.-J. {P{\'e}ters}, B.~{Steenput}, A.~{Rivoldini},
  E.~{{\"U}mit}, T.~{Van Hoolst}, and M.~{Yseboodt}.
\newblock {LaRa, an X-band coherent transponder ready to fly}.
\newblock In \emph{European Planetary Science Congress}, pages EPSC2022--1169,
  Sept. 2022.
\newblock \doi{10.5194/epsc2022-1169}.

\bibitem[{Le Maistre} et~al.(2023){Le Maistre}, {Rivoldini}, {Caldiero},
  {Yseboodt}, {Baland}, {Beuthe}, {Van Hoolst}, {Dehant}, {Folkner}, {Buccino},
  {Kahan}, {Marty}, {Antonangeli}, {Badro}, {Drilleau}, {Konopliv}, {Péters},
  {Plesa}, {Samuel}, {Tosi}, {Wieczorek}, {Lognonné}, {Panning}, {Smrekar},
  and {Banerdt}]{Lem23}
S.~{Le Maistre}, A.~{Rivoldini}, A.~{Caldiero}, M.~{Yseboodt}, R.-M. {Baland},
  M.~{Beuthe}, T.~{Van Hoolst}, V.~{Dehant}, W.~{Folkner}, D.~{Buccino},
  D.~{Kahan}, J.-C. {Marty}, D.~{Antonangeli}, J.~{Badro}, M.~{Drilleau},
  A.~{Konopliv}, M.-J. {Péters}, A.-C. {Plesa}, H.~{Samuel}, N.~{Tosi},
  M.~{Wieczorek}, P.~{Lognonné}, M.~{Panning}, S.~{Smrekar}, and B.~{Banerdt}.
\newblock {Spin state and deep interior structure of Mars from InSight radio
  tracking}.
\newblock \emph{Nature}, 2023.
\newblock \doi{s41586-023-06150-0}.

\bibitem[{Marty} et~al.(2011){Marty}, {Loyer}, {Perosanz}, {Mercier},
  {Bracher}, {Legresy}, {Portier}, {Capdeville}, {Fund}, and {Lemoine}]{Mar11}
J.~{Marty}, S.~{Loyer}, F.~{Perosanz}, F.~{Mercier}, G.~{Bracher},
  B.~{Legresy}, L.~{Portier}, H.~{Capdeville}, F.~{Fund}, and J.~{Lemoine}.
\newblock Gins: The cnes/grgs gnss scientific software.
\newblock In \emph{3rd Int. Coll. Sci. Fundam. Asp. Galileo Program}, ESA
  proceedings WPP326 31, pages 8--10, 2011.

\bibitem[{Reasenberg} and {King}(1979)]{Rea79}
R.~D. {Reasenberg} and R.~W. {King}.
\newblock {The rotation of Mars.}
\newblock \emph{\jgr}, 84:\penalty0 6231--6240, Oct. 1979.
\newblock \doi{10.1029/JB084iB11p06231}.

\bibitem[{Roosbeek}(1999)]{Roo99}
F.~{Roosbeek}.
\newblock {Analytical Developments of Rigid Mars Nutation and Tide Generating
  Potential Series}.
\newblock \emph{Celestial Mechanics and Dynamical Astronomy}, 75:\penalty0
  287--300, Jan. 1999.

\bibitem[{Sasao} et~al.(1980){Sasao}, {Okubo}, and {Saito}]{Sas80}
T.~{Sasao}, S.~{Okubo}, and M.~{Saito}.
\newblock {A Simple Theory on Dynamical Effects of Stratified Fluid Core upon
  Nutational Motion of the Earth}.
\newblock In R.~L. {Duncombe}, editor, \emph{Nutation and the Earth's
  Rotation}, volume~78, page 165, Jan. 1980.

\bibitem[{Simon} et~al.(2013){Simon}, {Francou}, {Fienga}, and {Manche}]{Sim13}
J.~L. {Simon}, G.~{Francou}, A.~{Fienga}, and H.~{Manche}.
\newblock {New analytical planetary theories VSOP2013 and TOP2013}.
\newblock \emph{\aap}, 557:\penalty0 A49, Sept. 2013.
\newblock \doi{10.1051/0004-6361/201321843}.

\bibitem[{Van den Acker} et~al.(2002){Van den Acker}, {Van Hoolst}, {de Viron},
  {Defraigne}, {Forget}, {Hourdin}, and {Dehant}]{Van02}
E.~{Van den Acker}, T.~{Van Hoolst}, O.~{de Viron}, P.~{Defraigne},
  F.~{Forget}, F.~{Hourdin}, and V.~{Dehant}.
\newblock {Influence of the seasonal winds and the CO$_{2}$ mass exchange
  between atmosphere and polar caps on Mars' rotation}.
\newblock \emph{Journal of Geophysical Research (Planets)}, 107\penalty0
  (E7):\penalty0 5055, July 2002.
\newblock \doi{10.1029/2000JE001539}.

\bibitem[{Yoder} and {Standish}(1997)]{Yod97}
C.~F. {Yoder} and E.~M. {Standish}.
\newblock {Martian precession and rotation from Viking lander range data}.
\newblock \emph{\jgr}, 102\penalty0 (E2):\penalty0 4065--4080, Jan. 1997.
\newblock \doi{10.1029/96JE03642}.

\bibitem[{Yoder} et~al.(2003){Yoder}, {Konopliv}, {Yuan}, {Standish}, and
  {Folkner}]{Yod03}
C.~F. {Yoder}, A.~S. {Konopliv}, D.~N. {Yuan}, E.~M. {Standish}, and W.~M.
  {Folkner}.
\newblock {Fluid Core Size of Mars from Detection of the Solar Tide}.
\newblock \emph{Science}, 300\penalty0 (5617):\penalty0 299--303, Apr. 2003.
\newblock \doi{10.1126/science.1079645}.

\bibitem[{Yseboodt} et~al.(2017){Yseboodt}, {Dehant}, and {P{\'e}ters}]{Yse17}
M.~{Yseboodt}, V.~{Dehant}, and M.-J. {P{\'e}ters}.
\newblock {Signatures of the Martian rotation parameters in the Doppler and
  range observables}.
\newblock \emph{\planss}, 144:\penalty0 74--88, Sept. 2017.
\newblock \doi{10.1016/j.pss.2017.05.008}.

\end{thebibliography}

\end{document}